%% file: New_BANYAN.tex
\documentclass[twocolumn]{aastex61}
\usepackage{amsmath,color,textcomp,url,graphicx,subfigure}
\usepackage{morefloats}

\newif\ifdisplaylongtables
\displaylongtablesfalse

\newcommand{\nasso}{27}
\newcommand{\kms}{\hbox{km\,s$^{-1}$}}
\newcommand{\mjup}{$M_{\mathrm{Jup}}$}

\newcommand{\msol}{$M_{\odot}$}
\newcommand{\masyr}{$\mathrm{mas}\,\mathrm{yr}^{-1}$}

\DeclareMathOperator\erfc{erfc}

\usepackage{color}
\definecolor{myred}{RGB}{200,0,0}

\received{October 26, 2017}
\revised{January 12, 2018}
\accepted{January 26, 2018}

\submitjournal{ApJS}

\shorttitle{THE BANYAN~$\Sigma$ ALGORITHM TO IDENTIFY YOUNG ASSOCIATION MEMBERS}
\shortauthors{Gagn\'e et al.}

\begin{document}

\title{BANYAN. XI. THE BANYAN~$\Sigma$ MULTIVARIATE BAYESIAN ALGORITHM TO IDENTIFY MEMBERS OF YOUNG ASSOCIATIONS WITHIN 150\,\lowercase{pc}}

\author[0000-0002-2592-9612]{Jonathan Gagn\'e}
\affiliation{Carnegie Institution of Washington DTM, 5241 Broad Branch Road NW, Washington, DC~20015, USA}
\affiliation{NASA Sagan Fellow}
\email{jgagne@carnegiescience.edu}
\author[0000-0003-2008-1488]{Eric E. Mamajek}
\affiliation{Jet Propulsion Laboratory, California Institute of Technology, 4800 Oak Grove Drive, Pasadena, CA 91109, USA}
\affiliation{Department of Physics and Astronomy, University of Rochester, Rochester, NY 14627, USA}
\author[0000-0002-8786-8499]{Lison Malo}
\affil{Institute for Research on Exoplanets, Universit\'e de Montr\'eal, D\'epartement de Physique, C.P.~6128 Succ. Centre-ville, Montr\'eal, QC H3C~3J7, Canada}
\author[0000-0003-1645-8596]{Adric Riedel}
\affil{Space Telescope Science Institute, 3700 San Martin Drive, Baltimore, MD 21218, USA}
\author[0000-0003-1286-5231]{David Rodriguez}
\affil{Space Telescope Science Institute, 3700 San Martin Drive, Baltimore, MD 21218, USA}
\author[0000-0002-6780-4252]{David Lafreni\` ere}
\affil{Institute for Research on Exoplanets, Universit\'e de Montr\'eal, D\'epartement de Physique, C.P.~6128 Succ. Centre-ville, Montr\'eal, QC H3C~3J7, Canada}
\author[0000-0001-6251-0573]{Jacqueline K. Faherty}
\affiliation{Department of Astrophysics, American Museum of Natural History, Central Park West at 79th St., New York, NY 10024, USA}
\author[0000-0003-2005-5626]{Olivier Roy-Loubier}
\affil{Institute for Research on Exoplanets, Universit\'e de Montr\'eal, D\'epartement de Physique, C.P.~6128 Succ. Centre-ville, Montr\'eal, QC H3C~3J7, Canada}
\author[0000-0003-3818-408X]{Laurent Pueyo}
\affil{Space Telescope Science Institute, 3700 San Martin Drive, Baltimore, MD 21218, USA}
\author[0000-0001-8654-9499]{Annie C. Robin}
\affil{Institut UTINAM, OSU THETA, Univ. Bourgogne Franche-Comt\'e, Besan\c{c}on, France}
\author[0000-0001-5485-4675]{Ren\'e Doyon}
\affil{Institute for Research on Exoplanets, Universit\'e de Montr\'eal, D\'epartement de Physique, C.P.~6128 Succ. Centre-ville, Montr\'eal, QC H3C~3J7, Canada}

\begin{abstract}

BANYAN~$\Sigma$ is a new Bayesian algorithm to identify members of young stellar associations within 150\,pc of the Sun. It includes \nasso\ young associations with ages in the range $\sim$\,1--800\,Myr, modelled with multivariate Gaussians in 6-dimensional $XYZUVW$ space. It is the first such multi-association classification tool to include the nearest sub-groups of the Sco-Cen OB star-forming region, the IC~2602, IC~2391, Pleiades and Platais~8 clusters, and the $\rho$~Ophiuchi, Corona Australis, and Taurus star-formation regions. A model of field stars is built from a mixture of multivariate Gaussians based on the Besan\c con Galactic model. The algorithm can derive membership probabilities for objects with only sky coordinates and proper motion, but can also include parallax and radial velocity measurements, as well as spectrophotometric distance constraints from sequences in color-magnitude or spectral type-magnitude diagrams. BANYAN~$\Sigma$ benefits from an analytical solution to the Bayesian marginalization integrals over unknown radial velocities and distances that makes it more accurate and significantly faster than its predecessor BANYAN~II. A \deleted{receiver operating characteristic}\added{contamination versus hit rate} analysis is presented and demonstrates that BANYAN~$\Sigma$ achieves a better classification performance than other moving group tools available in the literature, especially in terms of cross-contamination between young associations. An updated list of bona fide members in the \nasso\ young associations, augmented by the Gaia-DR1 release, as well as all parameters for the 6D multivariate Gaussian models for each association and the Galactic field neighborhood within 300\,pc are presented. This new tool will make it possible to analyze large data sets such as the upcoming Gaia-DR2 to identify new young stars. IDL and Python versions of BANYAN~$\Sigma$ are made available with this publication, and a more limited online web tool is available at \url{www.exoplanetes.umontreal.ca/banyan/banyansigma.php}.
\end{abstract}


\section{INTRODUCTION}

Coeval associations of stars that formed from a single molecular cloud are valuable benchmarks to study how the properties of stars evolve with time (e.g., \citealp{2004ARAA..42..685Z,2008hsf2.book..757T}). While precisely measuring the age of an individual star is challenging, the simultaneous study of a large ensemble of stars can provide age measurements with precisions down to a few Myr (e.g., \citealt{2015MNRAS.454..593B}). A small number of young associations in the Solar neighborhood that were identified to date are of particular interest in part because their lower-mass members can be studied easily. They are also co-eval populations, making them valuable to measure the initial mass function and serve as age calibrators for members across all masses.

Recent surveys (e.g., \citealp{2014ApJ...783..121G,2015ApJ...798...73G,2015AJ....150..182K,2015ApJS..219...33G,2016ApJ...821..120A,2016ApJS..225...10F,2016ApJ...833...96L}) have started uncovering the substellar and planetary-mass members of nearby young stellar associations, which will make it possible to understand how their fundamental properties evolve with time.

Because young associations of the Solar neighborhood are sparse and span up to $\sim$\,20\,pc in size, the distribution of their members can cover wide areas on the celestial sphere, and in some cases cover it almost entirely (e.g., the AB~Doradus and $\beta$~Pictoris moving groups; \citealp{2001ApJ...562L..87Z,2004ApJ...613L..65Z}). As a consequence, selection criteria based on sky coordinates and photometry alone are problematic.

As they formed recently and have not yet been perturbed significantly by other stars in the Galaxy, the members of a given young association still share similar space velocities $UVW$, with typical velocity dispersions below $\sim$\,3\,\kms. This provides a way to identify the members of a young association; however, measuring their full kinematics requires not only proper motions, but also absolute radial velocities and parallaxes for every star. This constitutes the most challenging aspect in identifying their faint, low-mass members, as obtaining such measurements for a large sample of faint objects is prohibitive.

Various methods have been developed to identify members of young associations when only sky coordinates and proper motion are available. These include the convergent point tool (e.g., \citealt{2005ApJ...634.1385M,2006AA...460..695T,2011ApJ...727...62R}), various goodness-of-fit metrics (e.g. \citealp{2014AJ....147..146K,2017AJ....153...18B,2017AJ....154...69S}), and the ``good box'' method \citep{2004ARAA..42..685Z}. \cite{2013ApJ...762...88M} developed BANYAN (Bayesian Analysis for Nearby Young AssociatioNs), an algorithm based on Bayesian inference where moving groups are modelled with unidimensional Gaussian distributions in Galactic coordinates $XYZ$ and space velocities $UVW$. More complex algorithms, such as BANYAN~II \citep{2014ApJ...783..121G} and LACEwING \citep{2017AJ....153...95R} have more recently been developed, where associations are modelled with freely rotating tridimensional Gaussian ellipsoids in position and velocity spaces. BANYAN~I included distance constraints from field and young sequences in a $M_J$ versus $I_C - J$ color-magnitude diagram in the 2MASS \citep{2006AJ....131.1163S} and Cousins (see \citealt{2013ApJ...762...88M} for details) systems, and were defined for spectral classes K and M. BANYAN~II included similar constraints based on two color-magnitude diagrams ($J-K_S$ versus $M_{W1}$ and $H-W2$ versus $M_{W1}$) in the 2MASS and WISE \citep{2010AJ....140.1868W} systems, and were defined for spectral types later than M5.

These tools made it possible to identify hundreds of candidate members in nearby associations of stars, spanning the planetary to stellar-mass domains (e.g., \citealp{2013ApJ...762...88M,2014ApJ...788...81M,2015ApJS..219...33G,2016ApJS..225...10F,2016ApJ...833...96L}). The majority of these classification tools include only the seven youngest ($\sim$\,10--200\,Myr) and nearest ($\lesssim$\,100\,pc) moving groups, with the exception of LACEwING, which includes 16 associations and open clusters with a larger age range ($\sim$\,5--800\,Myr).

The upcoming data releases of the Gaia mission \citep{2016AA...595A...1G} will mark a new era in the study of young associations, as they will provide precise parallax measurements for a billion stars in the Galaxy, covering the full members of all associations within 150\,pc down to late-M spectral types \citep{2017MNRAS.469..401S}. This advancement in the census of members will improve kinematic models, lead to detailed measurements of initial mass functions, and open the door to the discovery of new sparse associations (e.g., see \citealt{2017AJ....153..257O}). The first release of the Gaia mission (Gaia-DR1; \citealt{2016AA...595A...2G}) has already provided two million parallax measurements for stars in the Tycho-2 catalog \citep{2000AA...355L..27H}, which have not yet been used to improve the membership classification tools described above.

This work presents BANYAN~$\Sigma$, the next generation of the BANYAN tool based on Bayesian inference, which includes \nasso\ associations with ages in the range $\sim$\,1--800\,Myr, completing the sample of known and well-defined associations within $\sim$\,150\,pc \footnote{The Lupus star-forming region is slightly above this limit at $\sim$\,155\,pc \citep{2008AA...480..785L}, and the distance of the Chamaeleon star-forming region has recently been revised above 150\,pc; \citep{2017arXiv171004528V}}. BANYAN~$\Sigma$ includes a significantly improved Gaussian mixture model of the Galactic disk which captures a larger fraction of field interlopers, and updated models of young associations that benefit from the most recent Gaia-DR1 parallax measurements. The models are also advanced to six-dimensional multivariate Gaussians that capture full correlations in the $XYZUVW$ distribution of members, including those in mixed spatial-kinematic coordinates. Two versions of the BANYAN~$\Sigma$ code (IDL\footnote{Interactive Data Language; see \url{https://github.com/jgagneastro/banyan_sigma_idl}} and python\footnote{See \url{https://github.com/jgagneastro/banyan_sigma}}) are made publicly available \citep{zenodobanyansigmapython,zenodobanyansigmaidl}, and a web portal is made available for single-object queries\footnote{\url{www.exoplanetes.umontreal.ca/banyan/banyansigma.php}}.

Most previously available classification tools rely on time-consuming algorithms, such as numerical integrals, which make it challenging to analyze large datasets such as the upcoming full Gaia release, and use various approximations in converting observables and kinematic models to probabilities that affect their classification performance. In BANYAN~$\Sigma$, most of these approximations are removed, and Bayesian marginalization integrals are solved analytically. As a consequence, the tool is $\approx$\,80\,000 times faster than its predecessor BANYAN~II, making it easier to analyze very large data sets. BANYAN~$\Sigma$ is also the first classification tool to include the Taurus, $\rho$~Ophiuchi and Corona Australis star-forming regions (e.g., \citealp{2000AA...359..181W,2008hsf2.book.....R}), the  nearest OB association Sco-Cen, composed of the three subgroups Upper~Scorpius, Lower-Centaurus~Crux and Upper~Centaurus~Lupus (e.g., \citealp{1946PGro...52....1B,1999AJ....117..354D,2016MNRAS.461..794P}). It is also the first such tool to include the IC~2602 (e.g., \citealp{1961MNRAS.123..245W,2009AA...498..949M}),  IC~2391 (e.g., \citealp{2007AA...461..509P}), and Platais~8 \citep{1998AJ....116.2423P} clusters, and one of the first to include the Pleiades cluster. For the latter, \cite{2014AA...563A..45S} presented a multivariate Gaussian mixture model to assign membership probabilities based on kinematic and photometric observables. Their model uses a larger number of free parameters, made possible by the large number of known Pleiades members, but it does not include other young associations.

In Section~\ref{sec:framework}, the framework of BANYAN~II is described, which serves as a starting point for BANYAN~$\Sigma$, described in detail in Section~\ref{sec:banyansigma}. An updated list of bona fide members for \nasso\ young associations within 150\,pc is presented in Section~\ref{sec:members}, and is used to build the multivariate Gaussian kinematic models of BANYAN~$\Sigma$ in Section~\ref{sec:ymgmodels}. Section~\ref{sec:fieldmodel} presents a multivariate Gaussian mixture model of field stars based on the Besan\c{c}on Galactic model. A choice of Bayesian priors that ensures fixed recovery rates in all associations when using a $P = 90$\% Bayesian probability threshold is described in Section~\ref{sec:priors}. A performance analysis of BANYAN~$\Sigma$ is presented in Section~\ref{sec:performance}, and is compared to other tools available in the literature. The membership of stars previously considered as ambiguous are revisited in Section~\ref{sec:ambig} using BANYAN~$\Sigma$. This work is concluded in Section~\ref{sec:conclusion}.

\section{The BANYAN~II algorithm}\label{sec:framework}

In this section, the Bayesian framework behind the BANYAN~II tool \citep{2014ApJ...783..121G} is described. The framework of BANYAN~$\Sigma$ will start from the same principles, but will include several improvements that are described in Section~\ref{sec:banyansigma}.

BANYAN~II is a Bayesian \deleted{model selection}\added{classification} algorithm, which uses the direct kinematic observables of a star $\{O_i\}$, namely its sky position ($\alpha$,$\delta$), proper motion ($\mu_\alpha$,$\mu_\delta$), radial velocity ($\nu$) and distance ($\varpi$), to determine the probability that the star belongs to a population described by a hypothesis $H_k$, corresponding to either the Galactic field or one of several young associations. This is done by applying Bayes' theorem:
\begin{align}\label{eqn:bayesrule}
	P(H_k|\{O_i\}) = \frac{P(H_k)\,P(\{O_i\}|H_k)}{P(\{O_i\})},
\end{align}
\noindent where the likelihood $P(\{O_i\}|H_k)$ is the probability that a member of $H_k$ displays the observables $\{O_i\}$, the prior $P(H_k)$ is the probability that a star belongs to hypothesis $H_k$ irrespective of its kinematic properties, and the fully marginalized likelihood $P(\{O_i\})$ is the probability that a star displays observables $\{O_i\}$ irrespective of its membership. Once the priors and likelihoods are determined, the fully marginalized likelihood can be obtained with:
\begin{align}
	P(\{O_i\}) = \sum_k P(\{O_i\}|H_k) P(H_k).\label{eqn:margeq}
\end{align}

It is often the case that the radial velocity ($\nu$) and/or distance ($\varpi$) of a star are not known, preventing a direct calculation of the likelihood $P(\{O_i\}|H_k)$. The case where both measurements are missing will be considered here. In this scenario, a likelihood probability can still be obtained by marginalizing over both missing parameters:
\begin{align}\label{eqn:princint}
	P(\{O_i\}|H_k) = \int_{-\infty}^{\infty} \int_{0}^{\infty} \mathcal{P}_o(\{O_i\}|H_k)\,\mathrm{d}\varpi\,\mathrm{d}\nu,
\end{align}
\noindent where the symbol $\mathcal{P}_o$ is used to distinguish probability \emph{densities} from probabilities $P$ that are free of physical units.

\cite{2014ApJ...783..121G} demonstrated that young associations can be well described with Gaussian distributions by working in the Galactic position $XYZ$ and space velocity $UVW$ frame of reference $\{Q_i\}$. The distributions of direct observables $\{O_i\}$ for the members of a young association would be accurately described only by complex functions in this coordinate frame. For this reason, BANYAN~II approximated the likelihood by computing it directly in the $\{Q_i\}$ parameter space:
\begin{align}\label{eqn:princint2}
	P(\{O_i\}|H_k) = \int_{-\infty}^{\infty} \int_{0}^{\infty} \mathcal{P}_o(\{Q_j(\{O_i\}^\prime,\nu,\varpi)\}|H_k)\,\mathrm{d}\varpi\,\mathrm{d}\nu,
\end{align}
\noindent where $\{O_i\}^\prime$ represents the set of observables excluding the radial velocity $\nu$ and distance $\varpi$. Equation~\eqref{eqn:princint2} inherently ignores the Jacobian of the transformation $\{O_i\} \rightarrow \{Q_i\}$, discussed further in Section~\ref{sec:banyansigma}. In \cite{2014ApJ...783..121G}, both integrals were solved numerically on a uniform grid of $500\times 500$ points over $\nu$ and $\varpi$, covering $-35$ to $35$\,\kms\ and $0.1$ to $200$\,pc, respectively. On each point ($\nu$,$\varpi$) of the grid, the observables $\{O_i\}^\prime$ were transformed to  the $\{Q_i\}$ frame of reference, and compared with a model of hypothesis $H_k$ to derive the probability density $\mathcal{P}_o(\{Q_j(\{O_i\}^\prime,\nu,\varpi)\}|H_k)$. The sum of all probability densities on the grid were then taken as an approximation of the likelihood $P(\{O_i\}|H_k)$. These approximations were mostly limiting in that they prevented the inclusion of high-velocity ($|\nu| > 35$\,\kms) or distant ($\varpi > 200$\,pc) stars, but they also required 250\,000 probability density calculations for each star.

As measurement errors on the sky position of a star are always small enough to have a negligible contribution to $XYZUVW$ compared to those on proper motion, radial velocity or distance measurements, they were ignored completely in BANYAN~II, and will still be ignored in BANYAN~$\Sigma$. Properly including the measurement errors on proper motion would require the introduction of two additional integrals to obtain a modified likelihood that takes error bars into account:
\begin{align}
	P_e(\{O_i\}|H_k) = \int_{-\infty}^{\infty} \int_{-\infty}^{\infty} P(\{O_i\}|H_k) \mathcal{P_\mathrm{m}}(\mu_\alpha,\mu_\delta)\,\mathrm{d}\mu_\alpha\,\mathrm{d}\mu_\delta,\notag
\end{align}
\noindent where $\mathcal{P_\mathrm{m}}(\mu_\alpha,\mu_\delta)$ is a probability density function describing the proper motion measurement, such as the product of two Gaussian distributions centered on the measured values with the appropriate characteristic widths. Because numerically solving this likelihood would be impractical, BANYAN~II approximated their effect by using a propagation error formula to obtain error bars on $U$, $V$ and $W$, and added them in quadrature to the characteristic widths of the Gaussian models describing each hypothesis $H_k$.

The kinematic models of BANYAN~II are based on Gaussian ellipsoids in $XYZ$ and $UVW$ space, which are freely rotated along any axis. Models for 7 young associations were included:  TW~Hya (TWA; \citealp{1989ApJ...343L..61D,1997Sci...277...67K}), $\beta$~Pictoris ($\beta$PMG; \citealp{2001ApJ...562L..87Z}), Tucana-Horologium (THA; \citealp{2000AJ....120.1410T,2001ApJ...559..388Z}), Carina (CAR; \citealp{2008hsf2.book..757T}), Columba (COL; \citealp{2008hsf2.book..757T}), Argus (ARG; \citealp{2000MNRAS.317..289M}) and AB~Doradus (ABDMG; \citealp{2004ApJ...613L..65Z}). The model of field stars was obtained by fitting a spatial and a kinematic Gaussian ellipsoid to the Besan\c con Galactic model within 200\,pc.

\section{BANYAN~$\Sigma$: AN IMPROVED ALGORITHM}\label{sec:banyansigma}

The framework of BANYAN~$\Sigma$ improves on BANYAN~II by (1) using the analytical solution to the marginalization integrals over radial velocity and distance, (2) using multivariate Gaussian models for the young associations and a mixture of multivariate Gaussians to model the Galactic field, (3) removing several approximations in the calculation of the Bayesian likelihood (4) accounting for parallax motion, and (5) including a larger number of young associations. The algorithm of BANYAN~$\Sigma$ is described in this section, and additional improvements with respect to the kinematic models are described in Sections~\ref{sec:members}, \ref{sec:ymgmodels} and \ref{sec:fieldmodel}.\added{ The models of BANYAN~$\Sigma$ are built from a `training set' consisting of a set of bona fide or high-likelihood members of young associations compiled from the literature. The models are therefore not built statistically, and the algorithm of BANYAN~$\Sigma$ is analogous to a Bayesian classification algorithm with a Gaussian mixture model (e.g., see \citealt{2007JEI....16d9901B} and \citealt{McLachlan:2000uw}).}

The multivariate Gaussian models of BANYAN~$\Sigma$ are described in Section~\ref{sec:kinmodels},  and the change of coordinates from the direct observables $\{O_i\}$ to the Galactic frame of reference $\{Q_i\}$ is described in Section~\ref{sec:coordchange}. The analytical solution of the Bayesian likelihood is presented in Section~\ref{sec:solve}, and determination of the radial velocity $\nu$ and distance $\varpi$ that maximize the Bayesian likelihood (Equation~\ref{eqn:princint}) are developed in Section~\ref{sec:optimal}. Section~\ref{sec:pm_errors} presents a method to approximate the effect of measurement errors on proper motion. Section~\ref{sec:plxmotion} details a new improvement to BANYAN~$\Sigma$, allowing it to apply a parallax motion correction when the distance of a star is not known and its proper motion measurement is based on two epochs only. Sections~\ref{sec:addobs} to \ref{sec:uvwonly} describe additional options to the BANYAN~$\Sigma$ algorithm, to include measurements of radial velocity and/or distance, constraints from spectrophotometric observables, or to ignore the spatial distribution of young associations.

\subsection{Kinematic Models}\label{sec:kinmodels}

The distribution of stars in young stellar associations and the Galactic field are modelled with multivariate Gaussian distributions in $XYZUVW$ space. This is a generalization over the freely rotating individual $XYZ$ and $UVW$ Gaussian models used in both BANYAN~II \citep{2014ApJ...783..121G} and LACEwING \citep{2017AJ....153...95R}, as it models correlations between mixed spatial and kinematic coordinates.\deleted{ For a visual representation of such a model, see Figure~\ref{fig:twa}.} The kinematic model corresponding to hypothesis $H_k$ can \deleted{therefore }be written as:
\begin{align}
	&\mathcal{P}_M\left(\bar Q,\bar \tau, \bar{\bar\Sigma}\right) = \frac{e^{-\frac{1}{2}\mathcal{M}^2}}{\sqrt{\left(2\pi\right)^{6}\left|\bar{\bar\Sigma}\right|}},\notag\\
	\mbox{where }&\mathcal{M} = \sqrt{\left(\bar{Q}-\bar\tau\right)^T\bar{\bar{\Sigma}}^{-1}\left(\bar{Q}-\bar\tau\right)}\label{eqn:mahal}
\end{align}
\noindent is the Mahalanobis distance and the bar (e.g., $\bar v$) and double-bar (e.g., $\bar{\bar M}$) symbols are used to indicate vectors and matrices, respectively, in $XYZUVW$ space. \added{$\bar{Q}$ is a 6-dimensional vector built from $\{Q_i\}$, which correspond to the $XYZUVW$ coordinates of an object. The Mahalanobis distance is a generalization of the concept of measuring how many standard deviations a data point is from the center of a gaussian distribution, and is applied to multivariate Gaussian distributions in the present work. A Mahalanobis distance has no units and accounts for correlations in the multivariate Gaussian probability density function \citep{Mahalanobis:1936va}.}

The multivariate Gaussian model includes a total of 27 free parameters: six are stored in the  $\bar\tau$ vector and indicate the center of the association; six are stored in the diagonal of the covariance matrix $\bar{\bar\Sigma}$ and indicate the 6D size of the association; and 15 more are stored in the independent off-diagonal elements of $\bar{\bar\Sigma}$ and indicate the orientation of the ellipsoid in 6D space, or equivalently the correlations between each combination of coordinates. $\bar x^T$ indicates a vector transposition and $\bar{\bar s}^{\ -1}$ a matrix inverse.

The Bayesian likelihoods in the Galactic frame of reference $\{Q_i\}$ can thus be written as:
\begin{align*}
 	\mathcal{P}_q(\left\{Q_i\right\}|H_k) &= \mathcal{P}_M\left(\bar Q, \bar\tau, \bar{\bar\Sigma}\right),
 \end{align*}
\noindent where the dependencies of $\bar\tau$ and $\bar{\bar\Sigma}$ on the association index $k$ are implicit.

A simple approach for obtaining the parameters of a kinematic model is to calculate the average position $\bar\tau$ of the members in $XYZUVW$ space and their variances and covariances to build the $\bar{\bar\Sigma}$ matrix. A method that is more robust to outliers, and accounts for individual measurement errors, is presented in Section~\ref{sec:ymgmodels}.

\subsection{Change of Coordinates}\label{sec:coordchange}

Solving the Bayesian likelihood in Equation~\eqref{eqn:princint} requires applying a change of coordinates from the observables frame of reference $\{O_i\}$ to the Galactic frame of reference $\{Q_i\}$. The equations for this transformation are detailed by \cite{1987AJ.....93..864J}, where the components of $\bar{\bar Q}_i$ can be written as a linear combination of \deleted{$\nu$ and $\varpi$}\added{the radial velocity $\nu$ and the distance $\varpi$}:
\begin{align}\label{eqn:defQ}
	\bar Q = \bar\Omega\nu + \bar\Gamma\varpi;
\end{align}
\noindent the components of $\bar\Omega$ and $\bar\Gamma$ are:
\begin{alignat}{2}
	\bar\Omega &=  \left(0,0,0,M_0,M_1,M_2\right) & = \left(\boldsymbol{0},\boldsymbol{M}\right),\notag\\
	\bar\Gamma &= \left(\lambda_0,\lambda_1,\lambda_2,N_0,N_1,N_2\right) & = \left(\boldsymbol{\lambda},\boldsymbol{N}\right),\notag
\end{alignat}
\noindent and symbols in bold represent 3D vectors or matrices in $XYZ$ or $UVW$ space.

The vectors $\boldsymbol\lambda$, $\boldsymbol M$ and $\boldsymbol N$ transform the sky position, proper motion, radial velocity and distance to $XYZUVW$ following:
\begin{align}
	\boldsymbol{\lambda} &= \left(\cos b\cos l, \cos b\sin l, \sin b\right),\notag\\
	\boldsymbol{M} &= \boldsymbol{\mathcal{T}}\boldsymbol{\mathcal{A}}\,\boldsymbol{m},\notag\\
	\boldsymbol{N} &= \boldsymbol{\mathcal{T}}\boldsymbol{\mathcal{A}}\,\boldsymbol{n},\label{eqn:bigndef}\\
	\boldsymbol{\mathcal{A}} &= \left[\begin{array}{ccc}
    	\cos\alpha\cos\delta & -\sin\alpha & -\cos\alpha\sin\delta\\
    	\sin\alpha\cos\delta & \cos\alpha & -\sin\alpha\sin\delta\\
    	\sin\delta & 0 & \cos\delta\\
    \end{array}\right],\notag\\
    \boldsymbol{m} &= \left(1,0,0\right),\notag\\
    \boldsymbol{n} &= \kappa\left(0,\mu_\alpha\cos\delta,\mu_\delta\right),\notag
\end{align}
\noindent where  $l$ and $b$ are the Galactic longitude and latitude, $\alpha$ and $\delta$ are the right ascension and declination, $\mu_\alpha\cos\delta$ and $\mu_\delta$ are the proper motion, and $\kappa \approx 4.74\cdot 10^{-3}$ corresponds to $10^{-3}$\,AU/yr so that proper motions are expressed in \masyr.

The $\boldsymbol{\mathcal{T}}$ matrix is a combination of rotation matrices involving the equatorial position of the North Galactic Pole, and is detailed by \cite{1987AJ.....93..864J}. We however use a definition of $\boldsymbol{\mathcal{T}}$ where the first row has the opposite sign of that defined by \cite{1987AJ.....93..864J}, so that $U$ points towards the galactic center and $UVW$ forms a right-handed system\footnote{This display of  the $\boldsymbol{\mathcal{T}}$ matrix follows the \emph{column major} convention (e.g., IDL or Forftran). Because Python follows the \emph{row major} matrix convention, $\boldsymbol{\mathcal{T}}$ must be transposed in that language. In IDL, the $\boldsymbol{\mathcal{T}}$ matrix is encoded with \protect{T = [[-0.0548$\cdots$,0.494$\cdots$,$\cdots$],[-0.873$\cdots$,$\cdots$],[$\cdots$]]}. In Python, it is encoded with \protect{T = np.array([[-0.0548$\cdots$,-0.873$\cdots$,$\cdots$],[0.494$\cdots$,$\cdots$],[$\cdots$]])}.}:
\begin{align}\notag
	\boldsymbol{\mathcal{T}} &= \left[\begin{array}{ccc}
    	-0.054875560 & -0.87343709 & -0.48383502\\
	0.49410943 & -0.44482963 & 0.74698224\\
	-0.86766615 & -0.19807637 & 0.45598378\\
    	\end{array}\right].
\end{align}

Equation~\eqref{eqn:defQ} can be used to express $\mathcal{P}_q$ as a function of the observables $\{O_i\}$, but solving the marginalization integrals of Equation~\eqref{eqn:princint} requires applying a change of coordinates (from the $\{Q_i\}$ frame to the $\{O_i\}$ frame) to the probability density function itself $\mathcal{P}_q \rightarrow \mathcal{P}_o$. This step is detailed in Appendix~\ref{app:jacobian_demo}, and yields:
\begin{align*}
	\mathcal{P}_o\left(\{O_i\}|H_k\right) = \varpi^4\,\mathcal{P}_q\left(\{Q_i\}|H_k\right).
\end{align*}

Inserting the coordinate transformation (Equation~\ref{eqn:defQ}) into the kinematic model defined in Equation~\eqref{eqn:mahal} yields:
\begin{align}
	\mathcal{M}_k^2(\nu,\varpi) = &\sum_{ij}\left[\sigma_{ij,k}^{-1}\Omega_i\Omega_j\,\nu^2 + \sigma_{ij,k}^{-1}\Gamma_i\Gamma_j\,\varpi^2\right.\notag\\
	&+\sigma_{ij,k}^{-1}\left(\Omega_i\Gamma_j + \Gamma_i\Omega_j\right)\,\nu\varpi\notag\\
	&-\sigma_{ij,k}^{-1}\left(\Omega_i\tau_{j,k} + \tau_{i,k}\Omega_i\right)\,\nu\notag\\
	&-\sigma_{ij,k}^{-1}\left(\Gamma_i\tau_{j,k} + \tau_{i,j}\Gamma_j\right)\,\varpi\notag\\
	&\left.+\sigma_{ij,k}^{-1}\tau_{i,k}\tau_{j,k}\right].\label{eqn:mahal_dev1}
\end{align}

All terms in $\Omega_i$, $\Gamma_i$ and $\tau_{i,k}$ can be described as scalar products induced by the inverse covariance matrix $\sigma_{ij,k}^{-1}$, such as:
\begin{align}
	\left<\bar X,\bar Y\right>_k = \sum_{ij}\sigma_{ij,k}^{-1}X_i Y_j,\label{eqn:scalarprod}
\end{align}
\noindent where the $k$ index will be omitted in the remainder of this work for simplicity.

Equation~\eqref{eqn:mahal_dev1} can be further simplified from the fact that the covariance matrix is symmetric by definition. With these two simplifications we get:
\begin{align}
	\mathcal{M}^2(\nu,\varpi) = &\left<\bar\Omega,\bar\Omega\right> \nu^2 + \left<\bar\Gamma,\bar\Gamma\right> \varpi^2 + 2\left<\bar\Omega,\bar\Gamma\right> \nu\varpi\notag\\
	&-2\left<\bar\Omega,\bar\tau\right> \nu -2\left<\bar\Gamma,\bar\tau\right>\varpi+\left<\bar\tau,\bar\tau\right>\label{eqn:mahal_dev2}.
\end{align}

This completes the coordinate change of the Bayesian likelihood:
\begin{align}
	\mathcal{P}_o(\{O_i\}|H)\,\mathrm{d}\nu\,\mathrm{d}\varpi = \frac{\varpi^4e^{-\frac{1}{2}\mathcal{M}^2\left(\nu,\varpi\right)}}{\sqrt{\left(2\pi\right)^6\left|\bar{\bar\Sigma}\right|}}\,\mathrm{d}\nu\,\mathrm{d}\varpi.\label{eqn:bayesianlikelihood_q}
\end{align}

\subsection{Solving the Marginalization Integrals}\label{sec:solve}

A complete analytical solution to Equation~\eqref{eqn:princint} is developed in Appendix~\ref{app:solving} and yields:
\begin{align}
	\mathcal{P}(\{O_i\}|H) &= \frac{\mathcal{D}^\prime_{-5}\left(\gamma/\sqrt{2\beta}\right)e^{\gamma^2/4\beta-\zeta}}{\left|\bar\Omega\right|\sqrt{\pi^5\beta^5\left|\bar{\bar\Sigma}\right|}},\label{eqn:bayesoln}\\
	\mbox{with }\beta &= \frac{\left<\bar\Gamma,\bar\Gamma\right>}{2} - \frac{1}{2}\frac{\left<\bar{\Omega},\bar{\Gamma}\right>^2}{\left<\bar\Omega,\bar\Omega\right>},\label{eqn:beta}\\
	\gamma &= \frac{\left<\bar{\Omega},\bar{\Gamma}\right>\left<\bar{\Omega},\bar\tau\right>}{\left<\bar\Omega,\bar\Omega\right>} - \left<\bar\Gamma,\bar\tau\right>,\label{eqn:gamma}\\
	\zeta &= \frac{\left<\bar\tau,\bar\tau\right>}{2}-\frac{1}{2}\frac{\left<\bar{\Omega},\bar\tau\right>^2}{\left<\bar\Omega,\bar\Omega\right>},\label{eqn:zeta}\\
	\mathcal{D}^\prime_{-5}(x) =& \sqrt\frac{\pi}{2}\left(x^4+6x^2+3\right)\erfc\left(\tfrac{x}{\sqrt{2}}\right)\notag\\
	&-\left(x^3+5x\right)e^{-x^2/2},\notag
\end{align}
\noindent where $\mathcal{D}^\prime_{-5}(x)$ is a parabolic cylinder function \citep{Magnus:zYVRWLJr} modified for numerical stability.

A limitation of this development is that correlations between the measurement errors of the sky coordinates, proper motion and parallax cannot be accounted for. Properly accounting for such correlations would require performing much more CPU-intensive and less precise numerical integrals.\added{ However, we demonstrate in Section~\ref{sec:performance} that ignoring such correlations cause negligible biases in the Bayesian membership probabilities.}

In summary, obtaining the Bayesian probability for a given star and hypothesis $H$ requires calculating the components of the vectors $\bar\Omega$, $\bar\Gamma$, $\bar\tau$ (Equations~\ref{eqn:defQ} and \ref{eqn:bigndef}) and their various scalar products (Equations~\ref{eqn:beta}--\ref{eqn:zeta}) using Equation~\eqref{eqn:scalarprod}, then evaluating the non-marginalized Bayesian likelihood with Equation~\eqref{eqn:bayesoln}, and finally evaluating the Bayesian membership probability with Equation~\eqref{eqn:bayesrule}, which makes use of Equation~\eqref{eqn:margeq}. The priors for each hypothesis are also needed to evaluate Equation~\eqref{eqn:bayesoln}; these are defined in Section~\ref{sec:priors}. When a large number of stars is analyzed, it is possible to improve the efficiency in solving Equation~\eqref{eqn:bayesoln} by calculating the individual 6 components of the 6D vectors, as well as $\beta$, $\gamma$, $\zeta$ and $\mathcal{P}(\{O_i\}|H)$ for the full array of stars at once.

\subsection{Optimal Radial Velocity and Distance}\label{sec:optimal}

The optimal values for the radial velocity and distance $(\nu_\mathrm{o},\varpi_\mathrm{o})$ that maximize the non-marginalized Bayesian likelihood $\mathcal{P}_o(\{O_i\}|H)$ can be determined for each hypothesis $H$. They correspond to predictions for the radial velocity and distance of a star, \emph{assuming that the star is a true member of $H$} (e.g., \citealt{2016ApJ...833...96L} obtain BANYAN~II distance predictions with accuracies as low as $\sim$\,20\% by ignoring this fact).

The optimal radial velocity and distance are derived in Appendix~\ref{app:optimald}:
\begin{align*}
	\varpi_\mathrm{o} &= \frac{-\gamma + \sqrt{\gamma^2 + 32\beta}}{4\beta},\\
	\nu_\mathrm{o} &= \frac{4+\left<\bar\Gamma,\bar\tau\right>\varpi_\mathrm{o}-\left<\bar\Gamma,\bar\Gamma\right>\varpi_\mathrm{o}^2}{\left<\bar\Omega,\bar\Gamma\right>\varpi_\mathrm{o}},\\
	\sigma_\varpi &= |\bar\Gamma|^{-1},\\
	\sigma_\nu &= |\bar\Omega|^{-1},
\end{align*}
\noindent where $\sigma_\varpi$ and $\sigma_\nu$ represent statistical 1$\sigma$ error bars on the optimal values. These two values are given for each hypothesis and each star as an output of BANYAN~$\Sigma$, and can be taken as predictions on the radial velocity and distance measurements that would maximize the probability of a given hypothesis.

Because the optimal radial velocity and distance are intrinsically linked to the assumption that a star is a member of hypothesis $H$, adopting the values $(\nu_\mathrm{o},\varpi_\mathrm{o})$ for a low-probability hypothesis $H$ inherently carries a large risk of the true radial velocity and distance of a star of being several $\sigma$ away from the prediction. The values of $\sigma_\varpi$ and $\sigma_\nu$ will therefore be unreliable in such a situation.

\subsection{Approximating the Effect of Proper Motion Measurement Errors}\label{sec:pm_errors}

As discussed in Section~\ref{sec:framework}, including the effect of proper motion errors requires solving two additional marginalization integrals on the proper motion components, in addition to those on radial velocity and distance. Since obtaining an analytical solution to these four marginalization integrals is impractical, this section describes an analytical approximation for the effect of the proper motion measurement error.

The optimal radial velocity $\nu_\mathrm{o}$ and distance $\varpi_\mathrm{o}$ that maximize the non-marginalized Bayesian likelihood $\mathcal{P}_o(\{O_i\}|H)$ can be used to propagate the proper motion measurement errors to the galactic position $XYZ$ and space velocity $UVW$ in the vicinity of the maximum of the membership probability distribution function. The sky position, proper motion, $\nu_\mathrm{o}$, $\varpi_\mathrm{o}$ and proper motion errors are propagated in $XYZUVW$ space to obtain a Galactic error vector $\bar\sigma_q = \left(\sigma_X,\sigma_Y,\sigma_Z,\sigma_U,\sigma_V,\sigma_W\right)$.

It is then possible to add these errors in quadrature to the diagonal elements of $\bar{\bar\Sigma}$ without affecting the orientation of the multivariate Gaussian with:
\begin{align*}
	\bar{\bar\Sigma}^\prime &= \bar{\bar G}\,\bar{\bar\Sigma}\,\bar{\bar G},\\
	\bar{\bar G} &= \mathrm{diag}_{+}\sqrt{\frac{\mathrm{diag}_{-}\left(\bar{\bar\Sigma}\right) + \bar\sigma_q^2}{\mathrm{diag}_{-}\left(\bar{\bar\Sigma}\right)}},
\end{align*}
\noindent where $\mathrm{diag}_{-}\left(\bar{\bar M}\right)$ extracts the diagonal elements of matrix $\bar{\bar M}$ and $\mathrm{diag}_{+}\left(\bar{v}\right)$ builds a diagonal matrix with vector $\bar{v}$.

The need for a time-consuming matrix inversion of $\bar{\bar\Sigma}^\prime$ for each star to evaluate Equation~\eqref{eqn:mahal_dev2}, can be avoided with:
\begin{align*}
	\bar{\bar\Sigma}^{\prime\ -1} &= \bar{\bar G}^{-1}\,\bar{\bar\Sigma}^{-1}\,\bar{\bar G}^{-1},\\
	\bar{\bar G}^{-1} &= \mathrm{diag}_{+}\sqrt{\frac{\mathrm{diag}_{-}\left(\bar{\bar\Sigma}\right)}{\mathrm{diag}_{-}\left(\bar{\bar\Sigma}\right) + \bar\sigma_q^2}}.
\end{align*}

Including an approximated effect of the proper motion measurement errors will thus require a calculation of (1) the $\bar\Omega$, $\bar\Gamma$ and $\bar\tau$ vectors and their scalar products, (2)  $\nu_\mathrm{o}$ and $\varpi_\mathrm{o}$, (3) the $\bar\sigma_q$ vector and the corresponding inflated matrix $\bar{\bar\Sigma}^\prime$, and (4) all quantities from the beginning of Section~\ref{sec:coordchange} obtained with the updated scalar product based on $\bar{\bar\Sigma}^{\prime\ -1}$. \added{The approximation described here does not make the assumption that a star is a member of any young association or the field. Instead, the proper motion errors are propagated to $XYZUVW$ independently for each Bayesian hypothesis, ensuring that it is valid near the peak of the probability distribution of each hypothesis. In other words, the steps described above are carried out independently when calculating the membership probability of each hypothesis.}

\subsection{Parallax Motion}\label{sec:plxmotion}

When the proper motion of a nearby star is measured based on two epochs only, the measurement may be contaminated in part by parallax motion. \added{This is true because the measurement of a star's displacement between two epochs include the compounding effects of its true proper motion (i.e., its space velocity projected on the celestial sphere) with that of its displacement along its parallactic ellipse, the latter of which is caused by a change in the observer's point of view as the Earth progresses on its orbit around the Sun. These two effects can however be decoupled in the BANYAN~$\Sigma$ formalism, as long as the parallax factors $(\psi_\alpha,\psi_\delta)$, described below, are measured for the star. The parallax factors physically represent the motion of the star purely due to the Earth's motion between the two epochs if the star was placed at exactly 1\,pc from the Sun.}\deleted{This can be accounted for in the BANYAN~$\Sigma$ formalism, provided that the parallax factors $(\psi_\alpha,\psi_\delta)$, described below, are also calculated.}

The parallax motion $\left(\Delta_{\alpha\pi},\Delta_{\delta\pi}\right)$ of a star is given by \citeauthor{1977tsa..book.....S} (\citeyear{1977tsa..book.....S}, Chap. 9, p. 221):
\begin{align}
	\Delta_{\alpha\pi}\cos\delta = &\frac{\cos\alpha\cos{e}\sin{\ell_s}-\sin\alpha\cos{\ell_s}}{\varpi},\\
	\Delta_{\delta\pi} = &\frac{\cos\delta\sin{e}\sin{\ell_s}-\cos\alpha\sin\delta\cos{\ell_s}}{\varpi}\notag\\
	&-\frac{\sin\alpha\sin\delta\cos{e}\sin{\ell_s}}{\varpi},
\end{align}
\noindent where $e$ is the obliquity of the ecliptic of the Earth's orbit and $\ell_s$ is the ecliptic longitude of the Sun at a given epoch\footnote{$e$ and $\ell_s$  can be computed from the sky position $\alpha$, $\delta$ and the julian date using the IDL astrolib routine \emph{sunpos.pro}.}. These equations can be simplified by grouping all epoch-dependent terms into $(\phi_\alpha,\phi_\delta)$:
\begin{align}
	\Delta_{\alpha\pi}(\alpha,t)\cos\delta &= \frac{\phi_\alpha(\alpha,t)}{\varpi},\\
	\Delta_{\delta\pi}(\alpha,\delta,t) &= \frac{\phi_\delta(\alpha,\delta,t)}{\varpi},
\end{align}
\noindent where $t$ is the epoch.

The apparent motion $\left(\mu_\alpha^\prime,\mu_\delta^\prime\right)$ of a star between epochs $t_1$ and $t_2$ will thus be given by:
\begin{align}
	\mu_\delta^\prime &= \frac{\left(\delta(t_2) + \Delta_{\delta\pi}(t_2)\right)-\left(\delta(t_1)+\Delta_{\delta\pi}(t_1)\right)}{t_2-t_1}\notag\\
	&= \mu_\delta + \frac{1}{\varpi}\frac{\phi_\delta(t_2) - \phi_\delta(t_1)}{t_2-t_1}\notag\\
	&= \mu_\delta + \frac{\psi_\delta}{\varpi},\ \mathrm{and\ similarly}:\notag\\
	\mu_\alpha^\prime\cos\delta &= \mu_\alpha\cos\delta + \frac{\psi_\alpha}{\varpi}.\notag
\end{align}

Since Equation~\eqref{eqn:defQ} is linear in proper motion components, it can be expressed as a function of apparent motion with the form:
\begin{align}
	\bar Q &= \bar\Omega\nu + \bar\Gamma^\prime\varpi  - \bar\Phi,\label{eqn:defQ2}\\
	\bar\Phi &= \left(\boldsymbol 0,\boldsymbol{\mathcal{T}}\boldsymbol{\mathcal{A}}\,\boldsymbol{\psi}\right),\notag\\
	\boldsymbol{\psi} &= \kappa\left(0,\psi_\alpha,\psi_\delta\right),\notag\\
	\psi_\alpha &= \frac{\varpi\cos\delta\left(\Delta_{\alpha\pi}(t_2)-\Delta_{\alpha\pi}(t_1)\right)}{t_2-t_1}\notag,\\
	\psi_\delta &= \frac{\varpi\left(\Delta_{\delta\pi}(t_2)-\Delta_{\delta\pi}(t_1)\right)}{t_2-t_1}\notag,
\end{align}
\noindent where $\bar\Gamma^\prime$ is a function of apparent motion $(\mu_\alpha^\prime\cos\delta,\mu_\delta^\prime)$, the quantity that is directly measured, instead of true proper motion $(\mu_\alpha\cos\delta,\mu_\delta)$.

Because $\bar Q$ only appears in the Bayesian likelihood as relative to the center of the moving group model $\tau$, the effect of parallax motion can be fully accounted for by shifting the $UVW$ center of the young association kinematic model by $+\boldsymbol{\mathcal{T}}\boldsymbol{\mathcal{A}}\,\boldsymbol{\psi}$, which is equivalent to shifting $\bar\tau$ by $+\bar\Phi$:
\begin{align*}
	\bar\tau^\prime = \bar\tau + \bar\Phi.
\end{align*}

As a consequence, the parallax motion can be accounted for by using the measured apparent motion as if it were a true proper motion, and replacing and $\bar\tau \rightarrow \bar\tau^\prime$ in the BANYAN~$\Sigma$ formalism. This requires measurements of the parallax factors $(\psi_\alpha,\psi_\delta)$ for each star in addition to measurements of their apparent motion. \added{In practice, this correction is applied by the BANYAN~$\Sigma$ software only when the keyword \emph{use\_psi} is explicitly used. This indicates that: (1) the proper motion that is input to BANYAN~$\Sigma$ was measured from two epochs only, (2) the effect of parallax motion was not corrected in the proper motion measurement and therefore it really is a measurement of apparent motion, and (3) the parallax factors are provided to BANYAN~$\Sigma$ using the same two epochs as those between which the proper motion was measured. If any of the above statements are not true, the parallax motion correction described above should not be used.}

\subsection{Additional Kinematic Observables}\label{sec:addobs}

\added{Only sky position and proper motion are required for BANYAN~$\Sigma$ to compute membership probabilities. However, radial velocities and/or distances can also be included as input measurements to obtain more accurate membership probabilities. This is similar to the functioning of BANYAN~I \citep{2013ApJ...762...88M} and BANYAN~II \citep{2014ApJ...783..121G}.} In cases where a radial velocity measurement is available, Equation~\eqref{eqn:princint} can be rewritten as:
\begin{align*}
	\mathcal{P}(\{O_i\}|H) &= \int_{-\infty}^{\infty}\mathcal{P}_m(\nu) \int_{0}^{\infty}\frac{\varpi^4e^{-\frac{1}{2}\mathcal{M}^2}}{\sqrt{\left(2\pi\right)^6\left|\bar{\bar{\Sigma}}^\prime\right|}}\,\mathrm{d}\varpi\mathrm{d}\nu\label{eqn:rv_int},
\end{align*}
\noindent where $\mathcal{P}_m(\nu)$ is the probability density function that represents the radial velocity measurement.

\begin{figure*}
	\centering
	\includegraphics[width=0.95\textwidth]{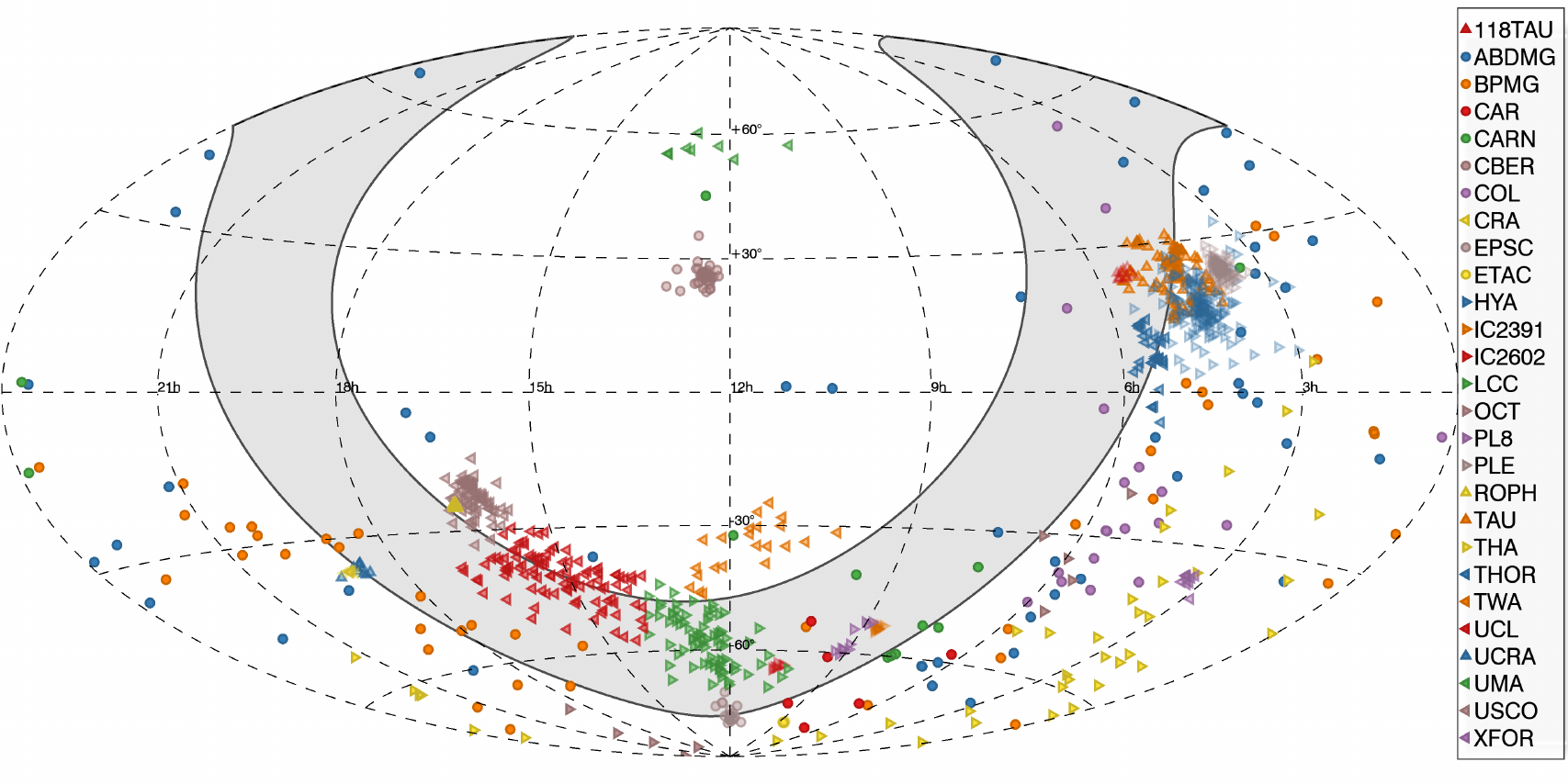}
	\label{fig:radec}
	\caption{Sky distribution of young association members that were used here to build the models of BANYAN~$\Sigma$. The Galactic plane ($|b| < 15$\textdegree) is designated with the gray region. The nearest young associations cover much larger fractions of the sky, which makes it harder to recognize their members without measuring their full 6D kinematics. Most of the young association members are located in the Southern hemisphere, with a few notable exceptions (UMA, CBER, PLE, HYA, TAU and 118TAU). See Section~\ref{sec:members} for more detail.}
\end{figure*}

Assuming that $\mathcal{P}_m(\nu)$ is a Gaussian distribution centered on $\nu_m$ with a characteristic width of $\sigma_{\nu,m}$, the equation above can be solved with a similar method to that described in Appendix~\ref{app:solving}, where the following scalar products are modified with:
\begin{align}
	\left<\bar\Omega,\bar\Omega\right> &\rightarrow \left<\bar\Omega,\bar\Omega\right> + \left(\sigma_{\nu,m}\right)^{-2},\notag\\
	\left<\bar\Omega,\bar\tau\right> &\rightarrow \left<\bar\Omega,\bar\tau\right> + \nu_m\left(\sigma_{\nu,m}\right)^{-2},\notag\\
	\left<\bar\tau,\bar\tau\right> &\rightarrow \left<\bar\tau,\bar\tau\right> + \nu_m^2\left(\sigma_{\nu,m}\right)^{-2}.\notag
\end{align}

The case where a distance measurement is available can be solved in a similar way, by replacing:
\begin{align}
	\left<\bar\Gamma,\bar\Gamma\right> &\rightarrow \left<\bar\Gamma,\bar\Gamma\right> + \left(\sigma_{\varpi,m}\right)^{-2},\notag\\
	\left<\bar\Gamma,\bar\tau\right> &\rightarrow \left<\bar\Gamma,\bar\tau\right> + \varpi_m\left(\sigma_{\varpi,m}\right)^{-2},\notag\\
	\left<\bar\tau,\bar\tau\right> &\rightarrow \left<\bar\tau,\bar\tau\right> + \varpi_m^2\left(\sigma_{\varpi,m}\right)^{-2}.\notag
\end{align}

The case where both radial velocity and distance measurements are available can be solved by combining all of the variable changes described above (the two changes on $\left<\bar\tau,\bar\tau\right>$ must be cumulated).

\subsection{Photometric Observables}

It is possible to constrain the distance of a star from its position in a color-magnitude or spectral type-magnitude diagram by comparing its absolute magnitude to a sequence of field stars, or to members of a young association, at a fixed color or spectral type. The position of a sequence in most of these diagrams is dependent on the age of its population, which translates to a different distance constraint for each Bayesian hypothesis.

Such photometric constraints can be included in the BANYAN~$\Sigma$ framework, in a similar way to the method described in Section~\ref{sec:addobs} for distance measurements, except that different values of the most likely distance $\varpi_m$ and its uncertainty $\sigma_{\varpi,m}$ must be used, one for each hypothesis, because they derive from different color-magnitude sequences. This is a consequence of the fact that the different Bayesian hypotheses correspond to populations of stars at different ages.

In the absence of a trigonometric distance measurement, users can create custom color-magnitude diagrams and determine values of $(\varpi_m,\sigma_{\varpi,m})$ for the field and each young association, and provide them to BANYAN~$\Sigma$ for a full inclusion of these constrainst in the Bayesian probabilities. Multiple color-magnitude diagrams can also be combined into single measurements of $(\varpi_m,\sigma_{\varpi,m})$ for a given star and young association, but failing to account for covariances between different photometric bands in a given stellar population would result in artificially small values of $\sigma_{\varpi,m}$. Such unrealistically precise constraints on the distance of a star would hinder the ability of BANYAN~$\Sigma$ to correctly identify the candidate members of a young association. \added{Both the IDL and Python implementations of BANYAN~$\Sigma$ can accept these photometric distance constraints through the \emph{constraint\_dist\_per\_hyp} and \emph{constraint\_edist\_per\_hyp} keywords, which are detailed in the documentation of the code. In summary, distinct color-magnitude diagrams for each hypothesis can be included by providing BANYAN~$\Sigma$ with distinct photometric constraints on the distance of a star.}

In the cases where both parallax and photometric measurements are available, they can be included in a more straightforward way to BANYAN~$\Sigma$ through the Bayesian priors: the vertical distances in a color-magnitude diagram between the measured absolute magnitude and the sequence of field or young objects at a fixed color can be transformed to a probability for each association using Bayes' theorem, and \deleted{these }\added{the natural logarithm of these photometric }probabilities can be included in\deleted{ the} BANYAN~$\Sigma$ \deleted{prior probabilities with an input keyword available in the code.}\added{with the \emph{ln\_priors} keyword available in both the IDL and Python implementations of the code.}

Other age-dating observables, such as X-ray, UV, H$\alpha$, rotation and lithium abundance measurements can similarly be translated to a membership probability at the age of each young association (e.g. by comparing measurements with the X-ray luminosity distributions of \citealt{2014ApJ...788...81M}), and can also be included in the BANYAN~$\Sigma$ prior probabilities. This framework allows users to add observables in the BANYAN~$\Sigma$ membership determination without needing to change its algorithm, and remains accurate as long as no kinematic measurements are used to assign prior probabilities. \added{The inclusion of such age indicators must rely on a user-specified method to translate each measurements into a probability that a given star is a member of each hypothesis, given the age of each young association. A compounded probability that each star is a member of each young association must then be calculated, based on only these youth indicators and no kinematics (e.g., by multiplying together the probabilities obtained from independent age indicators). The natural logarithm of these probabilities must then be input to BANYAN~$\Sigma$ with the keyword \emph{ln\_priors}. These data are passed to BANYAN~$\Sigma$ using a Python dictionary or an IDL structure depending on which version of the code is used, and we refer the reader to the respective documentations, which are provided as additional material to this manuscript, for more detail.}

No color-magnitude sequences are provided here with the first version of BANYAN~$\Sigma$, but they will be provided in future work as they are developed to target specific types of members.

\input{table_young_pars3.tex}
\ifdisplaylongtables
	
\else
	\input{table_bonafide_short.tex}
\fi
\input{table_bonafide_refs.tex}

\subsection{Ignoring the Galactic Position $XYZ$}\label{sec:uvwonly}

It is possible that the full spatial extent of some young associations have not yet been completely explored, especially at larger distances not covered by the Hipparcos survey. This possibility has been hypothesized by \cite{2017AJ....153...18B} among others, and the fact that the BANYAN tools rely on $XYZ$ as well as $UVW$ prevents an exploration of that potentially missing population of members.

The BANYAN~$\Sigma$ framework can be adapted to rely on $UVW$ only, by artificially setting the first three diagonal elements of the covariance matrices $\bar{\bar\Sigma}$ to a large value (e.g. $10^9$\,pc$^2$), and setting all other elements that contain at least one spatial component to zero. This approach is equivalent to using very large spatial widths for all models of young associations and the field in the BANYAN~$\Sigma$ formalism, and the general solution presented in Equation~\eqref{eqn:bayesoln} remains unchanged. An option is provided in BANYAN~$\Sigma$ to ignore the spatial $XYZ$ coordinates, and can be used to locate young association members that are spatially distant to the locus of known members. However, the rate of contamination from field stars is $\sim$\,100 times larger when using this option (see Section~\ref{sec:performance}), and we therefore recommend extreme caution when using it.

\section{BONA FIDE MEMBERS OF YOUNG ASSOCIATIONS WITHIN 150\,\lowercase{pc}}\label{sec:members}

In this section, a list of bona fide members of young associations within 150\,pc is compiled. This list will constitute the training set for the Gaussian models used in BANYAN~$\Sigma$ (see Sections~\ref{sec:kinmodels} and \ref{sec:ymgmodels} for more detail on the models). Each association considered in this work is listed in Table~\ref{tab:young_pars3} with its age estimate and the total number of bona fide members that were compiled. In the literature, stars are typically considered bona fide members when they benefit from signs of youth and full kinematic measurements that allow them to be placed in $XYZUVW$ space \citep{2013ApJ...762...88M,2014ApJ...783..121G}. The indicators of youth used in the literature depend on the spectral type of the stars, and include isochronal age determinations through color-magnitude positions, lithium measurements, UV or X-ray luminosity, rotational velocity, H$\alpha$ emission and rotational velocity; see \cite{2014prpl.conf..219S} for a review of these age-dating methods. Here we require the same measurements for bona fide members of \deleted{most}{the nearest or most well-studied} young associations.\added{ Nine of the 27 associations that are further away than $\sim$\,90\,pc do not have enough members with full 6D kinematics to require them in the construction of their model. In these cases, an average radial velocity or parallax (or both) for the association are used instead of individual measurements.}

\added{In Section~\ref{sec:asso_fullkin}, we provide a summarized description of the 18 young associations for which models are built using the individual 6D kinematics of all members. The 9 associations with incomplete kinematics are described in Section~\ref{sec:asso_partkin} and their adopted average distances and radial velocities are listed in Table~\ref{tab:partialkin}. A description the Argus association, included in BANYAN~II but excluded here, is provided in Section~\ref{sec:asso_rej}. A few individual objects that require further attention are discussed in Section~\ref{sec:asso_indivobjects}. The addition of Gaia-DR1 data to several previously recognized high-probability candidate members of the young associations studied here makes them new bona fide members: a description of these new members is provided in Section~\ref{sec:asso_newbfide}. The method that we used to calculate 6D kinematics and their error bars from kinematic observables is described in Section~\ref{sec:asso_calc}.}

All members\added{ compiled in this section} were cross-matched with the 2MASS, AllWISE \citep{2010AJ....140.1868W,2014ApJ...783..122K} and the Gaia-DR1 catalogs. When available, sky positions, proper motions and parallaxes from the Gaia-DR1 catalog were preferred to literature measurements. Targets with no radial velocity measurements reported in the literature membership lists were cross-matched with various catalogs that provide radial velocities \citep{1996PASP..108...64U,1997AJ....113.1458H,2006AA...460..695T,2006ARep...50..733B,2007AN....328..889K,2007AJ....133.2524W,2008AA...480..735F,2011ApJ...727....6S,2012arXiv1207.6212C,2013AJ....146..134K,2014ApJ...788...81M}. When authors did not report radial velocity measurement errors, we adopted those calculated by \citeauthor{2017AJ....153...95R} (\citeyear{2017AJ....153...95R}; see their Table~6).

\figsetstart
\figsetnum{1}
\figsettitle{Multivariate Gaussian models of young associations included in BANYAN~$\Sigma$}
\newcount\tmpnum
\def\storedata#1#2{\tmpnum=0 \edef\tmp{\string#1}\storedataA#2\end}
\def\storedataA#1{\advance\tmpnum by1
   \ifx\end#1\else
      \expandafter\def\csname data:\tmp:\the\tmpnum\endcsname{#1}%
      \expandafter\storedataA\fi
}
\def\getdata[#1]#2{\csname data:\string#2:#1\endcsname}
\storedata\mydata{{118TAU}{ABDMG}{BPMG}{CAR}{CARN}{CBER}{COL}{CRA}{EPSC}{ETAC}{HYA}{IC2391}{IC2602}{LCC}{OCT}{PL8}{PLE}{ROPH}{TAU}{THA}{THOR}{TWA}{UCL}{UCRA}{UMA}{USCO}{XFOR}}
\newcount\ii
\loop
\ifnum\ii<27
\advance\ii by 1
\figsetgrpstart
\figsetgrpnum{1.\ii}
\figsetgrptitle{\getdata[\the\ii]\mydata}
\figsetplot{\getdata[\the\ii]\mydata_params.pdf}
\figsetgrpnote{Multivariate Gaussian model of \getdata[\the\ii]\mydata. The 1, 2 and 3$\sigma$ projected contours of the multivariate Gaussian model are displayed as orange lines, and black points represent individual bona fide members. Unidimensional distributions (i.e., histograms) of bona fide members are displayed as green bars. In the tridimensional projection figures, a single 1$\sigma$ contour of the multivariate Gaussian model is displayed. Projections of the bona fide members positions on the three axis planes are displayed with blue spheres to facilitate viewing. See Section~\ref{sec:kinmodels} for more detail.}
\figsetgrpend
\repeat
\figsetend
\begin{figure*}[p]
	\centering
	\includegraphics[width=0.98\textwidth]{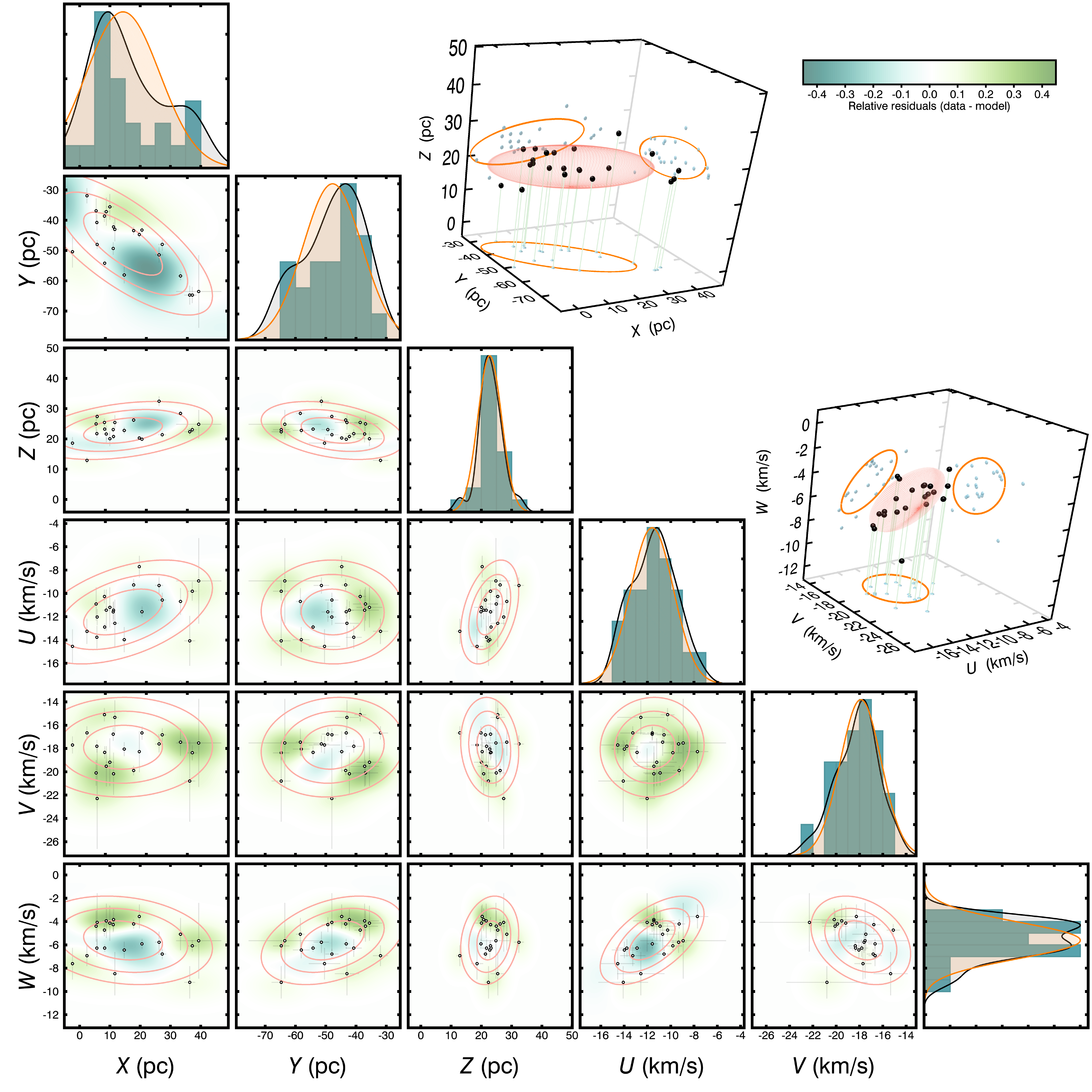}
	\caption{Multivariate Gaussian model of TWA. The 1, 2 and 3$\sigma$ projected contours of the multivariate Gaussian model are displayed as orange lines, and black points represent individual bona fide members. The residuals resulting from the difference of a 2D kernel density estimate distribution using Silverman's rule of thumb and the multivariate Gaussian models are displayed in the background of each 2D projection. Green shades indicate an over-density of members compared to the model, and blue shades indicate an under-density. Unidimensional distributions (i.e., histograms) of the bona fide members are displayed as green bars. The thick black lines represents a 1D kernel density estimate using Silverman's rule, and the thick orange line represents the projection of the multivariate Gaussian model.  In the tridimensional projection figures, a single 1$\sigma$ contour of the multivariate Gaussian model is displayed. Projections of the bona fide members positions on the three axis planes are displayed with blue spheres to facilitate viewing. The complete figure set (\nasso\ images), one for each young association, is available in the online journal. See Section~\ref{sec:kinmodels} for more detail.}
	\label{fig:twa}
\end{figure*}

\subsection{Associations With Full Kinematics}\label{sec:asso_fullkin}

\added{In this section, we provide a short description  of the 18 young associations included in BANYAN~$\Sigma$ for which the models will be built only from their \emph{classical} bona fide members, in the sense that only members with signs of youth and full 6D kinematics are included.}

The list of bona fide members presented in \cite{2014ApJ...783..121G}, which was largely based on that of \cite{2013ApJ...762...88M}, was used as a starting point in this work. The \cite{2014ApJ...783..121G} list includes members of TWA, $\beta$PMG, THA, CAR, COL, ARG and ABDMG. \cite{2017ApJS..228...18G} compiled an updated list of TWA members with new available data, and rejected contaminants from more distant associations, and was therefore adopted here. \added{We refer the reader to \cite{2004ARAA..42..685Z} and \cite{2008hsf2.book..757T} for a detailed description of these associations.}

\textbf{Carina-Near} (CARN) has been identified as a co-moving group of $\sim$\,200\,Myr-old stars by \cite{2006ApJ...649L.115Z}, which includes a core of eight members and ten members that are part of a spatially larger stream. Few studies have focused on this moving group since its discovery, likely because it is among the older ones. \cite{2017ApJ...841L...1G} used a preliminary version of BANYAN~$\Sigma$ to show that the variable T2.5 brown dwarf SIMP~J013656.5+093347 is likely a $\sim$\,13\,\mjup\ member of the CARN stream, and \cite{2017AJ....153...95R} included CARN in a young association classification tool (LACEwING) for the first time.

The \textbf{Ursa~Major cluster} (UMA; e.g., \citealt{1992AJ....104.1493E}) is a well-studied population of co-moving stars, consisting of a core of co-eval young stars, and a stream of stars with heterogeneous compositions and ages. \cite{1993AJ....105..226S} estimated an age of $\sim$\,300\,Myr for the core population, and \cite{2015ApJ...813...58J,2017AAS...22913105J} estimated an age of $414 \pm 23$\,Myr based on interferometric measurements of its A-type members. The core membership lists of \cite{2003AJ....125.1980K} was adopted for this work, and the stream was not included in BANYAN~$\Sigma$ because of its heterogeneous nature.

The \textbf{Hyades cluster} (HYA) is a nearby (40--50\,pc) and relatively young (600--800\,Myr; \citealt{1998AA...331...81P}) cluster that has been extensively studied in the literature (e.g., \citealp{1998AA...331...81P,2004ARAA..42..685Z}). The membership list of \cite{1998AA...331...81P} was adopted here.

The \textbf{Upper Scorpius}, (USCO), \textbf{Upper~Centaurus-Lupus} (UCL) and \textbf{Lower~Centaurus-Crux} (LCC) groups are part of the Sco-Cen star-forming region \citep{1946PGro...52....1B,1999AJ....117..354D}, which consists of 5--30\,Myr stars located at distances of $\sim$\,110--150\,pc. The membership lists of \cite{2011MNRAS.416.3108R}, \cite{2016MNRAS.461..794P} and \cite{2017arXiv171000909D} were adopted here. \cite{2011MNRAS.416.3108R} only list membership probabilities for the Sco-Cen region, and do not classify their members in the three subgroups. Their list was therefore cross-matched with that of \cite{1999AJ....117..354D} to assign the correct sub-group, but all members of Sco-Cen that were newly discovered by \cite{2011MNRAS.416.3108R} were not included at this stage. Once completed, the BANYAN~$\Sigma$ tool can be used to assign these new members to the correct subgroup; this is done in Section~\ref{sec:ambig}. Several radial velocities for USCO, UCL and LCC cataloged by \cite{2007AN....328..889K} -- which seem to be mistakenly listed as originating from \cite{2006AstL...32..759G}\footnote{Listed in the \cite{2007AN....328..889K} catalog as `Index of Radial Velocity Catalogues' = 2 in table {III/254/crvad2} of VizieR.} -- are astrometric radial velocities assuming moving group membership and an average $UVW$ velocity. These measurements were rejected from our compilation.

The \textbf{Octans association} (OCT; \citealt{2008hsf2.book..757T}) is a group of young stars at $\approx$\,120\,pc from the Sun, which has not been characterised as well as other young associations mainly due to its sky position located far in the Southern hemisphere (declinations between $-87$ and $-20$\textdegree). \cite{2015MNRAS.447.1267M} performed a survey of its low-mass members and determined a lithium age of 20--40\,Myr for this group. The members of OCT were compiled from \cite{2015MNRAS.447.1267M}.

The \textbf{Pleiades cluster} (PLE; \citealp{1921PASP...33..214C,1989ApJ...344L..21S}) is one of the best-studied clusters in the Solar neighborhood. It is located at a distance of $\sim$\,130\,pc and recent estimates of its age based on its lithium depletion boundary are in the range $\sim$\,110--120\,Myr \citep{2015ApJ...813..108D}. The membership lists of \cite{2007ApJS..172..663S} and \cite{2017AA...598A..48G} were adopted here. \cite{2014AA...563A..45S} presented a Bayesian method to identify members of the pleiades based on multivariate Gaussians mixture models. This method differs from BANYAN~$\Sigma$ in that it does not consider other young associations, includes various photometric colors, works directly in proper motion space, and does not consider sky position because they study stars in the direction of the cluster only. The larger number of free parameters introduced by a mixture of multivariate Gaussians makes it possible to model the proper motion and color-magnitude distribution of the Pleiades members, which are not well represented by a single Gaussian distribution. The large number of known Pleiades members allows such a highly parametrized model, but it would likely be challenging to apply this method to sparser or nearby young associations.

\textbf{Coma Berenices} (CBER; also called Melotte~111 and Collinder~256; e.g. \citealt{2006MNRAS.365..447C}) is a massive and well-studied open cluster located at $\sim$\,85\,pc. \cite{2014AA...566A.132S} estimate an age of $560_{-80}^{+100}$\,Myr based on the Hertzsprung-Russell diagram position of its Ap-type stars. The membership lists of \cite{2006MNRAS.365..447C}, \cite{2007AJ....134.2340K} and \cite{2014MNRAS.441.2644C} were used here.

\textbf{IC~2602} (Melotte~102; \citealt{1961MNRAS.123..245W}) is one of the nearest open clusters, and is located near the Sco-Cen OB region. Its members are located at a distance of $\approx$\,150\,pc \citep{2009AA...497..209V} and the cluster has a lithium depletion boundary age of $46_{-5}^{+6}$\,Myr \citep{2010MNRAS.409.1002D}. The list of members published by \cite{2014AA...566A.132S} and \cite{2009AA...498..949M} were adopted as a starting point for BANYAN~$\Sigma$.

\textbf{IC~2391} (Omicron Velorum; \citealt{2007AA...461..509P}) is a $\sim$\,$50 \pm 5$\,Myr-old cluster \citep{2004ApJ...614..386B} located at $\approx$\,150\,pc, and is also in the vicinity of the Sco-Cen OB region. The membership list of \cite{2017AA...601A..19G} was used here.

\subsection{Associations With Partial Kinematics}\label{sec:asso_partkin}

\added{This section describes the 9 young associations that do not have enough known members with full 6D kinematics to build their kinematic models based on only such members. Instead, an average radial velocity or distance (or both) are adopted for the young association itself. Thes average distances are obtained by calculating the weighted average of all members with measured distances, where the weights are set to the inverse square of the measurement errors. The error bars on the average distance correspond to an estimate of the intrinsic distance dispersion of the members, rather than a proper measurement error of the average, and was obtained using an unbiased weighted standard deviation\footnote{See \url{https://www.gnu.org/software/gsl/manual/html_node/Weighted-Samples.html}} of the individual members' distance measurements, with the same weights as described above. The average radial velocities are calculated with the same method, but spectral binaries were avoided in their calculation. All average distances and/or radial velocities that were measured in this section and used in the construction of the kinematic models are listed in Table~\ref{tab:partialkin}.}

\added{Instead of assigning the exact same average distance or radial velocity to each members with a missing observable, artifical values were drawn from a random distribution (limited to $\pm$1$\sigma$) with a characteristic width set to the measured intrinsic dispersion of the members (listed in Table~\ref{tab:partialkin}). This avoids artifically placing several members along lines in $XYZ$ and $UVW$ space -- or along planes in $UVW$ space when both radial velocity and distance are missing. The lack of full 6D kinematics for a significant number of candidates will result in a lower recovery rate of their true members by BANYAN~$\Sigma$, and a larger number of contaminants from field stars. Quantifying this effect in terms of exact true-positive and false-positive rates is however not currently possible given our lack of information on the true shapes and sizes of their spatial and kinematic distributions , however Gaia-DR2 will allow us to greatly refine the kinematic models of these associations.}

\input{table_avg_observables.tex}

\textbf{$\boldsymbol\epsilon$~Chamaeleontis} (EPSC; \citealp{2000ApJ...544..356M,2003ApJ...599.1207F,2013MNRAS.435.1325M}) is a young (3--5\,Myr) and relatively distant (100--120\,pc) association that is part of the Chamaeleon molecular cloud complex \citep{2008ApJ...675.1375L}. \cite{2013MNRAS.435.1325M} refined the age of EPSC to $3.7_{-1.4}^{+4.6}$\,Myr by comparing its members with the Dartmouth isochrones of \cite{2008ApJS..178...89D}. The membership list of \cite{2013MNRAS.435.1325M} was adopted here.\deleted{ Only 8 members have full kinematic measurements, and an additional 17 high-likelihood members only lack a parallax measurement. The average distance of the 8 bona fide members weighted by the inverse square of the measurement error ($102.3 \pm 5.7$\,pc) was assigned to the 17 high-likelihood members in constructing the kinematic model of EPSC. The error bar on this average distance corresponds to an estimate of the intrinsic distance distribution rather than a measurement error on the average, and was obtained with an unbiased weighted standard deviation\footnote{See \url{https://www.gnu.org/software/gsl/manual/html_node/Weighted-Samples.html}}, where the weights are also set to the inverse square of measurement errors. All average radial velocities and trigonometric distances are obtained in the same way in the remainder of this section. Instead of assigning a distance of exactly $102.3 \pm 5.7$\,pc to each member with a missing parallax, artificial distances were drawn from a Gaussian random distribution (limited to $\pm$1$\sigma$) with a characteristic width of $5.7$\,pc and centered on $102.3$\,pc, each with an error bar of $5.7$\,pc. This avoids artifically placing several members along lines in $XYZ$ and $UVW$ space, or along planes in $UVW$ space when both radial velocity and distance are missing.}

The \textbf{$\boldsymbol\eta$~Chamaeleontis cluster} (ETAC; \citealt{1999ApJ...516L..77M}) is a group of young ($11 \pm 3$\,Myr; \citealt{2015MNRAS.454..593B}) stars located at a distance of $\sim$\,100\,pc, and in the vicinity of the Sco-Cen OB association. The membership lists of \cite{2000ApJ...544..356M} and \cite{2004MNRAS.355..363L} were adopted in this work.\deleted{ Only three members ($\eta$ Cha, EG~Cha and RS~Cha) have both distance and radial velocity measurements available in the literature. Two additional likely members (EO~Cha and EQ~Cha) have a radial velocity measurement only, and one other likely member (HD~75505), has a distance measurement only. The average of all available distances and radial velocities weighted by the square inverse of the measurement errors ($\varpi_{\rm avg} = 94.4 \pm 2.0$\,pc, $\nu_{\rm avg} = 20.0 \pm 3.1$\,\kms) were adopted for the missing data of the 14 likely members of ETAC. Spectral binaries were avoided in calculating the average radial velocity of the bona fide members of ETAC, and other associations described below.}

The \textbf{32~Orionis group} (THOR; \citealp{2007IAUS..237..442M,2010AAS...21542822S,2017MNRAS.468.1198B}) is a young group of $\sim$\,25\,Myr-old stars located at $\sim$\,96\,pc. The age of THOR was revised to $22_{-3}^{+4}$\,Myr by \cite{2015MNRAS.454..593B} from a comparison of its members to model isochrones. \cite{2016ApJ...820...32B} recently identified the first substellar candidate member of THOR, with an estimated mass of 14\,\mjup\ near the planetary-mass boundary. The membership list of \cite{2017MNRAS.468.1198B} was adopted here\added{.}\deleted{ , but only four members of THOR have both distance and radial velocity measurements (32~Ori~AB, HR~1807, HD~35714 and TYC~112--1486--1). Their average weighted by the square inverse of the measurement error ($\varpi_{\rm avg} = 96.2 \pm 3.5$\,pc) was adopted for the 32 likely members that only lack a distance measurement.}

\input{table_xyzuvw_cuts.tex}

The $\boldsymbol\chi^{\boldsymbol 1}$~\textbf{For} association (XFOR; also called Alessi~13) was identified by \cite{2002AA...389..871D}, and \cite{2013AA...558A..53K} estimated an age of $\sim$\,525\,Myr based on the main-sequence turnoff. However, \cite{2015IAUS..314...21M} argue that it could be as young as $\sim$\,30\,Myr due to the saturated X-ray emission of its members. Further studies will be required to address this discrepancy. Only one member of XFOR has been published with full kinematics (the triple star $\chi^{1}$~For). In order to identify more members, the \cite{2014AA...564A..79D} list of 4\,102 XFOR candidates was cross-matched with the Gaia-DR1. Of the 261 matches with a parallax measurement, only 9 are located within 10\,pc of the $\chi^{1}$~For system in $XYZ$ space. This indicates that the \cite{2014AA...564A..79D}  sample seems highly contaminated by background stars and we therefore recommend caution in its use. A literature search was performed to identify one additional radial velocity measurement for CD--37~1263, thus completing its kinematic measurements and making it a new likely bona fide member of XFOR (although its age is not investigated here). Six of the eight additional potential members are located within 5\,\kms\ of the star $\chi^{1}$~For in $UVW$ space if we assume the same radial velocity measurement, and they are therefore included as high-likelihood candidate members. Two additional XFOR members were identified by Alessi et al. (priv. comm.): HD~21434 and HD~17864. Both have a parallax measurement in Gaia-DR1, but only HD~17864 also has a radial velocity measurement available in the literature.\deleted{ The average radial velocity of high-likelihood and bona fide members weighted by the squared inverse of their measurement error ($18.8 \pm 1.4$\,\kms) was adopted for high-likelihood members with no available radial velocities in the construction of the XFOR kinematic model.}

\textbf{Platais~8} (or a~Car; PL8) is a $\sim$\,60\,Myr-old cluster of stars at a distance of $\sim$\,130\,pc identified by \cite{1998AJ....116.2423P}. It has since received very little attention in the literature, and only 4 of its members benefit from full kinematic measurements (a~Car, HD~76230, H~Vel, OY~Vel).\deleted{ An additional 7 high-likelihood candidate members only lack a radial velocity measurement: the average radial velocity of the 4 members, weighted by the inverse square of the measurement errors ($21.9 \pm 2.0$\,\kms) was assigned to them in the construction of its kinematic model.}

\textbf{$\boldsymbol\rho$~Ophiuchi} (ROPH) is the nearest star-forming cloud complex to the Sun. It has been the subject of extensive studies in the past decades (e.g. see \citealp{1991ESOSR..11....1R,2008hsf2.book..351W,2008hsf2.book.....R}). Its age is estimated at $<$\,2\,Myr, and includes embedded clusters with stars believed to be as young as $\sim$\,0.1\,Myr \citep{1999ApJ...525..440L}. Because this group is too distant for its members to have been directly detected by the Hipparcos mission, \cite{2008AN....329...10M} used the measured parallaxes of Hipparcos stars illuminating the Lynds~1688 dark cloud, which is part of ROPH, to estimate its distance at $131 \pm 3$\,pc. They also estimate an average radial velocity of $-6.3 \pm 0.3$\,\kms\ from individual radial velocity measurements of its members. The membership lists of \cite{2008hsf2.book..351W} and \cite{2017AA...597A..90D} were adopted here, and the average radial velocity and distance of \cite{2008AN....329...10M} were adopted for all members with missing measurements. Only the 194 out of 340 members that have a proper motion measurement were used in the construction of the BANYAN~$\Sigma$ kinematic model of ROPH. A cross-match of these with Gaia~DR1 yielded 84 matches, indicating that the second data release will provide a wealth of new information on the distances and proper motions of ROPH members.

The \textbf{Corona Australis} (CRA) star-forming region is located at a distance of $\sim$150\,pc \citep{2008hsf2.book..735N,2008hsf2.book.....R}, and includes the well-studied R~CrA dark cloud (e.g., see \citealt{1992ApJ...397..520W}). \cite{2012MNRAS.420..986G} estimated the age of the eclipsing binary system TY~CrA between $3.8_{-0.2}^{+2.7}$\,Myr and $5.2_{-0.7}^{+3.1}$\,Myr, based on a comparison of the dynamical masses of its components with a set of warm and cold PISA pre-main sequence models \citep{2011AA...533A.109T}, respectively. Here we therefore adopt an age of $\sim$4--5\,Myr for CRA. The membership list of \cite{2008hsf2.book..735N} was adopted here.\deleted{ We measured the average distance and radial velocity of CRA members, weighted by the square inverse of measurement errors, and obtained $\varpi_{\rm avg} = 139.4 \pm 6.1$\,pc and $\nu_{\rm avg} = -0.4 \pm 1.2$\,\kms: these values were assigned to all members with missing measurements.}

Several stars in the vicinity of CRA discovered by \cite{2000AAS..146..323N} were found to be located between CRA and the Sco-Cen region, and at similar distances to both of these regions. This population likely constitutes of stars that formed along a filament between CRA and Sco-Cen $\sim$\,10\,Myr ago. We included them in the models of BANYAN~$\Sigma$, and tentatively name this population \textbf{Upper~CrA} (UCRA hereafter).\deleted{ We assigned the distance and radial velocity averages of its members, weighted by their squared inverse measurement errors ($\varpi_{\rm avg} = 148 \pm 3$\,pc and $\nu_{\rm avg} = -2.5 \pm 2.4$\,\kms), to the members missing these measurement.}

\input{table_new_bonafide_objects.tex}

The \textbf{Taurus-Auriga} (TAU)  star-forming region is a complex of several dark clouds located at $\sim$\,130\,pc, and composed of stars with ages $\sim$1--2\,Myr that share similar kinematics (e.g., see \citealt{1995ApJS..101..117K,2008hsf2.book.....R}). The membership lists of \cite{2009ApJ...703..399L} and \cite{2014ApJ...784..126E} were adopted here, without differenciating the sub-groups. Measurement errors were not provided for the TAU radial velocities measured by \cite{2000AA...359..181W}, but they report two sets of measurements from two distinct instruments. We measured the standard deviations of the radial velocity differences for the 25 stars in their sample that were observed with both instruments, ignoring 10 spectral binaries and 2 stars with significantly different measurements ($> 8$\,\kms). We adopted this standard deviation of 2\,\kms\ as their radial velocity measurement errors.\deleted{ The averages of the distances and radial velocities of the TAU members, weighted by the squared inverse of their error bars ($\varpi_{\rm avg} = 126 \pm 16$\,pc and $\nu_{\rm avg} = 17.0 \pm 2.9$\,\kms), were assigned to the members with missing kinematics.}

\cite{mamajek118tau}\footnote{Available on Figshare at \url{https://figshare.com/articles/A_New_Candidate_Young_Stellar_Group_at_d_121_pc_Associated_with_118_Tauri/3122689}} identified 11 stars in the vicinity of TAU that share a larger proper motion and a closer distance to the Sun than the rest of the group (see the discussions of \citealt{2017ApJ...836L..15C} and \citealt{2017ApJ...838..150K}). This group, named after its brightest member \textbf{118~Tau} (118TAU hereafter),  seems to display a slightly younger age than TAU, at $\sim$\,10\,Myr. The membership list of \cite{mamajek118tau} was adopted here.\deleted{ We calculated the average distances and radial velocities of 118TAU members weighted by their squared inverse measurement errors ($\varpi_{\rm avg} = 112.4 \pm 5.6$\,pc and $\nu_{\rm avg} = 14.7 \pm 1.1$\,\kms) and assigned them to members with missing measurements.}

\subsection{Rejected Associations}\label{sec:asso_rej}

The \textbf{Argus association} was removed entirely from the models of BANYAN~$\Sigma$, as \cite{2015MNRAS.454..593B} demonstrated that it is either largely contaminated, or composed of objects that do not form a coeval association (see also \citealt{2015IAUS..314...21M}). In addition, the \textbf{Octans-Near} \citep{2013ApJ...778....5Z} and \textbf{Hercules-Lyra} associations \citep{1998PASP..110.1259G,2004AN....325....3F,2006ApJ...643.1160L,2013AA...556A..53E} were not included in BANYAN~$\Sigma$, as they were also demonstrated to be likely composed of non-coeval stars \citep{2014ApJ...786....1B,2015IAUS..314...21M,2017AJ....153...95R}.

\subsection{Discussion on Individual Objects}\label{sec:asso_indivobjects}

\added{Individual stars that require more detailed considerations are discussed in this section. In addition to this, we note that several }\deleted{Several} stars\deleted{ were} listed by different authors as bona fide members of different associations (e.g., V570~Car, CP--68~1388 and CD--69~1055)\deleted{. These cases} are excluded from the BANYAN~$\Sigma$ kinematic models and are listed in Table~\ref{tab:excluded}.

AB~Pic was incorrectly listed by \cite{2014ApJ...783..121G} as a bona fide member of both THA and CAR. This was a consequence of \cite{2004ARAA..42..685Z} listing it as a bona fide member of THA and \cite{2008hsf2.book..757T} revising it to a bona fide member of CAR.  Since its UVW position is at $0.7 \pm 0.7$\,\kms\ from the core of CAR and at $5.5 \pm 2.1$\,\kms\ from that of THA, here it was included in the list of CAR members (see also the discussions of \citealt{2015MNRAS.454..593B}, Section B2.3, and \citealt{2013ApJ...762...88M}, Section 9.1.5).

DK~Leo was listed by \cite{2013ApJ...762...88M} as an ambiguous candidate member between $\beta$PMG, COL and ABDMG, because of contradictory measurements for its radial velocity \citep{2007AN....328..889K,2001MNRAS.328...45M,2010AA...514A..97L}, and \cite{2014ApJ...783..121G} incorrectly listed it as a bona fide member of both $\beta$PMG and CAR. Here it is excluded from the list of bona fide members until more radial velocity measurements become available.

HD~23524 was defined as a bona fide member of THA by \cite{2011ApJ...732...61Z}, and \cite{2013ApJ...762...88M} defined it as a bona fide member of COL because it lies nearer in $XYZ$ space, although they note that its membership is ambiguous. Here HD~23524 is excluded from the BANYAN~$\Sigma$ kinematic models.

HIP~3556 was noted as a radial velocity variable and a spectroscopic double-lined binary by \cite{2017ApJ...841...73K}. Until a full radial velocity curve is available, this object is excluded from the BANYAN~$\Sigma$ kinematic models.

2MASS~J06085283--2753583 was identified by \cite{2010ApJ...715L.165R} as a candidate member of $\beta$PMG and it now has full kinematic measurements, but \cite{2016ApJS..225...10F} showed that it is an ambiguous member, which is probably due to its small proper motion. This object was therefore not included in the kinematic models of BANYAN~$\Sigma$.

\subsection{New Bona Fide Members}\label{sec:asso_newbfide}

\added{New bona fide members that could be defined as such based on Gaia-DR1 data are described in this section, and are listed in Table~\ref{tab:newbonafide}.}

The candidate members of THA identified by \cite{2014AJ....147..146K} were cross-matched with Gaia-DR1. Seven were found to have a parallax measurement: three of them did not match any moving group in BANYAN~II and were therefore rejected (2MASS~J02000918--8025009, 2MASS~J02105538--4603588 and 2MASS~J05332558--5117131), and the other 4 were confirmed as new bona fide members of THA (2MASS~J02303239--4342232, 2MASS~J04000382--2902165, 2MASS~J04000395--2902280, 2MASS~J04021648--1521297).

A similar cross-match of the \cite{2014ApJ...788...81M} candidate members missing a distance measurement with Gaia-DR1 yielded 21 matches, 16 of which were confirmed as new bona fide members (5 in $\beta$PMG, 8 in THA, 2 in ABDMG and one in COL). The $UVW$ position of 2MASS~J20395460+0620118 (an ABDMG candidate from \citealt{2014ApJ...788...81M}) is a better match to the \cite{2014ApJ...783..121G} position of ARG or $\beta$PMG than that of ABDMG. It was therefore categorized as an ambiguous member until it is studied in more detail. Furthermore, \cite{2014ApJ...788...81M} assign 2MASS~J02303239--4342232 in COL, whereas \cite{2014AJ....147..146K} call it a THA member, and it is therefore categorized as an ambiguous member in this compilation.

\cite{2017AJ....154...69S} performed a survey of new low-mass members in $\beta$PMG, and identified 39 new objects with signs of youth, sky position, proper motion and radial velocities that match $\beta$PMG. As only parallaxes are still needed for them to be included in our kinematic models, their sample was cross-matched with Gaia-DR1. Five objects were found to have a parallax measurement. Four of them (HD~337919, TYC~2136--2484--1, TYC~2658--31--1, TYC~1084--672--1) have Gaia-DR1 trigonometric distances above 200\,pc, preventing a credible membership in $\beta$PMG (they were also rejected as $\beta$PMG candidates by \citealt{2017AJ....154...69S}). The last object, TYC~2703--706--1, was designated as a $\beta$PMG candidate by \cite{2017AJ....154...69S}, but has a $UVW$ position that is located at 8.9\,\kms\ from the central position of $\beta$PMG, and only 3.4\,\kms\ of that of Columba, as defined by \cite{2014ApJ...783..121G}. We therefore categorize it as an ambiguous member between $\beta$PMG and COL until it is investigated further.

Six more objects in the \cite{2017AJ....154...69S} sample have a parallax measurement from other works in the literature \citep{2012ApJ...758...56S,2014AJ....147...85R}; 5/6 were already included in the list of bona fide members presented here. The remaining star, AF~Psc, has a parallax measurement by \cite{1995gcts.book.....V}, which seems to have been overlooked in previous studies. It was thus added to the list of bona fide members of $\beta$PMG.

\cite{2017ApJ...840...87R} presented several new M-type young moving group candidates, however none of them benefit from a parallax distance measurement either in Gaia-DR1 or elsewhere in the literature, hence they were not included to the list of bona fide members.

\subsection{Calculation of the 6D Kinematics}\label{sec:asso_calc}

The Galactic positions $XYZ$ and space velocities $UVW$, in a right-handed system where $U$ points toward the Galactic center, were calculated for all members by assuming Gaussian error bars in sky position, proper motion, radial velocity and parallax. A $10^4$-element Monte Carlo approach was used to propagate error bars in $XYZUVW$ space, by adopting the standard deviation of each coordinate as its measurement error, and therefore assuming that error bars are Gaussian in $XYZUVW$ space. The resulting list of bona fide members is presented in Table~\ref{tab:bonafide}. A list of new bona fide members confirmed in this work is given in Table~\ref{tab:newbonafide}\deleted{.}\added{, and their positions on the sky are displayed in Figure~\ref{fig:radec}.}

\section{KINEMATIC MODELS OF YOUNG ASSOCIATIONS}\label{sec:ymgmodels}

The bona fide members compiled in Table~\ref{tab:bonafide} were used to build the kinematic models of young associations considered in BANYAN~$\Sigma$. Objects with total error bars on their Galactic position ($\sqrt{\sigma_X^2+\sigma_Y^2+\sigma_Z^2}$) above 20\% of their distance or on their space velocity ($\sqrt{\sigma_U^2+\sigma_V^2+\sigma_W^2}$) above 8\,\kms\ were excluded. Companions in binary systems were ignored, and the spatial-kinematic position of binary systems was approximated as that of the primary star to avoid the need for model-dependent mass estimates.

A rejection algorithm based on minimum spanning trees (MSTs; e.g. see \citealt{2009MNRAS.395.1449A,2015ApJ...798...73G}) was used to ignore outliers in spatial and kinematic space. Groups that required adopting average radial velocities or distances for some high-likelihood members were exempted from this rejection step because of their small number of bona fide members (118TAU, CRA, EPSC, ETAC, PL8, ROPH, TAU, THOR, UCRA and XFOR). \added{If these groups are contaminated by outliers that originate from larger spatial or kinematic distributions, this exemption will result in models that are artifically biased to larger sizes, therefore increasing the rate of contamination that these groups may be subject to. The discovery of more members will be necessary to better assess and correct this effect.}A spanning tree is built by connecting each star of a young association in $XYZ$ or $UVW$ space with straight lines while avoiding loops; the MST is the spanning tree with the shortest total length. MSTs provide a measurement of scale that does not depend on the shape of a distribution, and therefore does not require making the assumption that the stars are normally distributed, or aligned with the $XYZUVW$ axes.

For each association containing a total of $N_k$ members, the algorithm of \cite{2004MNRAS.348..589C} was used to build $N_k+1$ MSTs, in both $XYZ$ and $UVW$ spaces separately. The first spatial and kinematic MSTs with respective total lengths $L_{\rm spa}$ and $L_{\rm kin}$ include all members, and the $N_k$ additional MSTs ignore one member at a time, and have respective total lengths $L_{i,\rm spa}$ and $L_{i,\rm kin}$.

\input{table_young_pars.tex}

Relative lengths $L_{i, \rm rel}$ of MSTs ignoring one star each were then calculated, and an arbitrary threshold $L_{\rm t}$ was set at a value 5\% smaller than the 90\% percentile value {$\left<L_{i, \rm rel}\right>_{\rm 90\%}$} of all the relative MST lengths:

\begin{align*}
	L_{i, \rm rel} &= \sqrt{\left(\frac{L_{i,\rm spa}}{L_{\rm spa}}\right)^2 + \left(\frac{L_{i,\rm kin}}{L_{\rm kin}}\right)^2},\\
	L_{\rm t} &= 0.95 \left<L_{i, \rm rel}\right>_{\rm 90\%}.
\end{align*}

Any individual MST with relative length $L_{i, \rm rel} < L_{\rm t}$ indicates that removing a single star $i$ significantly shortens the spatial and/or kinematic size of the young association, and therefore that star $i$ is an outlier. Such outliers were removed, and an iterative rejection process with a more conservative arbitrary threshold {$L_{\rm t} = 0.9 \left<L_{i, \rm rel}\right>_{\rm 90\%}$} was subsequently used until no stars are rejected.

In a few cases, the bona fide members that survived the MST rejection displayed a clumped distribution of members with some remaining outliers that were visually easy to recognize. They typically survived the MST selection cuts because they are located near at least one other outlier. As a consequence, the 1D projections of the multivariate Gaussian models compared to the location of bona fide members (see Figure~\ref{fig:twa}) were visually inspected to impose additional rejection criteria on the distribution of true members. These criteria are listed in Table~\ref{tab:xyzuvwcuts}.

The bona fide members that were rejected from the model construction are listed in Table~\ref{tab:excluded}.\added{ An average of 3 objects (typically less than 8 objects) were rejected in each young association. Only the PLE and HYA had more rejected members (10 and 15, respetively), but given their large number of members this represents less than 8\% of their populations.} These objects could still be true members of their respective associations with low-quality \added{or inaccurate }kinematic measurements, therefore we do not consider that they \deleted{must}\added{are} necessarily \deleted{be }non-members.\added{ The exclusion of such outliers will avoid biasing the spatial and kinematic sizes of the BANYAN~$\Sigma$ models to artificially large values, which would result in larger rates of contamination, as long as they are either non-members or suffer from inaccurate or low-quality kinematic measurements. If some of them are true members of a young association, it is likely that other objects with similar kinematics exist and are not currently known. In such a case, the gradual discovery of moving group members at the edges of the current models with BANYAN~$\Sigma$, or other methods to identify new moving groups (e.g., see \citealt{2017AJ....153..257O}) will make it possible to uncover them. Careful searches using BANYAN~$\Sigma$ with the kinematics-only mode described in Section~\ref{sec:uvwonly} will also make it possible to identify such groups of previously unrecognized members that are outside of the spatial dimensions of the current models, but not outside of their kinematic dimensions. Such investigations are left for future work, and all members that are rejected here will be ignored in what follows.}

The weighted averages $Q_{,\rm avg}$ and covariances $C_{ij}$ of all spatial-kinematic coordinates $\{Q_i\}$ were calculated, where the weight of a given star is set to the inverse square of its spatial and kinematic error bars $\sigma_{k,\rm spa}$ and $\sigma_{k,\rm kin}$ relative to the association averages $\sigma_{\rm avg, spa}$ and $\sigma_{\rm avg, kin}$, added in quadrature:
\begin{align*}
	Q_{i,\rm avg} &= \frac{\sum_k w_k Q_i}{w_{\rm tot}},\\
	C_{ij} &= \frac{\sum_k w_k \left(Q_i-Q_{i,\rm avg}\right)\left(Q_j-Q_{j,\rm avg}\right)}{w_{\rm tot}\left(w_{\rm tot}^2-\sum_k w_k^2\right)\left(\sum_k w_k^2\right)^{-1}},\\
	w_k &= \left(\left(\frac{\sigma_{k,\rm spa}}{\sigma_{\rm avg, spa}}\right)^2+\left(\frac{\sigma_{k,\rm kin}}{\sigma_{\rm avg, kin}}\right)^2\right)^{-1},\\
	w_{\rm tot} &= \sum_k w_k.
\end{align*}

The values of weights were set to a maximum of $w_k < 50$, corresponding to measurement errors 10 times more precise than the association average, to avoid the possibility of a very small number of precise $XYZUVW$ measurements bearing too much weight in the kinematic models.

The covariance matrix $\bar{\bar\Sigma}$ and center vector $\bar\tau$ of an association were then built from its components $Q_{i,\rm avg}$ and $C_{ij}$, and the covariance matrix was regularized\footnote{Here we use the term `regularized' in the sense where we ensure that the matrix is invertible, ie. that it has a positive non-zero determinant.} to avoid numerical problems. This was done through a singular value decomposition of the covariance matrix:
\begin{align*}
	\bar{\bar\Sigma} = \bar{\bar\Sigma}_U \bar{\bar\Sigma}_{\rm sv} \bar{\bar\Sigma}_V^{\rm T},
\end{align*}
\noindent where $\bar{\bar\Sigma}_{\rm sv}$ is a diagonal matrix containing the singular values. If the determinants $\left|\bar{\bar\Sigma}_U\right|$ or $\left|\bar{\bar\Sigma}_V\right|$ were found to be negative, random noise with a standard deviation equal to half the error bars was added to the $XYZUVW$ coordinates of all members until $\bar{\bar\Sigma}$ was found to be nonsingular. If an association contains less than 30 members, the three spatial and three kinematic singular values are forced to a minimum of 1\,pc and 0.2\,\kms, respectively.

Because the addition of noise in the regularization process can break the symmetry of the covariance matrix, the non-diagonal elements of the covariance matrix are forced to be symmetric by setting the values of both $\Sigma_{ij}$ and $\Sigma_{ji}$ to the average $\left(\Sigma_{ij}+\Sigma_{ji}\right)/2$. In a final step, the non-diagonal elements $\Sigma_{ij}$ were forced to values within the range $\pm \sqrt{\Sigma_{ii}\Sigma_{jj}}\left(1-10^{-5}\right)$ to respect the properties of a covariance matrix. These steps did not cause any of the covariance matrices to become singular again.

The regularization of the covariance matrix is necessary to ensure that the marginalization integrals solved in Section~\ref{sec:solve} converge. Ill-defined covariance matrices with a negative determinant would cause the analytical solution to diverge.

The resulting moving group models are displayed in Figure~\ref{fig:twa}, and their parameters are listed in Tables~\ref{tab:young_pars3} and \ref{tab:young_pars1}.\added{ The off-diagonal elements of the covariance matrices are provided in a FITS file with the BANYAN~$\Sigma$ algorithm as online-only material.}\added{ A kernel density estimate distribution was built for the 6D distribution of bona fide members using Silverman's rule of thumb, i.e., each data point is represented with a zero-covariance 6D multivariate Gaussian where the diagonal elements of the covariance matrix are given by:}
\begin{align}\notag
	\Sigma_{ii} &= \left(2 N_k\right)^{-1/5}\sigma_i^2,\\
	\Sigma_{ij} &= 0, i \neq j,
\end{align}
\noindent \added{where $\sigma_i$ is the standard deviation of the dimension $i$ of the members' positions, and $N_k$ is the total number of members. The 1D projections of the kernel density estimate distribution are shown in the panels of Figure~\ref{fig:twa} that display histograms of the members' positions, and the residual difference between the multivariate Gaussian models and the 2D projections of the kernel density estimate distributions are shown in blue- and green-shaded backgrounds with the 2D distributions of members. There are several cases (such as TWA) where a number of projections show over- and under-densities of members by up to $\approx$\,40\% of the multivariate Gaussian model, but we recommend against using multivariate Gaussian mixture models that would more correctly reproduce the distribution of known members until a large number of members are known (e.g., with the release of Gaia-DR2). Modelling the currently significantly incomplete distributions of association members with more complex models would negatively affect the ability of BANYAN~$\Sigma$ to recover the missing members that are located between the clumps of currently known members (i.e., in the blue-shaded regions in Figure~\ref{fig:twa}).}

The distance and radial velocity distributions of young moving group models were built by drawing $10^5$ synthetic objects from their spatial multivariate Gaussian model. The average distance or radial velocity was taken as the peak value of the resulting distribution and asymmetric characteristic widths covering half of 68\% of the area under the curve on each side were measured. The resulting distance and radial velocity distributions as a function of the association ages are listed in Table~\ref{tab:young_pars3} and displayed in Figures~\ref{fig:ymg_distances} and \ref{fig:ymg_rvs}. These figures illustrate the range in distances and radial velocities where new moving group members of a given age can likely be discovered using BANYAN~$\Sigma$.

In Figures~\ref{fig:ymg_sizes} and \ref{fig:ymg_scatters}, the characteristic spatial and kinematic scales $S_{\rm spa}$ and $S_{\rm kin}$ are displayed for different associations as a function of age, with:
\begin{align*}
	S_{\rm spa} &= \sqrt{|\boldsymbol\Sigma_{\rm spa}|^{1/3}},\\
	S_{\rm kin} &= \sqrt{|\boldsymbol\Sigma_{\rm kin}|^{1/3}},
\end{align*}
\noindent where $\boldsymbol\Sigma_{\rm spa}$ and $\boldsymbol\Sigma_{\rm kin}$ are 3$\times$3 matrices that contain only the purely spatial or kinematic terms of the covariance matrix, respectively. The values of $S_{\rm spa}$ and $S_{\rm kin}$ are listed for each young association in Table~\ref{tab:young_pars3}.

Figure~\ref{fig:ymg_sizes} illustrates how clusters are spatially much smaller than other types of associations, which become more dispersed as they age. The velocity dispersion of associations included in BANYAN~$\Sigma$, as displayed in Figure~\ref{fig:ymg_scatters}, does not show a clear correlation with age. TAU is a clear outlier in both figures because a single multivariate Gaussian model is used to represent all of the TAU sub-groups. 

The distribution of associations in the Galactic plane is displayed in Figures~\ref{fig:XY_groups} and \ref{fig:XZ_groups}. The spatial location of the associations are defined as the contour that encompasses 68\% of the projected multivariate Gaussian model, with an arbitrary minimum minor axis set at 4\,pc for display. This figure illustrates how several associations are spatially close to each other, and how some of the nearest ones ($\beta$PMG, ABDMG) encompass the Sun. OCT is spatially the largest association because its members are spatially distributed in two distinct clumps, even though they share the same kinematics. Here we leave OCT as a single group because this may allow for the identification of new OCT members located spatially between the two clumps of known members.

\begin{figure*}[p]
	\centering
	\subfigure[Distance versus age]{\includegraphics[width=0.49\textwidth]{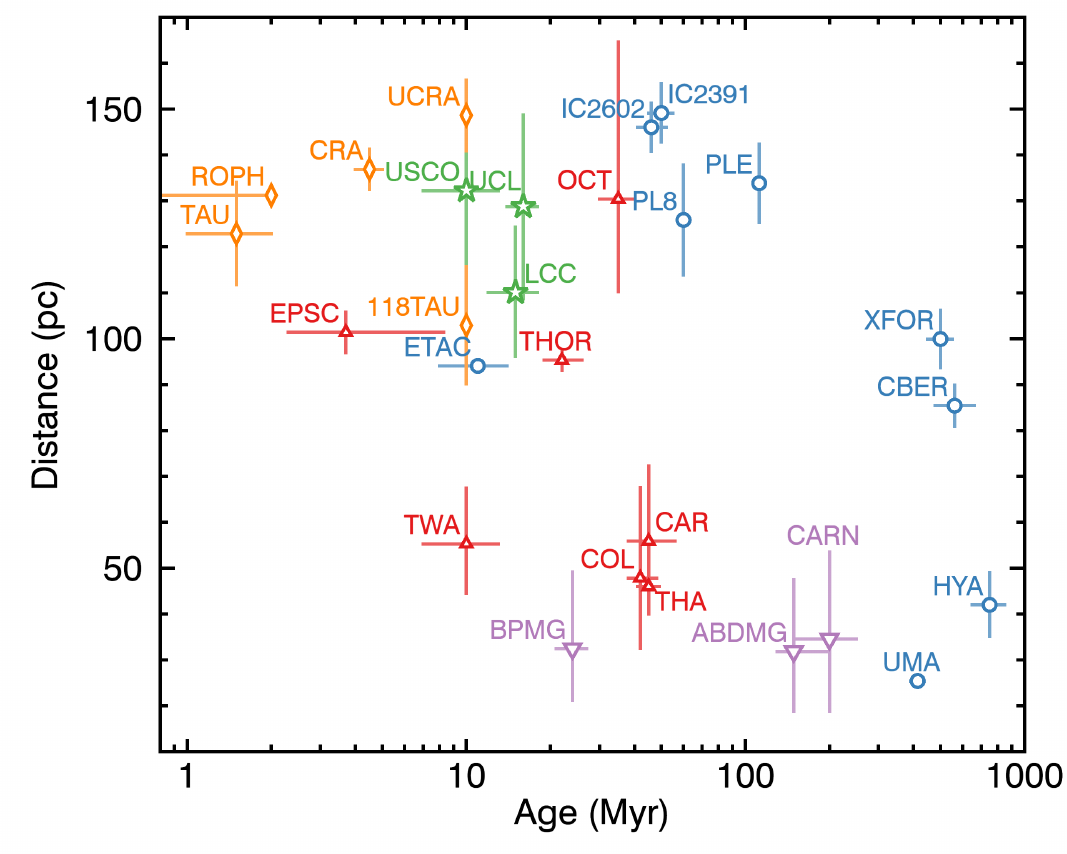}\label{fig:ymg_distances}}
	\subfigure[Radial velocity versus age]{\includegraphics[width=0.49\textwidth]{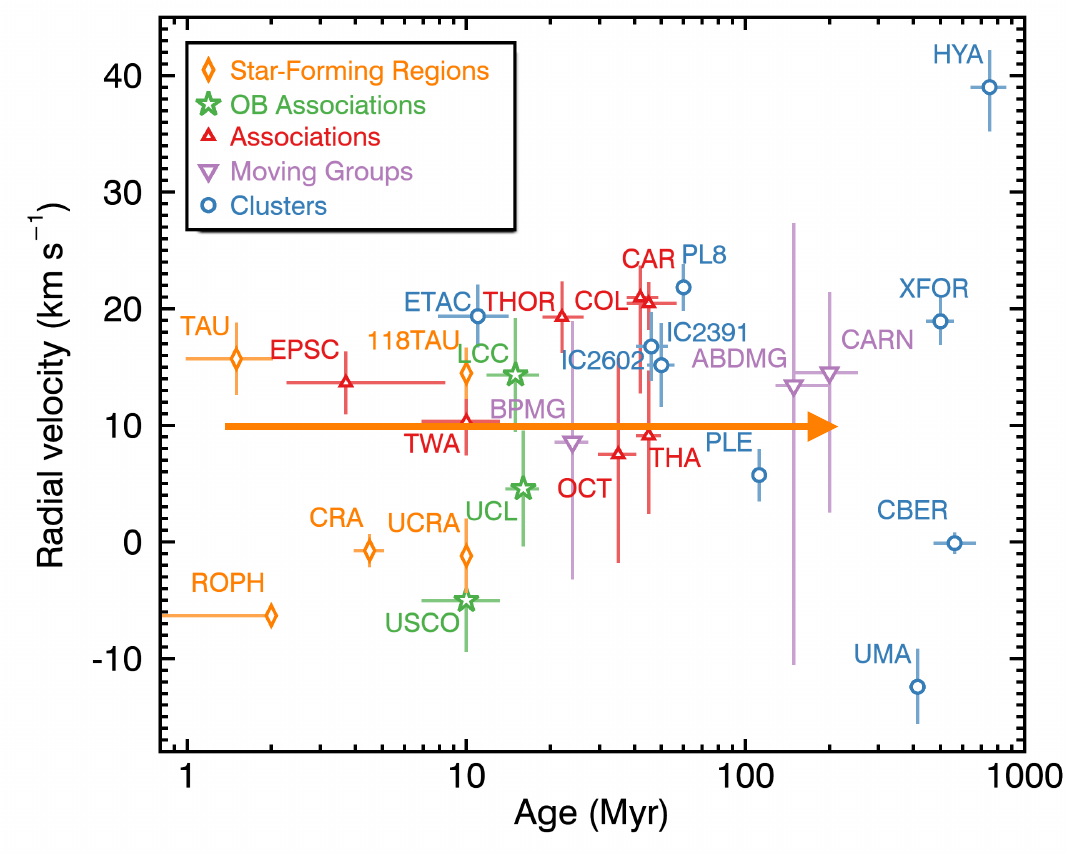}\label{fig:ymg_rvs}}
	\subfigure[Spatial size versus age]{\includegraphics[width=0.49\textwidth]{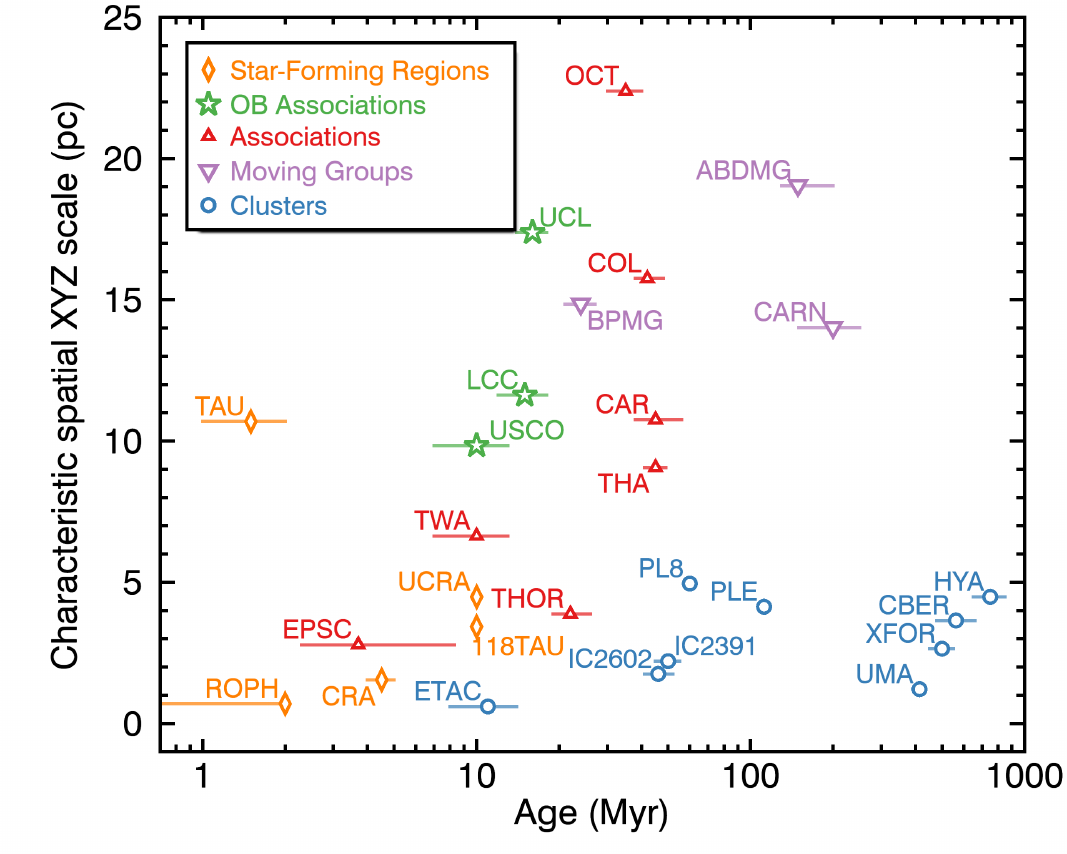}\label{fig:ymg_sizes}}
	\subfigure[Velocity dispersion versus age]{\includegraphics[width=0.49\textwidth]{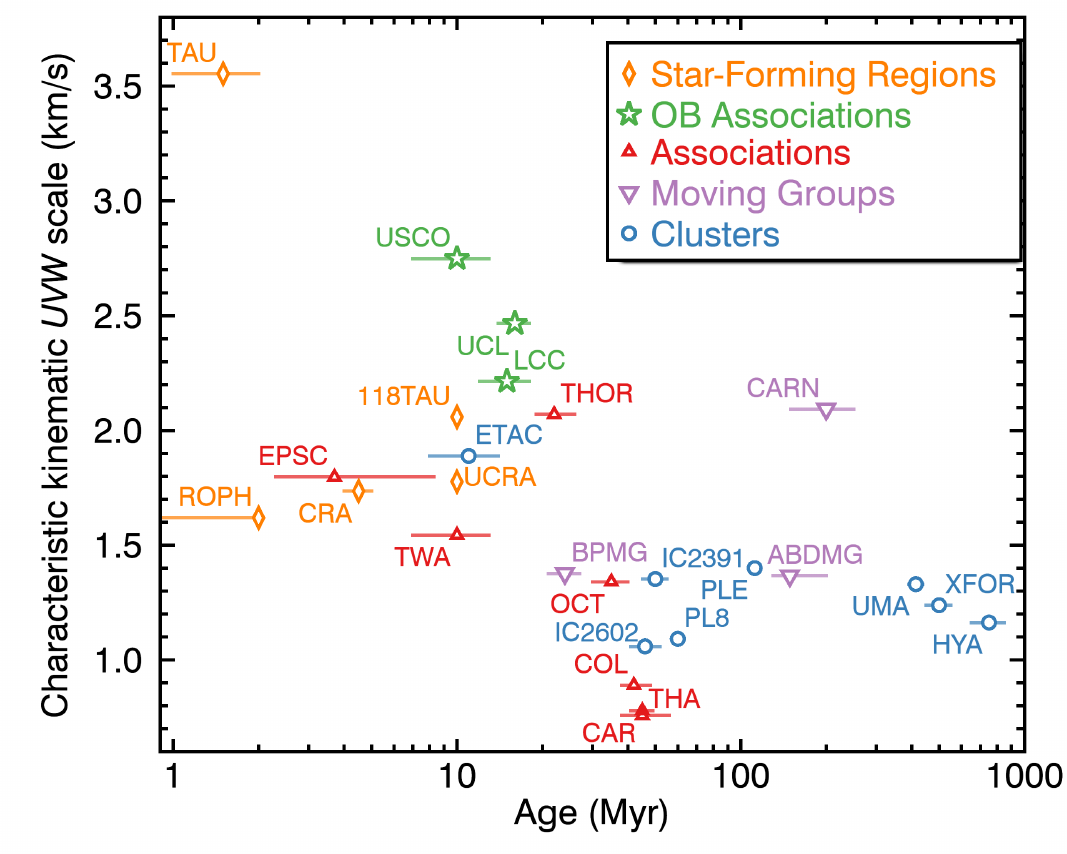}\label{fig:ymg_scatters}}
	\caption{Distance, radial velocity, spatial size and kinematic scatter of BANYAN~$\Sigma$ association models as a function of age. Associations of the Solar neighborhood provide individual epochs that cover a large period of ages, relevant to disk evolution, planetary formation and brown dwarfs atmospheric cooling. Associations in our sample seem to display an increasing spatial size as a function of age, except for the denser open clusters. TAU is also an exception because its model includes several sub-groups. See Section~\ref{sec:ymgmodels} for more detail.}
	\label{fig:ymg_ages}
\end{figure*}

\begin{figure*}[p]
	\centering
	\includegraphics[width=0.98\textwidth]{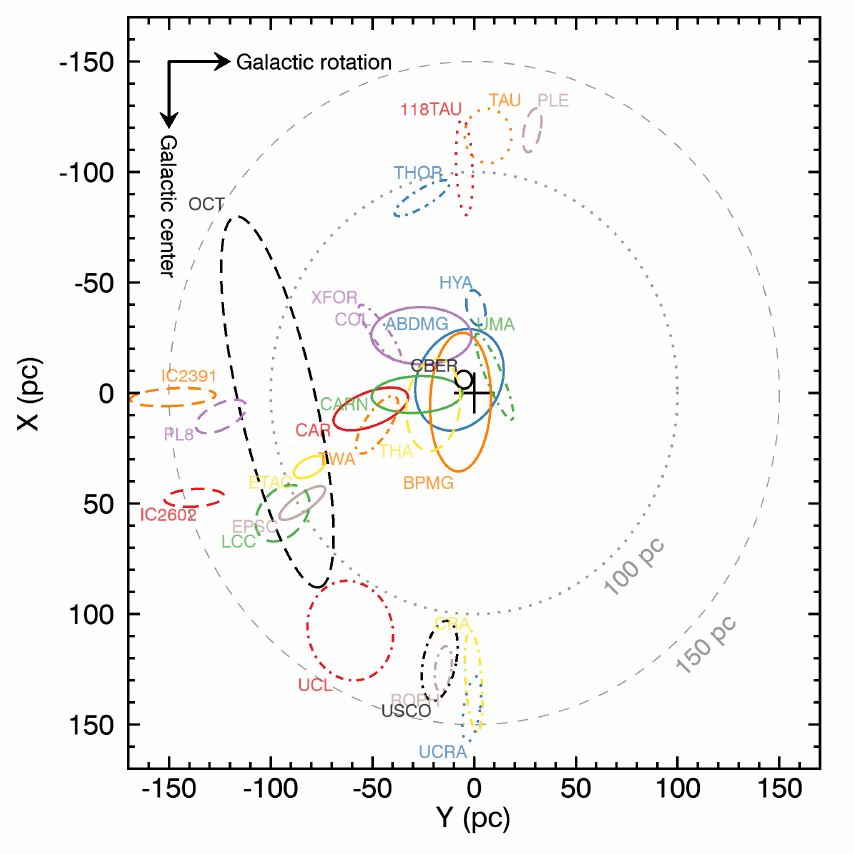}
	\caption{Distribution in Galactic coordinates $X$ and $Y$ of all moving group models constructed in Section~\ref{sec:kinmodels}. The models included in BANYAN~$\Sigma$ cover all known associations and star-forming regions within 150\,pc. This figure is an update of Figure~8 in \cite{2011ASPC..448..481R}, although it is limited to 150\,pc instead of 200\,pc. See Section~\ref{sec:ymgmodels} for more detail.}
	\label{fig:XY_groups}
\end{figure*}

\begin{figure*}[p]
	\centering
	\includegraphics[width=0.98\textwidth]{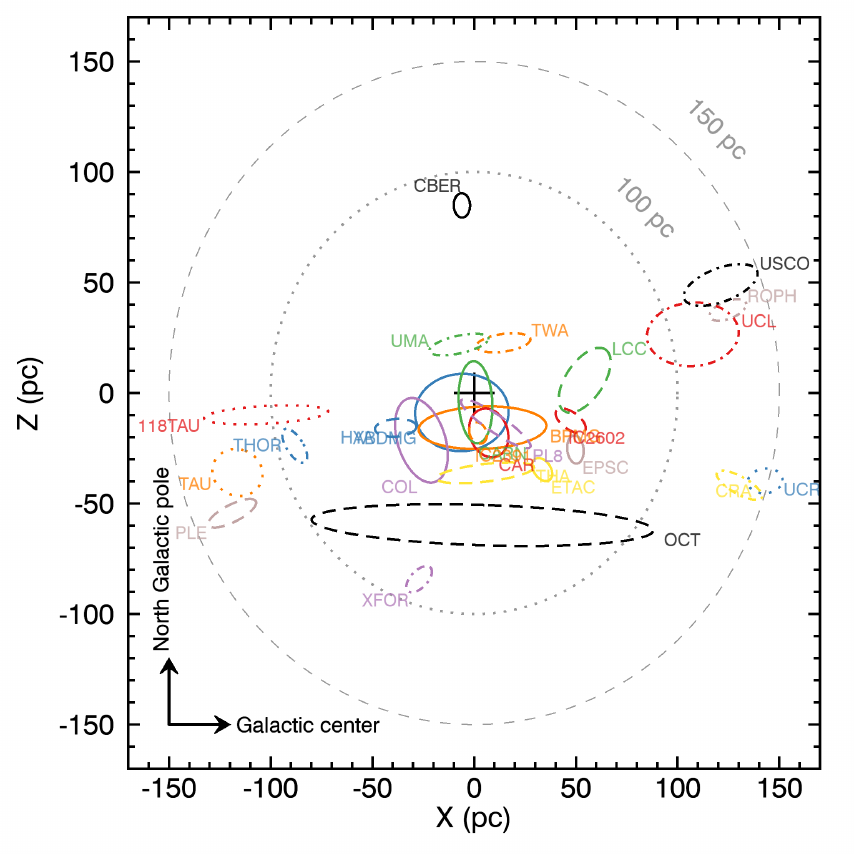}
	\caption{Distribution in Galactic coordinates $X$ and $Z$ of all moving group models constructed in Section~\ref{sec:kinmodels}. The color and linestyle coding is the same as that of  Figure~\ref{fig:XY_groups}. See Section~\ref{sec:ymgmodels} for more detail.}
	\label{fig:XZ_groups}
\end{figure*}

\section{A MODEL OF FIELD STARS IN THE SOLAR NEIGHBORHOOD}\label{sec:fieldmodel}

This section describes the construction of a kinematic model for the 
field hypothesis. It is based on the Besan\c{c}on model \citep{1996AA...305..125R,2003AA...409..523R,2012AA...538A.106R,2014AA...569A..13R,2017AA...605A...1R}\footnote{Available at \url{http://model2016.obs-besancon.fr/}} of the Galactic disk in the Solar neighborhood (with a very small contribution from the Galactic halo), and uses the multivariate Gaussian formalism described in Section~\ref{sec:kinmodels} for it to be compatible with the solution of the marginalization integrals developed in Section~\ref{sec:solve}.

The Besan\c{c}on Galactic Model version used here follows the scheme described by \cite{2014AA...564A.102C} for the thin disk population, which is based on their Model~B (see their Table~5). In summary, thin disk stars are generated from a 3-slopes initial mass function and a decreasing star formation rate, and follow the evolutionary tracks of \cite{1994AAS..106..275B,2008AA...484..815B,2009AA...508..355B} for masses larger than $0.7$\msol, and from \cite{1997AA...327.1039C} for lower masses. Companions in binary systems are generated with a probability function that depends on the spectral type of the primary, and follow empirical mass-ratio and semi-major axis distributions, as described by \cite{2011AIPC.1346..107A}. The thick disk and halo populations are simulated with the best-fitting parameters obtained in the analysis of \cite{2014AA...569A..13R} based on SDSS \citep{2015ApJS..219...12A} and 2MASS data, and the isochrones of \cite{1992ApJS...81..163B}.

There are two complications that prevent a correct modeling of the field star kinematics with a simple multivariate Gaussian model: (1) the distribution in $Z$ is similar to a hyperbolic secant function, which has wider wings than a Gaussian distribution; and (2) the distributions in $X$ and $Y$ are approximately uniform in the Solar neighborhood.

The first problem can be addressed by modelling the field hypothesis with a mixture of $N$ multivariate Gaussian distributions:
\begin{align*}
	\mathcal{P}_\mathrm{field} &= \sum_{j=1}^N c_j\,\mathcal{P}_{j,\mathrm{field}},\\
	\mathcal{P}_{j,\mathrm{field}} &= \frac{1}{\sqrt{\left(2\pi\right)^{6}\left|\bar{\bar\Sigma}_j\right|}} \exp{\left(-\frac{1}{2}\left(\bar{Q}-\tau_j\right)^T\bar{\bar{\Sigma}}_j^{-1}\left(\bar{Q}-\bar\tau_j\right)\right)},\\
	\sum_{j=1}^N c_j &= 1,
\end{align*}
\noindent which yields the solution described in Equation~\eqref{eqn:bayesoln} for the probability $\mathcal{P}(\{O_i\}|H_{j,\mathrm{field}})$ associated with field component $j$. The resulting field probability will then be:
\begin{align*}
	\mathcal{P}(\{O_i\}|H_{\mathrm{field}}) = \sum_{j=1}^N c_j \mathcal{P}(\{O_i\}|H_{j,\mathrm{field}}).
\end{align*}

The second problem of the approximately uniform $X$ and $Y$ distributions can be mitigated by artificially inflating the two diagonal elements of all covariance matrices $\bar{\bar\Sigma}_j$ corresponding to the $X$ and $Y$ dimensions by a factor much larger than the typical distances which will be involved in using BANYAN~$\Sigma$. The density of the field model will however need to be re-adjusted to avoid affecting the stellar density in the Solar neighborhood.

The very small covariance between all $XYZUVW$ coordinates of field stars compared to their variances (the Pearson correlation coefficient of all dimensions is smaller than 0.1) makes the problem of fitting a mixture of multivariate Gaussians significantly easier. The covariance matrices $\bar{\bar{\Sigma}}_j$ can be assumed diagonal and the fitting can be done simultaneously in four one-dimensional spaces $Z$, $U$, $V$ and $W$ instead of a single 4-dimensional space. The $X$ and $Y$ coordinates are ignored in a first step, as they will be approximated as uniform by using large Gaussian widths.

Multivariate Gaussians mixtures with $N=1$ to 10 components were fitted to the $Z$, $U$, $V$ and $W$ distributions of field stars using a Levenberg-Marquardt least-squares fit. The best-fitting models as well as the individual components of the $N = 10$ model are shown in Figure~\ref{fig:field_fits}. Models with $N > 6$ provide a good visual fit to all $Z$, $U$, $V$ and $W$ components.

In Figure~\ref{fig:chi2_ngauss_field}, the reduced $\chi^2$ as a function of the number of mixture components $N$ is displayed for each dimension, and for the global fit across $ZUVW$. This figure shows that the goodness-of-fit does not improve significantly at $N > 7$ for the kinematic dimensions $UVW$, but $Z$ keeps improving up to $N = 9$--10. The $N = 10$ components model was adopted, as it represents a good balance between accuracy and usability.

This simpler approach was preferred to one based on the Bayesian information criterion, as the very large number of field stars would have allowed for an arbitrarily large number of mixture components, which would make BANYAN~$\Sigma$ impractical to use while causing very little difference in the calculated probabilities. The simplicity of the least-squares fitting also allowed the identification of a good general solution without needing to compute resources-intensive likelihood functions based on a very large number of field stars.

The $X$ and $Y$ components of the field model were arbitrarily set to a large characteristic width of 1\,500\,pc. To ensure that this did not affect the density of stars in the Solar neighborhood, a Monte Carlo simulation was used to draw field objects until a ratio of stars within 300\,pc to the total number of stars could be counted with a precision of less than 1\%, assuming Poisson error bars. This required a total of $10^9$ synthetic stars to be drawn and yielded a ratio of $1.69 \times 10^{-5}$, which was divided to the total number of objects within 300\,pc in the Besan\c{c}on model ($7.15 \times 10^{6}$) to re-normalize the field model. This ensures that the multivariate Gaussian mixture model has the same density of stars as the Besan\c{c}on model within 300\,pc. The adopted field model parameters are provided as a FITS file containing all input data used by the BANYAN~$\Sigma$ IDL and Python codes \citep{zenodobanyansigmapython,zenodobanyansigmaidl}.

\begin{figure*}
	\centering
	\subfigure[Galactic position $Z$]{\includegraphics[width=0.488\textwidth]{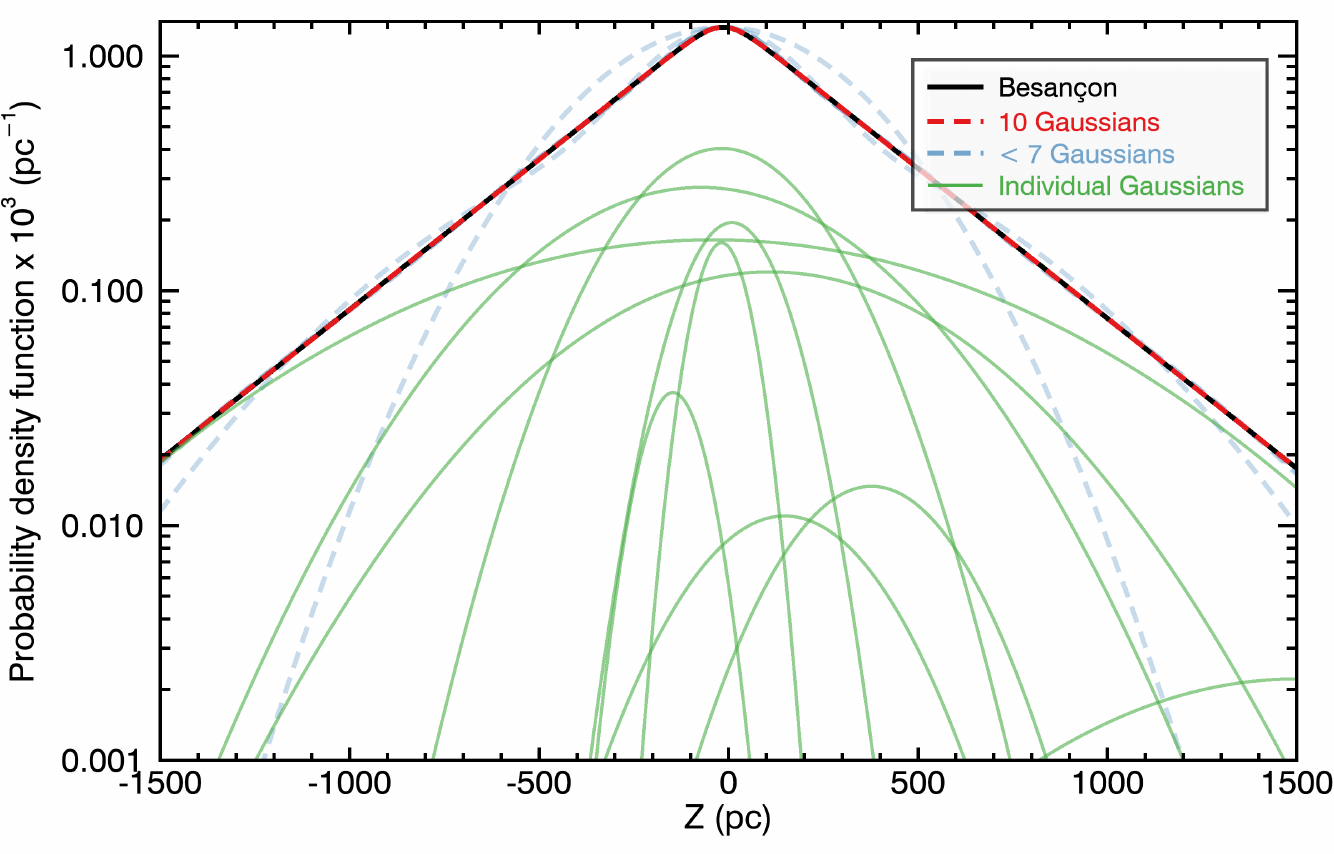}\label{fig:fieldZ}}
	\subfigure[Space velocity $U$]{\includegraphics[width=0.488\textwidth]{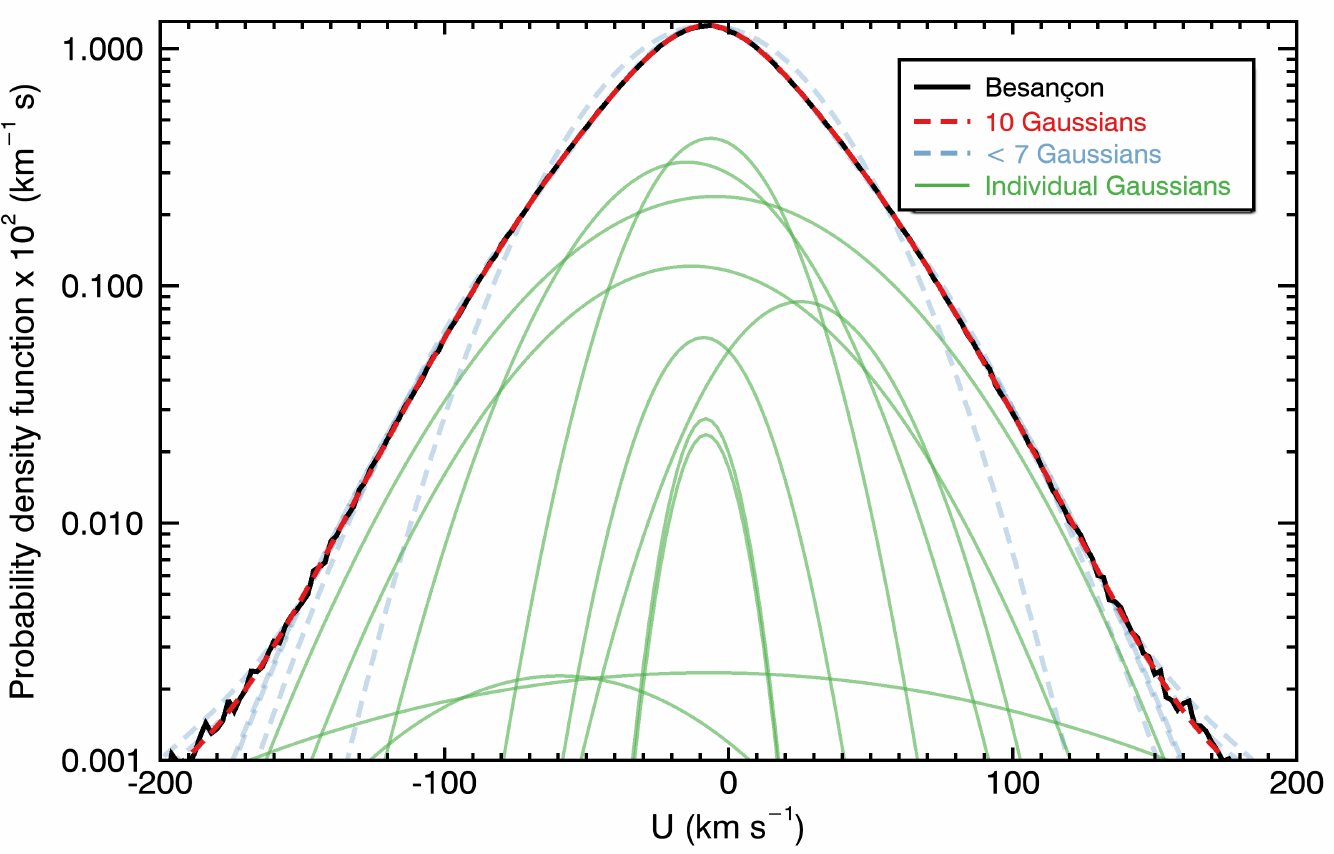}\label{fig:fieldU}}
	\subfigure[Space velocity $V$]{\includegraphics[width=0.488\textwidth]{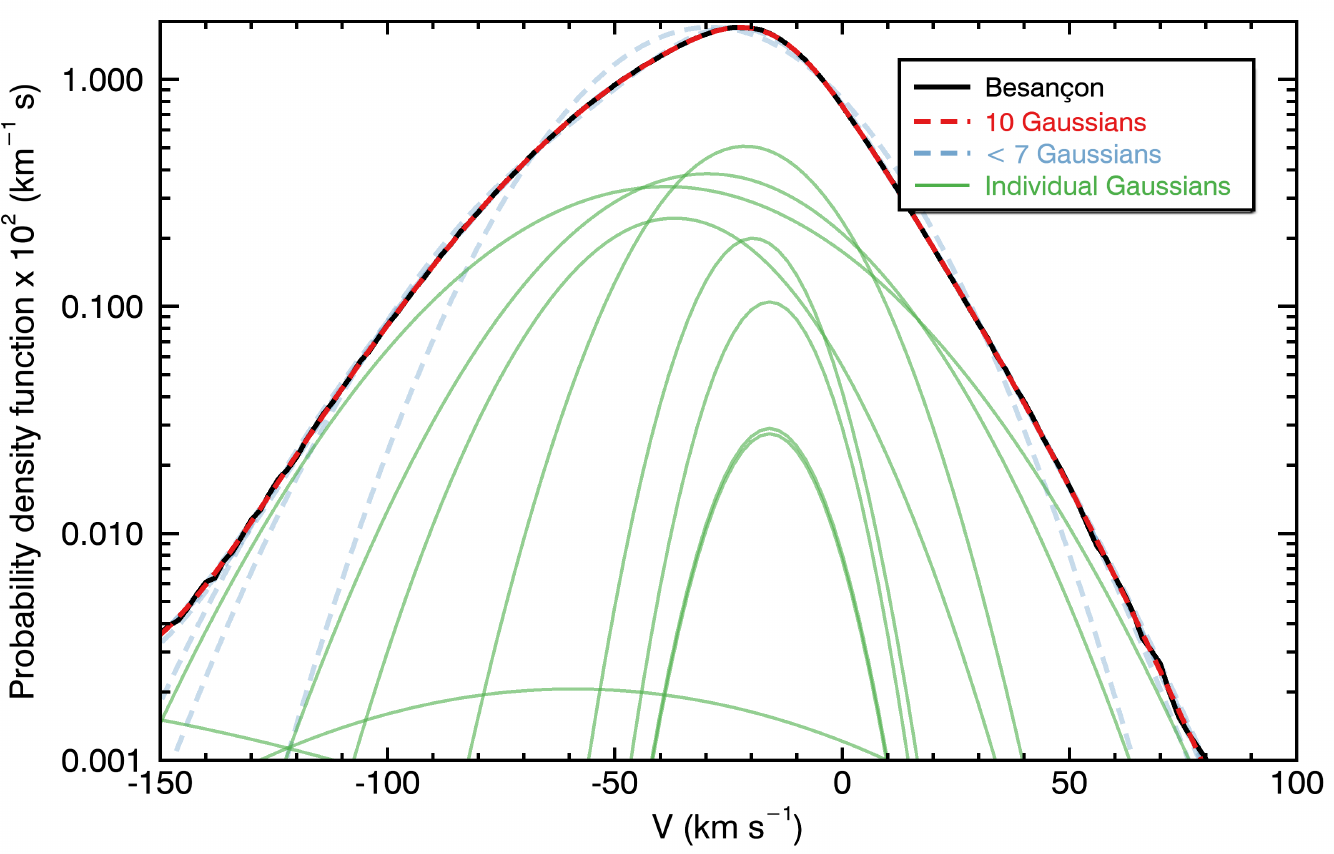}\label{fig:fieldV}}
	\subfigure[Space velocity $W$]{\includegraphics[width=0.488\textwidth]{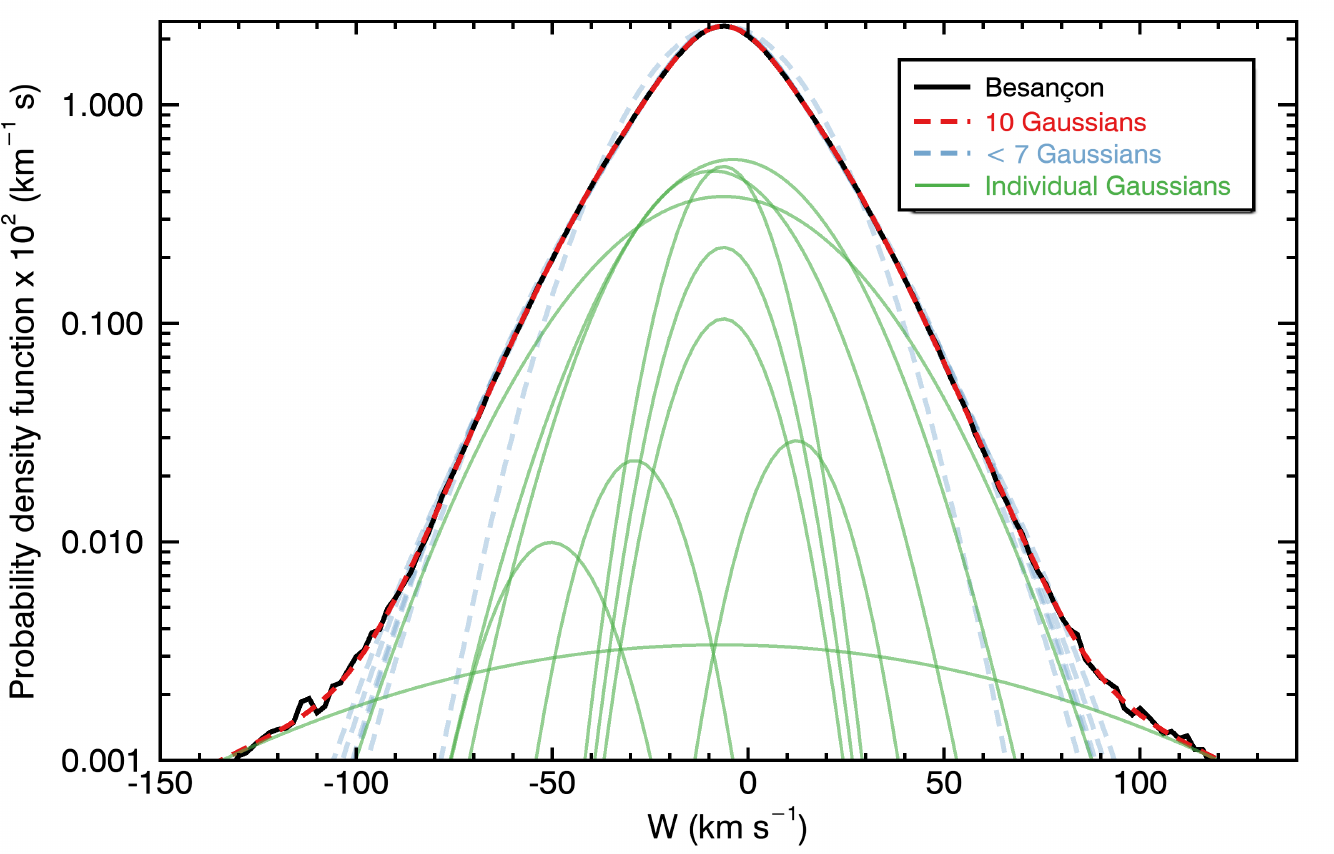}\label{fig:fieldW}}
	\caption{Unidimensional projections of the best-fitting ten-components multivariate Gaussians mixture models for the field stars (red dashed lines) compared to the distribution of stars in the Besan\c{c}on Galactic model (black solid lines). The individual components of the best-fitting models are displayed as green lines, and best-fitting models that include fewer than 7 Gaussian components are displayed in blue dashed lines. Space velocity distributions are slightly asymmetric and the distribution in $Z$ component of the Galactic position has much wider wings than a Gaussian model. The distributions in $X$ and $Y$ are approximated as uniform in the Solar neighborhood. See Section~\ref{sec:fieldmodel} for more detail.}
	\label{fig:field_fits}
\end{figure*}

\begin{figure}
	\centering
	\includegraphics[width=0.498\textwidth]{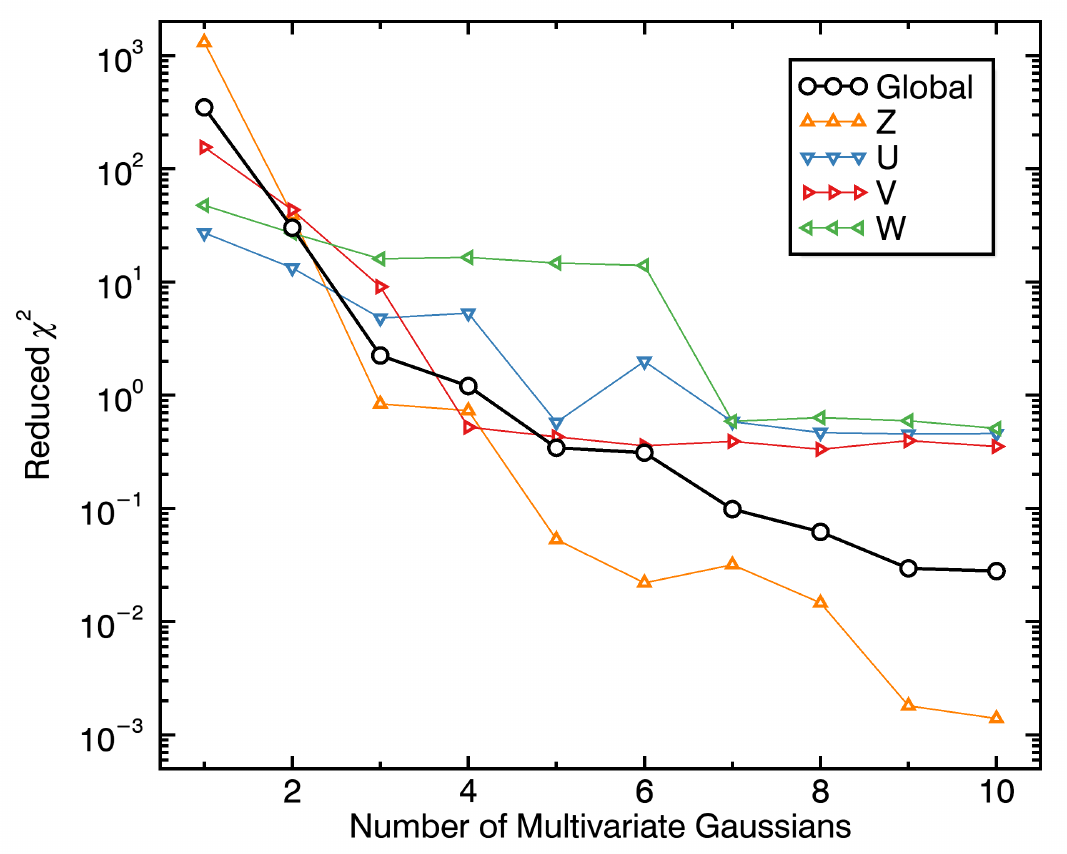}
	\caption{Reduced $\chi^2$ of the multivariate field model as a function of the number of included multivariate Gaussian components. The distribution in space velocities $UVW$ are slightly asymmetric and require $\gtrsim$\,7 Gaussian components to be properly modelled, but the distribution in $Z$ across the Galactic disk require more components as its wings are much wider than a Gaussian distribution. The distribution of stars in $X$ and $Y$ positions within the Galactic disk are approximated as uniform in the Solar neighborhood. See Section~\ref{sec:fieldmodel} for more detail.}
	\label{fig:chi2_ngauss_field}
\end{figure}

\subsection{The Spatial Size of Proper Motion and Galactic Latitude-Limited Stellar Samples}

One common way to eliminate distant stars from a sample is to impose a lower limit on the total proper motion and/or Galactic latitude. The model of the nearby Galactic disk developed in this work was used to determine the efficiency of such proper motion and Galactic latitude cuts at selecting nearby stars.

A set of 200 thresholds on the magnitude of proper motion and 5 thresholds on the Galactic latitude were selected uniformly in the ranges 5--300\,\masyr\ and 0--40\textdegree, respectively. For each combination of thresholds, the $ZUVW$ coordinates of $10^7$ stars were drawn randomly from the multivariate Gaussians mixture model of the Galactic disk. Because the $X$ and $Y$ coordinates are approximated as uniform in the Solar neighborhood, they were drawn from a uniform random distribution bounded within a distance that produces a good sampling of the distance distribution of stars selected by the proper motion threshold. Bounds of  $\pm$\,10\,000\,pc $/ \mu$\,(\masyr)  on both $X$ and $Y$ were found to be adequate. All stars with proper motions and Galactic latitudes larger than the specific set of thresholds were selected, and the smallest distance that encompasses 90\% of the sample was calculated.

The resulting distances encompassing 90\% of a sample are displayed in Figure~\ref{fig:pmlimits} as a function of proper motion and Galactic latitude selection cuts. This figure demonstrates that agressive cuts on proper motion ($\mu > 100$\,\masyr) must be used to limit a sample to distances $\lesssim$\,500\,pc. The threshold on proper motion and Galactic latitude of the \emph{BANYAN All-Sky Survey} for members of young associations \citep{2015ApJ...798...73G} only limited their sample to distances of $< 800$\,pc, much larger than the 200\,pc distance limit that was used to build the model of field stars in BANYAN~II \citep{2014ApJ...783..121G}. This problem was mitigated by the fact that the survey focused on substellar objects well detected in 2MASS, which limited the sample to distances $\lesssim$\,200\,pc for $>$\,M5-type objects. However, this outlines that a model of field stars that remains valid at much larger distances, such as the one developed in this section, is necessary to limit the rate of false-positives in all-sky searches for stellar members of young associations.

\begin{figure}
	\centering
	\includegraphics[width=0.498\textwidth]{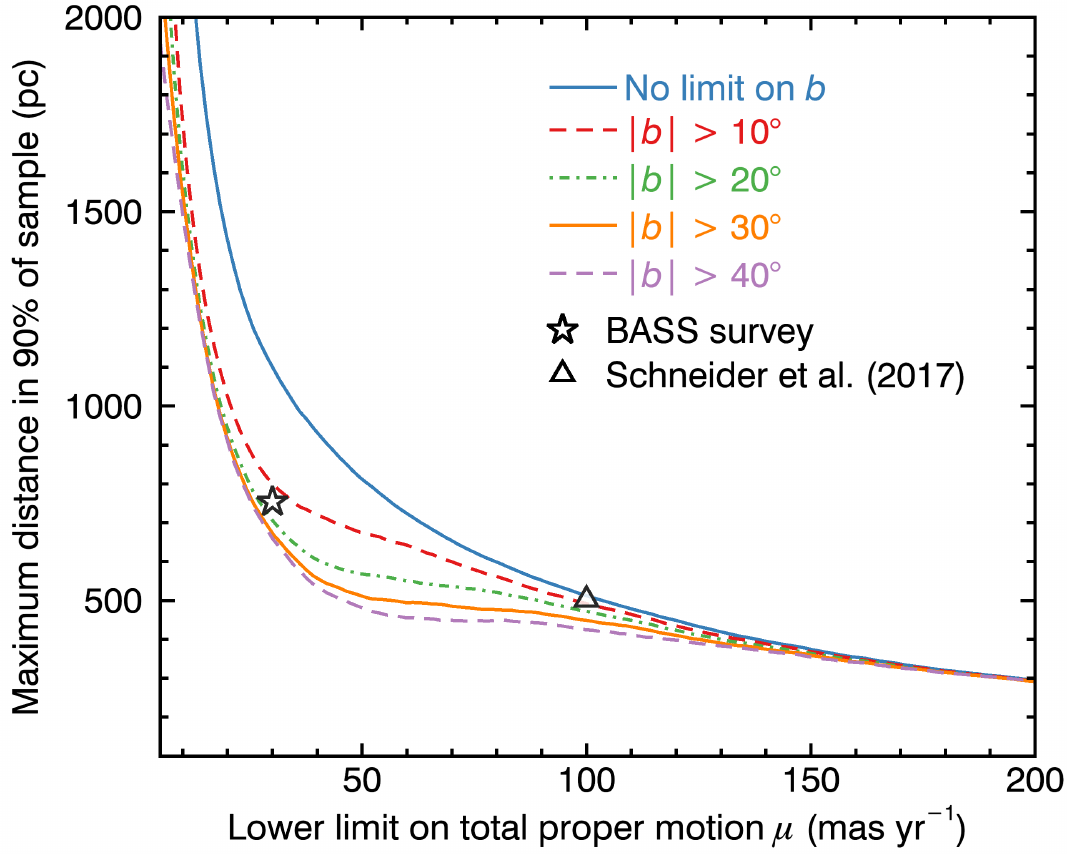}
	\caption{Largest distance encompassing 90\% of randomly selected field stars, as a function of lower cuts on total proper motion $\mu$ and absolute galactic latitude $|b|$. The criteria used in the BASS survey for young brown dwarfs \citep{2015ApJ...798...73G,2015ApJS..219...33G} and the search for red L dwarfs of \cite{2017arXiv170303774S} are displayed with a star and triangle symbols, respectively. This figure demonstrates that field interlopers with distances up to $\approx$\,750\,pc and $\approx$\,500\,pc are likely contaminating the two respective input samples. BANYAN~II has a limited capability of capturing contaminants at distances further than 200\,pc, as more distant stars were not included in its model of the Galactic field. See Section~\ref{sec:fieldmodel} for more detail.}
	\label{fig:pmlimits}
\end{figure}

\section{THE CHOICE OF BAYESIAN PRIORS}\label{sec:priors}

The contamination and recovery rates in a sample of candidate members selected with BANYAN~$\Sigma$ will be dependent on the young association where a given star is classified as a likely member. The more distant associations will provide a larger recovery rate of true association members at a fixed rate of contamination, because the members are distributed on a smaller region of the sky. As a consequence, using the same Bayesian probability threshold for the candidate members of all young associations will result in samples of vastly different sizes, completion, and contamination rates.

It is possible to adjust the Bayesian priors of the young associations in a way that equalizes the contamination or recovery rates of all young associations at an arbitrary Bayesian probability threshold.\added{ These priors will be used in the Bayesian membership probability determination (see Equation~\ref{eqn:bayesrule}).} The threshold $P = 90$\% was selected here so that BANYAN~$\Sigma$ approaches recovery rates $R_k$ of 50\% ($\mu$ only), 68\% ($\mu+\nu$), 82\% ($\mu+\varpi$), or 90\% ($\mu+\nu+\varpi$) in terms of the fraction of recovered bona fide members. This decision is arbitrary, but will allow translating the BANYAN~$\Sigma$ probabilities to survey completeness fractions simpler. This will therefore make BANYAN~$\Sigma$ easier to use in searches for new members across all \nasso\ associations.

This was done for each young association $H_k$ with a Monte Carlo method. The $XYZUVW$ coordinates of $10^7$ synthetic stars were drawn from the kinematic model of the association. All coordinates were transformed to sky position, proper motion, radial velocity and distance. Gaussian random error bars of $10$\,\masyr\ were added to each component of the proper motion, and radial velocity and distance were assumed to be missing. Simplified BANYAN~$\Sigma$ probabilities $P^\prime(H_k|\{O_i\})$ were determined for all synthetic objects by ignoring all other groups $H_j, j \neq k$:
\begin{align*}
	P^\prime(H_k|\{O_i\}) = \frac{P^\prime(H_k)\,P^\prime(\{O_i\}|H_k)}{P^\prime(H_\mathrm{field})\,P^\prime(\{O_i\}|H_\mathrm{field})},
\end{align*}
\noindent where all initial priors $P^\prime(H_k)$ and $P^\prime(H_{\rm field})$ were set to unity.

A set of $10^5$ probability thresholds $P_t$ were defined to explore how they affected the number of true positives ($N_{\rm TP}$; association members with $P^\prime(H_k|\{O_i\}) \geq P_t$) and false negatives ($N_{\rm FN}$; association members with $P^\prime(H_k|\{O_i\}) < P_t$).

The probability thresholds $P_t$ were distributed along:
\begin{align*}
	P_t &= \frac{1}{2}\left(\frac{\arctan{\left(\omega x\right)}}{\arctan{\left(\omega\right)}}+1\right),\\
	\omega &= 95,
\end{align*}
\noindent where x is an array of $10^5$ uniformly distributed values in the range $[-1,1]$. This produces an array of thresholds where $P_t \approx 0$ and $P_t \approx 1$ are especially well sampled as $\omega$ takes larger values.

Recovery rates $R_k$ (also called `true positive rates' or TPRs) are determined as a function of threshold $P_t$, with:
\begin{align*}
	R_k = \frac{N_{\rm TP}}{N_{\rm TP}+N_{\rm FN}},
\end{align*}
\noindent where $N_{\rm TP}$ is the number of synthetic stars originating from the model of association association $H_k$ with $P_k \geq P_t$, and $N_{\rm FN}$ is the number with $P_k < P_t$.

For each young association, the probability threshold $P_{t,\rm crit}$ that generates the desired recovery rate (50--90\% depending on observables) was then selected, and a multiplicative factor $\alpha_k$ that ensures $P_{t,\rm crit} = 90$\% when included to the Bayesian prior is then determined:
\begin{align*}
	\alpha_k = \frac{0.9\left(1-P_{t,\rm crit}\right)}{\left(1-0.9\right)P_{t,\rm crit}}.
\end{align*}

To avoid biasing the young association probabilities of ambiguous candidate members, an average of each young association factor $\alpha_k$, weighted by the individual membership probabilities $P(H_k|\{O_i\})$, is used to determine the field prior:
\begin{align*}
	P(H_k) &= 1,\\
	P(H_{\rm field}) &= \left(\frac{\sum^\prime_k \alpha_k P(H_k|\{O_i\})}{\sum^\prime_k P(H_k|\{O_i\})}\right)^{-1},
\end{align*}
\noindent where the sum $\sum^\prime_k$ excludes $k = \rm{field}$. The quantities $\alpha_k$ are fixed for all candidate stars, but $P(H_{\rm field})$ has to be computed for each star, since it depends on $P(H_k|\{O_i\})$.

This choice of priors ensures that the recovery rates are similar (between 50\% and 90\% depending on the available observables) across all young associations when a probability threshold $P = 90$\% is adopted, without biasing the relative young association membership probabilities. However, the false-positive rates at $P = 90$\% will be different for each association, and are discussed in Section~\ref{sec:performance}. The resulting $\alpha_k$ values are listed in Table~\ref{tab:young_pars3}.

\begin{figure*}
	\centering
	\subfigure[Performance for young stars recovery]{\includegraphics[width=0.488\textwidth]{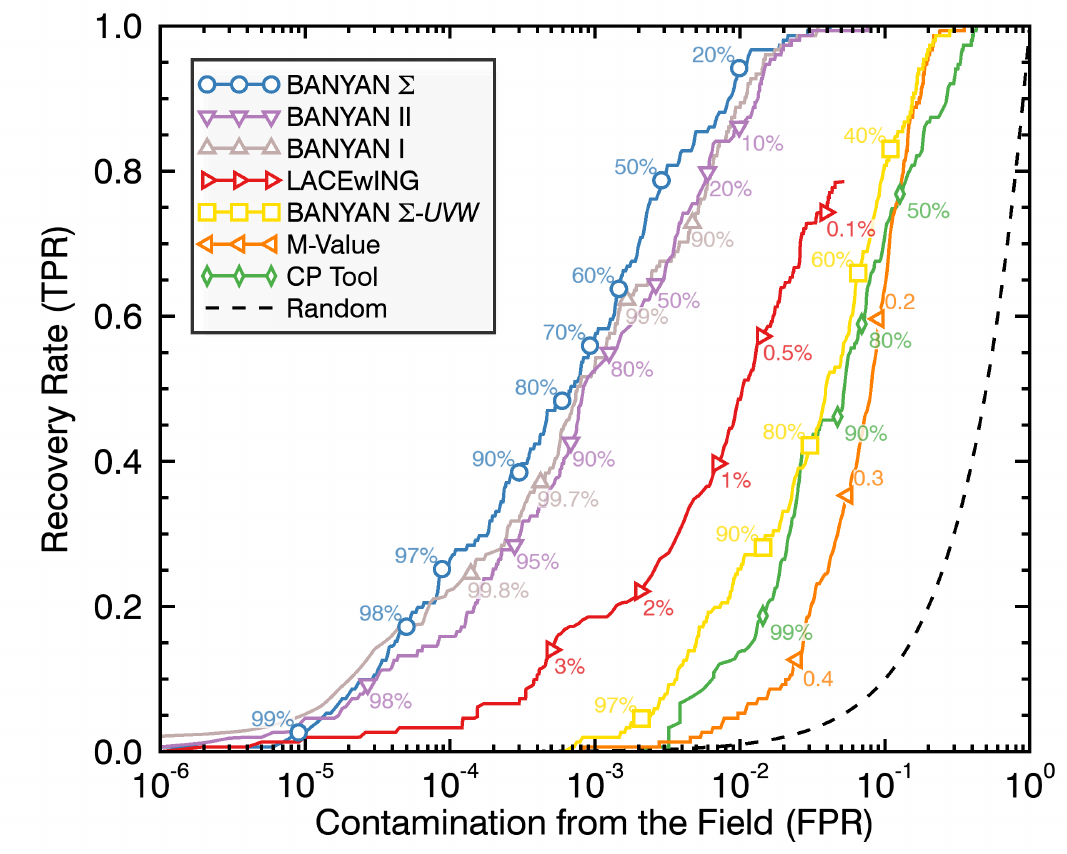}\label{fig:roc_curves}}
	\subfigure[Performance for association assignment]{\includegraphics[width=0.488\textwidth]{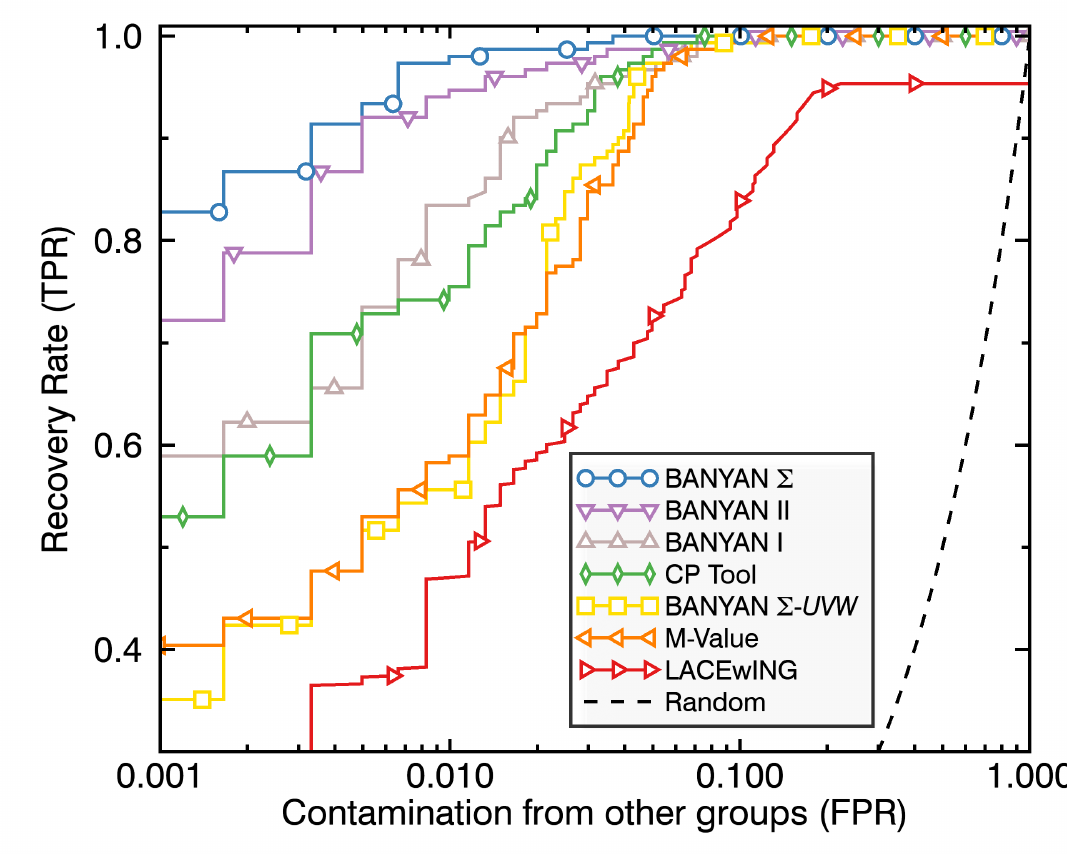}\label{fig:roc_cross_contam_curves}}
	\caption{Panel~(a): Receiver operating characteristic curves of different membership classification tools for distinguishing members of $\beta$PMG, THA, COL, TWA and ABDMG from field objects based on only sky position and proper motion, and ignoring cross-contamination between associations. CP designates the convergent point tool. BANYAN~$\Sigma$ achieves the best performance under this metric, especially in the region TPR $> 0.6$. Individual values of the probability or goodness-of-fit metric thresholds are indicated along ROC curves.\\
	Panel~(b): Receiver operating characteristic curves of different membership classification tools for distinguishing members of the same 5 associations, using only sky position and proper motion, and ignoring field contamination. Classification tools that use more complex kinematic models of young associations such as BANYAN~$\Sigma$ tend to perform better under this metric. See Section~\ref{sec:performance} for more detail.}
	\label{fig:rocs}
\end{figure*}
\begin{figure}
	\centering
	\includegraphics[width=0.488\textwidth]{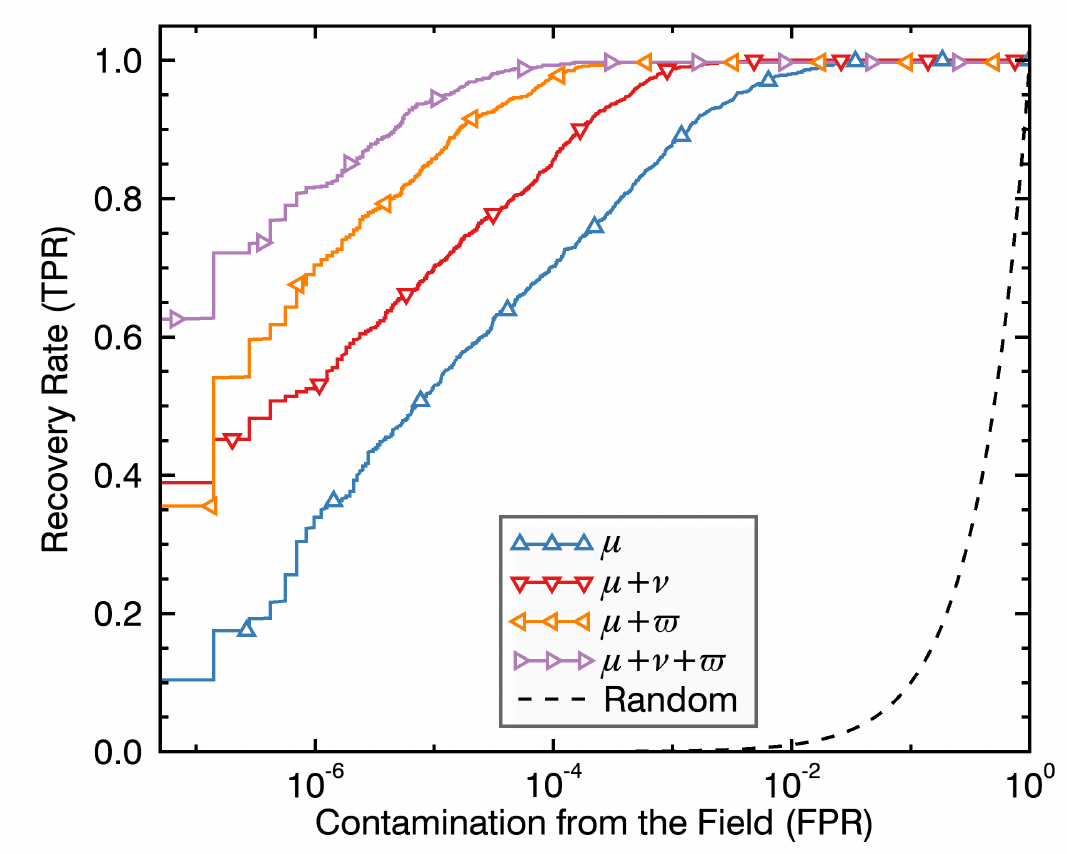}
	\caption{Receiver operating characteristic curve for BANYAN~$\Sigma$ as a function of the observables used, for all groups included here. Sky coordinates were used in all cases, and cross-contamination between the groups was ignored. The addition of radial velocity and distance cut down the rate of field contamination by factors of $\sim$\,10 and $\sim$\,100 respectively at a fixed recovery rate. See Section~\ref{sec:performance} for more detail.}
	\label{fig:roc_per_input}
\end{figure}
\begin{figure}
	\centering
	\includegraphics[width=0.488\textwidth]{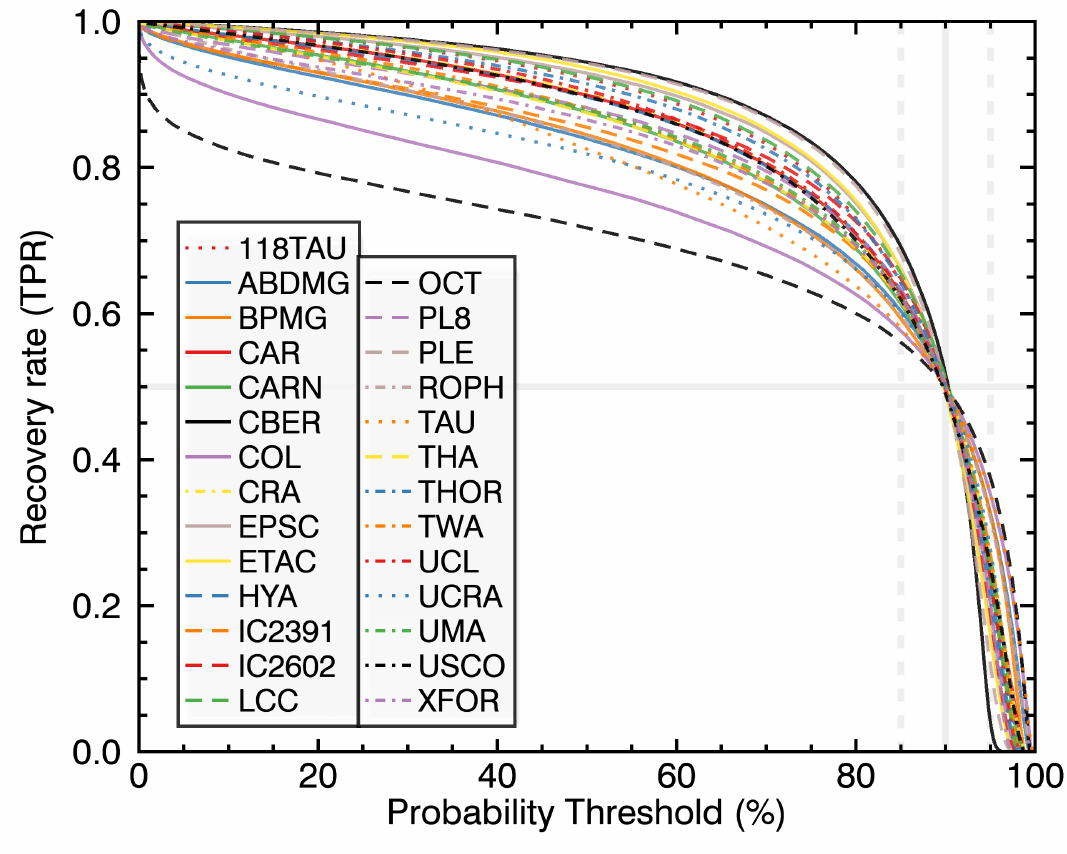}
	\caption{True positive rate (TPR) as a function of Bayesian probability threshold $P_t$ for different young associations, when only sky position and proper motion are considered. Bayesian priors were chosen for $P_t =$\,90\% to yield TPR\,$=$\,50\% in this scenario (e.g., see thick gray lines). The dashed thick gray lines represent the range in $P_t$ for which the range of recovery rates are reported in Table~\ref{tab:young_roc}. See Section~\ref{sec:performance} for more detail.}
	\label{fig:all_tpr}
\end{figure}

\section{THE PERFORMANCE OF BANYAN~$\Sigma$ AS A BAYESIAN CLASSIFIER}\label{sec:performance}

This section describes the classification performance of BANYAN~$\Sigma$ (in normal mode and in $UVW$-only mode), and compares it to those of the convergent point tool \citep{1971MNRAS.152..231J,1999MNRAS.306..381D,2005ApJ...634.1385M,2011ApJ...727...62R}, BANYAN~I \citep{2013ApJ...762...88M}, BANYAN~II \citep{2014ApJ...783..121G}, LACEwING \citep{2017AJ....153...95R} and the $\mathcal{M}$-value metric introduced by \cite{2017AJ....153...18B}. The probabilities or goodness-of-fit metrics of these different tools cannot be compared in absolute terms, however their rate of contamination as a function of the rate of true member recovery are physically meaningful and can be directly compared.

The $7.15 \times 10^{6}$ objects within 300\,pc from the Besan\c{c}on Galactic model and the bona fide members listed in Table~\ref{tab:bonafide} were used to determine the classification performance of each tool. In each case, the membership probability that each object belongs to a group $H_k$ or the field $H_{\rm field}$ were calculated using only sky position and proper motion, and the number of recovered true members $N_{k,\rm TP}$ from $H_k$ was counted for a range of thresholds $P_t$, with:
\begin{align}
	\frac{P_k}{P_k + P_{\rm field}} \geq  P_t,
\end{align}
\noindent or $P_k \geq  P_t$ for the tools that do not have a field hypothesis.

The number of false positives $N_{k,\rm FP}$ was defined as the number of Besan\c{c}on objects that respect the same criterion. This particular way of normalizing probabilities ignores cross-contamination between young groups, and instead focuses on the contamination from field stars. These quantities make it possible to build receiver operating characteristic (ROC) curves, defined as the true positives rate TPR $= N_{k,\rm TP}/N_k$ as a function of the false positives rate FPR $= N_{k,\rm FP}/N_{\rm field}$. The straight line defined by TPR $=$ FPR corresponds to the performance of a random classification, and a ROC curve that is farthest above TPR $>$ FPR corresponds to an optimal classification performance.

The ROC curve of each classification tool was built by taking the sum of $N_{k,\rm TP}$ for only the young associations that are common to all tools, interpolated on a fixed array of $N_{k,\rm FP}$. These young associations considered by all tools are $\beta$PMG, THA, COL, TWA and ABDMG. The resulting ROC curves are displayed in Figure~\ref{fig:roc_curves}. The ROC curves do not compare the particular thresholds of each different tool, which are defined in different ways, but rather compares their astrophysically meaningful TPRs as a function of FPRs, which can be directly compared in an informative way. BANYAN~$\Sigma$ achieves a performance slightly better than BANYAN~II and BANYAN~I, especially at large true-positive rates (TPR $> 0.6$). It is likely that the lack of a field model in LACEwING, the convergent point tool and the $\mathcal{M}$-value is the main reason they do not perform as well as the BANYAN tools under this metric. BANYAN~$\Sigma$ in $UVW$-only mode achieves a much lower performance under this metric, but performs better than the $\mathcal{M}$-value and the convergent point tool.

\added{The resulting FPRs of all young associations are displayed in Figure~\ref{fig:ymg_fpr} for each configuration of BANYAN~$\Sigma$. In Figure~\ref{fig:ymg_fpr_angsize}, the FPRs in the case where only sky position and proper motion are used are displayed as a function of the characteristic angular size of the young associations, defined as their characteristic spatial size $S_{\rm spa}$ (see Section~\ref{sec:ymgmodels}) divided by their distance. The FPRs are dominated by the characteristic angular size of young associations: the nearby and large associations that cover a significant area of the sky are much harder to distinguish from field stars, because there is a much larger set of field stars that can match their kinematics by pure chance. HYA and UMA suffer from much less contamination that would be expected given their characteristic angular size; this is due to their average kinematics that significantly differ from most field stars and from other young associations (see Table~\ref{tab:young_pars3}).}

Another set of ROC curves was built in a similar way to measure the cross-contamination performance, by ignoring the field hypothesis and defining false-positives as members recovered in association $H_k$ that originate from another association $H_l, l \neq k$. The resulting ROC curves are displayed in Figure~\ref{fig:roc_cross_contam_curves}. The classification tools that use more complex kinematic models tend to
perform better under this metric: BANYAN~$\Sigma$ achieves the best performance, followed by BANYAN~II, BANYAN~I, the convergent point tool, BANYAN~$\Sigma$ in $UVW$-only mode, the $\mathcal{M}$-value and LACEwING. The kinematic models of LACEwING are similar to those of BANYAN~II, and its lower performance may instead be related to approximations that are done when transforming the $N\sigma$ metrics taken in sky position and proper motion space to probabilities directly. In particular, most of the LACEwING cross-contamination is due to confusion between ABDMG and $\beta$PMG, the members of which span the widest distributions of sky positions and proper motions.

In order to characterize the classification gains that are obtained by adding more observables in BANYAN~$\Sigma$, field contamination ROC curves were built for each mode: (1) proper motion only, (2) proper motion and radial velocity, (3) proper motion and distance, or (4) proper motion, radial velocity and distance. These ROC curves were built from all associations available in BANYAN~$\Sigma$, and are displayed in Figure~\ref{fig:roc_per_input}. This figure demonstrates that using measurements of radial velocity and distance cut down field contamination by factors of $\sim$\,10 and $\sim$\,100 respectively at a fixed recovery rate. The fact that using only a distance measurement makes BANYAN~$\Sigma$ about a factor of ten better compared with only a radial velocity measurement is likely a consequence of distance being useful to constraint both $XYZ$ and $UVW$ sets of coordinates. Radial velocity only helps to constrain $UVW$. This observation was already made for BANYAN~II and LACEwING \citep{2014ApJ...783..121G,2017AJ....153...95R}.

\begin{figure*}
	\centering
	\subfigure{\includegraphics[width=0.49\textwidth]{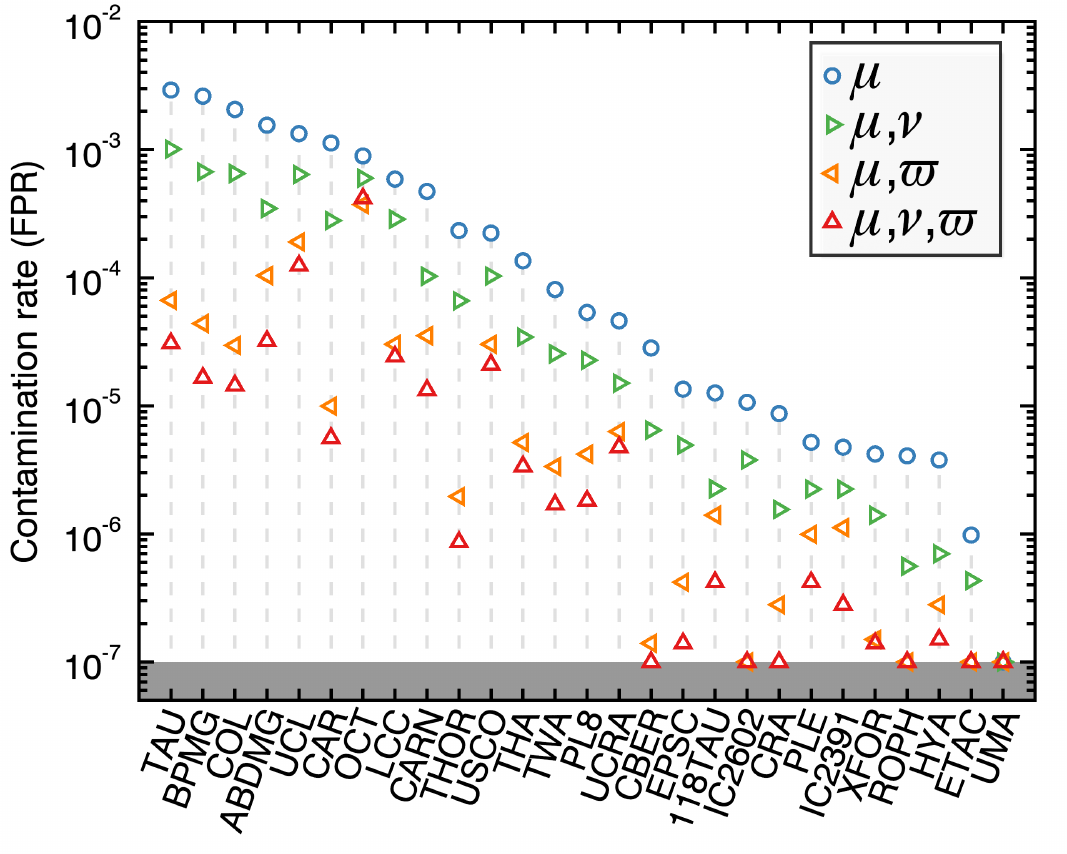}\label{fig:ymg_fpr}}
	\subfigure{\includegraphics[width=0.49\textwidth]{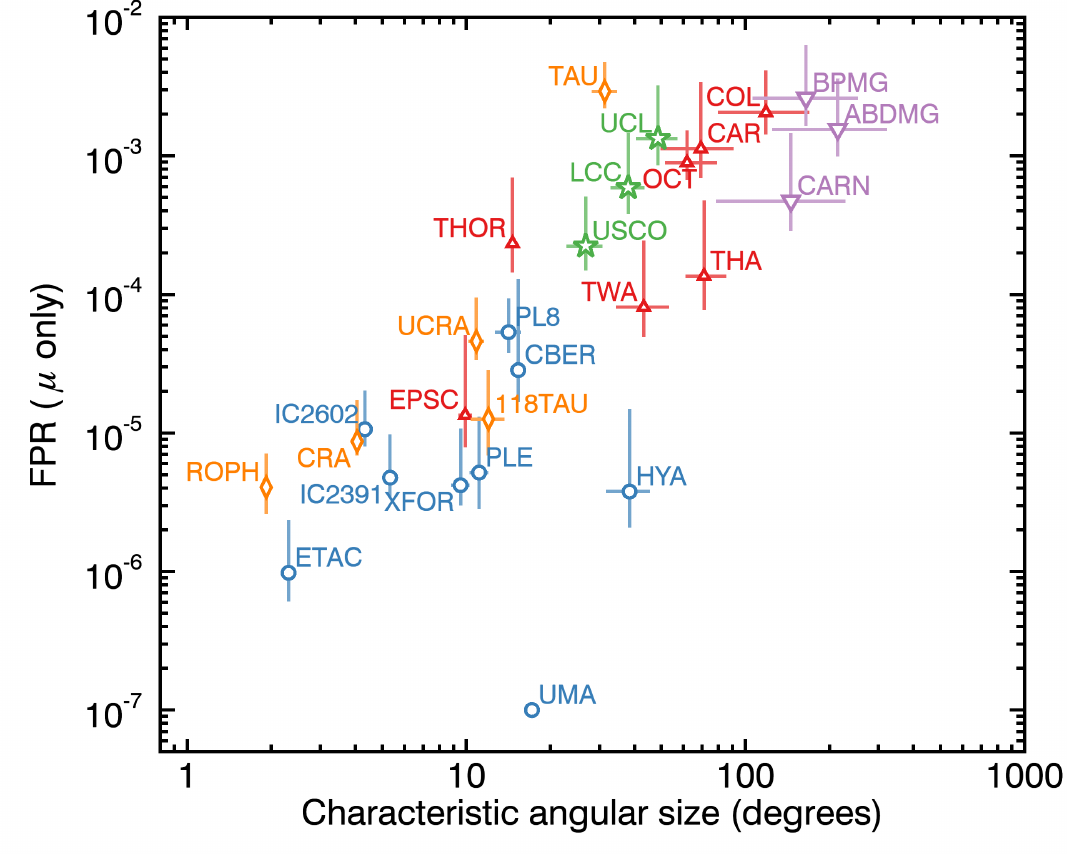}\label{fig:ymg_fpr_angsize}}
	\caption{Left panel: False-positive rates for all young associations considered here as a function of input observables, for a threshold $P_t$\,$=$\,90\%. The associations are sorted with decreasing false-positive rates in the case where only sky position and proper motion are considered. The grayed out region corresponds to the limit of our false-positive rates esimations, and corresponds to no false-positives in the full Besan\c{c}on Galactic model described in Section~\ref{sec:performance}. Right panel: False-positive rate as a function of the characteristic angular size of young associations, defined as the characteristic size of each young association $S_{\rm spa}$ (see Section~\ref{sec:ymgmodels}, Figure~\ref{fig:ymg_sizes} and Table~\ref{tab:young_pars3}) divided by its distance. The false-positive rates displayed in this figure correspond to the case where only sky position and proper motion are considered in BANYAN~$\Sigma$. There is a clear trend where associations with larger characteristic angular sizes suffer from more contamination from field stars. This effect alone dominates the false-positive rates, with only two exceptions: HYA and UMA are outliers with far fewer false-positives than would be expected from this trend. This is caused by their average $UVW$ kinematics that differ significantly from those of most field stars and other young associations (see Table~\ref{tab:young_pars3}). See Section~\ref{sec:performance} for more detail.}
	\label{fig:ymg_fpr_plots}
\end{figure*}

\input{table_riz2011.tex}

The models of young associations were also used to draw $10^5$ synthetic association members in order to obtain smooth TPRs as a function of probability threshold for each group in BANYAN~$\Sigma$, which are displayed in Figure~\ref{fig:all_tpr}. This figure illustrates the effect of the $\alpha_k$ thresholds described in Section~\ref{sec:priors}, causing the TPR curves of all young associations to meet at $P_t = 90$\% and TPR $=$ 50\%. Similar curves were built for FPRs and Matthews Correlation Coefficients (MCC; \citealt{1975AcCrA..31..480M}), defined in the range $\left[-1,1\right]$ indicate the quality of a Bayesian classifier: An MCC of $-1$ indicates a perfect mis-classification, a value of $0$ indicates a performance similar to a random classification, and a value of $+1$ indicates a perfect classifier. In these particular simulations, the number of true positives and false negatives (i.e., all synthetic objects originating from a young association model $H_k$) were scaled in the range $\left[0,N_k\right]$ instead of $\left[0,10^5\right]$, to give a realistic representations of the the FPR and MCC, where $N_k$ represents the number of stars in each young association. Because young associations with $N_k < 20$ are expected to be incomplete, we have set a minimal value of $N_k = 20$ in these situations. In a similar way, the number of true negatives and false positives (i.e., all synthetic objects originating from the Besan\c{c}on Galactic model) were scaled in the range $\left[0,217\,680\right]$, corresponding to the number of OBAFG-type stars in the model (out of $7.15 \times 10^{6}$ stars). In doing this, we assume that most young associations are complete at this fraction; this approximation only affects the FPR and MCC values reported in this work. The values $N_k$ are listed for each young association in Table~\ref{tab:young_pars3}.

The resulting TPR, FPR and MCC values for each young association, using each possible set of observables, are reported in Table~\ref{tab:young_roc}, as the range of possible values within $P_t$\,$\in$\,85--95\% and centered on $P_t$\,$=$\,90\%. They demonstrate how members of distant associations are much easier to distinguish from field objects because of their narrower distribution on the sky, with smaller FPRs and larger MCCs.\added{ Both the IDL and Python implementations of BANYAN~$\Sigma$ specify as an output for each star the TPR and FPR of a sample that would be constructed with only candidate members that have membership probabilities equal or higher than those of the star in question. This will allow users to interpret the Bayesian probabilities more easily without needing to rely on Table~\ref{tab:young_roc}.}

\added{We investigated the effect of ignoring covariances between the input observables (see Section~\ref{sec:solve}) with a $10^4$-elements Monte Carlo simulation. The median proper motion and parallax errors of all bona fide members in Gaia-DR1 (respectively 0.2 and 0.15\,\masyr, and 0.3\,mas) were assigned as the measurement errors to the observables of a typical member (TWA~1). The radial velocity of TWA~1 was ignored to maximize the effect of the covariances in the other measurements, and therefore to be as conservative as possible. Random measurements of its proper motion and parallax were taken from a 3D multivariate Gaussian distribution, where each dimension corresponds to the two components of proper motion and the parallax, and assuming a 99.9\% correlation between all dimensions. Each of these synthetic stars were attributed a vanishingly small error on its proper motion and distance (which we set to the median Gaia-DR1 values divided by one hundred). The membership probabilities of the $10^4$ synthetic stars were then calculated with BANYAN~$\Sigma$, and the average probability was calculated, which corresponds to the final probability marginalized over all values of proper motion and parallax. Ignoring the covariances in the observed measurements resulted in a negligible difference of $\sim$\,0.09\% in the membership probability, and similarly negligible differences of respectively 0.2\% and 0.01\% in the measured optimal distance and radial velocity. A similar Monte Carlo analysis was performed where the distance measurement was ignored, and yielded even smaller differences: $3\times 10^{-5}$\% in probability, $3\times 10^{-6}$\% in optimal distance and $8\times 10^{-6}$\% in optimal radial velocity. The effects of ignoring covariances between the components of sky position and other quantities would be even smaller, because the error bars on the sky position are always significantly smaller than those on proper motion and parallax.}

\section{CLASSIFYING PREVIOUSLY AMBIGUOUS MEMBERS WITH BANYAN~$\Sigma$}\label{sec:ambig}

In this section, BANYAN~$\Sigma$ is used to \deleted{tentatively lift the membership ambiguities of}\added{assign classifications to previously} ambiguous members encountered in the construction of the bona fide members lists (e.g., see the discussion in Section~\ref{sec:members}). None of the stars analyzed in this section were used to construct the young association models of BANYAN~$\Sigma$. Only objects with a $P \geq 95$\% Bayesian probability of belonging to young associations are discussed here.

\textbf{TWA~19~AB} is a young star that was defined as a candidate member of TWA by \cite{2005ApJ...634.1385M}, but later listed as a likely contaminant from LCC by \cite{2017ApJS..228...18G} based on a preliminary version of BANYAN~$\Sigma$. Here we obtain a membership probability of $P = 99$\% that it is a member of LCC.

\cite{2004ARAA..42..685Z} listed several tentative members of ABDMG as having a `questionable membership'. We find that 5 of them (HD~6569, HD~13482, HD~139751, HD~218860 and HD~224228) obtain a $P > 97$\% membership probability associated with ABDMG.

\textbf{2MASS~J05361998--1920396} is a young L2\,$\gamma$ substellar object that was listed as an ambiguous member of COL and $\beta$PMG by \cite{2016ApJS..225...10F}. Here, we obtain an unambiguous $P = 99.9$\% probability that it is a member of COL.

\textbf{CP--68~1388}, \textbf{GSC~09235--01702}, \textbf{CD--69~1055} and \textbf{MP~Mus} were all identified as ambiguous members of LCC or EPSC based on literature compilations. All but GSC~09235--01702 obtain an unambiguous $P > 98$\% LCC membership probability when analyzed with BANYAN~$\Sigma$. GSC~09235--01702 remains somewhat ambiguous with a 85\% EPSC membership probability and a 12.7\% LCC membership probability.

As mentioned in Section~\ref{sec:members}, there is a subset of new candidate members of the Sco-Cen region (consisting of the USCO, LCC and UCL sub-groups) discovered by \cite{2011MNRAS.416.3108R} that were not assigned to either of its sub-groups. Of these candidate members, 35 benefit from at least a radial velocity and/or parallax measurement, and obtain a $P > 95$\% young association membership probability. These objects are listed in Table~\ref{tab:riz2011} along with their respective probabilities in UCL, LCC and USCO.

Only seven of them remain ambiguous between more than one of the Sco-Cen subgroups, and they all obtain negligible membership probabilities for associations outside of the Sco-Cen region, with one exception: 2MASS~J13172895--4255587 (F3\,V; \citealt{1978mcts.book.....H}) obtains respective membership probabilities of 79.4\% for TWA, 10.5\% for UCL, 5.7\% for LCC, and 4.3\% for the field. This star has not been identified by \cite{2017ApJS..228...18G}, who used a preliminary BANYAN~$\Sigma$ model of TWA to identify new members based on Hipparcos. Further studies of this star may be able to confirm whether its age and radial velocity match those of TWA better than the Sco-Cen region.

\section{SUMMARY AND CONCLUSIONS}\label{sec:conclusion}

A new Bayesian algorithm to identify young association members was presented. It derives membership probabilities from the sky position and proper motion of an object, and optionally radial velocity, parallax and spectrophotometric distance constraints. It includes various improvements over its predecessor BANYAN~II \citep{2014ApJ...783..121G}, as it includes more associations, an updated list of bona fide members including Gaia-DR1 data, better spatial-kinematic models, a more accurate model of field stars, fewer approximations, new options and a significantly enhanced execution speed due to an analytical solution of the marginalization integrals. One limitation of BANYAN~$\Sigma$ is that it cannot account for correlations in measured error bars, such as those reported in Gaia\deleted{. Doing so would require a different structure when solving the marginalization integrals, and it is not guaranteed that an analytical solution would still be possible in this scenario.}\added{, but this results in biases of less than 0.1\% in membership probability and measurements of the optimal distance and radial velocity.}

The new BANYAN~$\Sigma$ tool includes all of the \nasso\ currently known young associations within 150\,pc, for which the current census of bona fide members is updated. It is also made publicly available in IDL and Python at \url{www.exoplanetes.umontreal.ca/banyan/banyansigma.php}. Additional figures and information on this work can be found on the website \url{www.astro.umontreal.ca/\textasciitilde gagne}.

We strongly encourage users to investigate the position of candidate members in appropriate color-magnitude diagrams using the BANYAN~$\Sigma$ optimal distance when no parallax measurements are used as inputs, as candidate lists generated without them suffer from $\sim$\,100 times more false positives in comparison (See Section~\ref{sec:performance} and Figure~\ref{fig:roc_per_input}). The BANYAN~$\Sigma$ algorithm can include distance constraints from population sequences in color-magnitude diagrams, and future versions will provide examples of such sequences.

This first version of BANYAN~$\Sigma$ will be the basis of a search for new isolated planetary-mass members in young associations based on 2MASS and AllWISE through the BASS-Ultracool survey (J.~Gagn\'e, J.~K.~Faherty, \'E.~Artigau et al., in preparation; see also \citealt{2017ApJ...841L...1G} and \citealt{2015ApJ...808L..20G} for preliminary results from BASS-Ultracool), as well as a search for new stellar members based on Gaia-DR1 (J.~Gagn\'e, O.~Loubier, J.~K.~Faherty et al., in preparation). The release of Gaia-DR2 will allow us to significantly improve the spatial and kinematic models of BANYAN~$\Sigma$, and identify new members of young associations that are not part of the Tycho catalog. A second version of the BANYAN~$\Sigma$ software with such improved models will be released in a future work that will aim at furthering the census of young association members based on Gaia-DR2.

\input{table_young_roc.tex}
\global\pdfpageattr\expandafter{\the\pdfpageattr/Rotate 0}

\acknowledgments

The authors would like to thank the anonymous referee and the AAS statistics consultant for valuable and detailed comments that significantly improved the quality of this paper. We thank Bruno~S.~Alessi and Eric~Bubar for sharing data. We thank No\'e Aubin-Cadot, Joel Kastner, Thierry Bazier-Matte, Simon G\'elinas, Jean-Fran\c{c}ois D\'esilets and Brendan Bowler for useful comments, as well as the anonymous referee of \cite{2015ApJS..219...33G}, who suggested the inclusion of a parallax motion correction in the BANYAN tools.

This research made use of: the SIMBAD database and VizieR catalog access tool, operated at the Centre de Donn\'ees astronomiques de Strasbourg, France \citep{2000AAS..143...23O}; data products from the Two Micron All Sky Survey (\emph{2MASS}; \citealp{2006AJ....131.1163S,2003yCat.2246....0C}), which is a joint project of the University of Massachusetts and the Infrared Processing and Analysis Center (IPAC)/California Institute of Technology (Caltech), funded by the National Aeronautics and Space Administration (NASA) and the National Science Foundation \citep{2006AJ....131.1163S}; data products from the \emph{Wide-field Infrared Survey Explorer} (\emph{WISE}; and \citealp{2010AJ....140.1868W}), which is a joint project of the University of California, Los Angeles, and the Jet Propulsion Laboratory (JPL)/Caltech, funded by NASA. This project was developed in part at the 2017 Heidelberg Gaia Sprint, hosted by the Max-Planck-Institut f\"ur Astronomie, Heidelberg. This work has made use of data from the European Space Agency (ESA) mission {\it Gaia} (\url{http://www.cosmos.esa.int/gaia}), processed by the {\it Gaia} Data Processing and Analysis Consortium (DPAC, \url{http://www.cosmos.esa.int/web/gaia/dpac/consortium}). Funding for the DPAC has been provided by national institutions, in particular the institutions participating in the {\it Gaia} Multilateral Agreement. Part of this research was carried out at the Jet Propulsion Laboratory, Caltech, under a contract with NASA. BGM simulations were executed on computers from the Utinam Institute of the Universit\'e de Franche-Comt\'e, supported by the R\'egion de Franche-Comt\'e and Institut des Sciences de l'Univers (INSU). EEM acknowledges support from the NASA NExSS program.

\emph{JG} designed BANYAN~$\Sigma$, compiled the bona fide members, wrote the IDL codes, wrote the manuscript, generated figures and led all analyses;\emph{EEM} shared parts of the young association literature data, characteristics and bona fide members, and provided general comments;\emph{ORL} wrote the initial Python translation of BANYAN~$\Sigma$; \emph{LM}, \emph{AR} and \emph{DR} performed the BANYAN~I, LACEwING and convergent point tool calculations used in Section~\ref{sec:performance}, respectively; \emph{AR} also provided help with the bona fide members compilation; \emph{DL} and \emph{JKF} shared ideas and comments; \emph{LP} shared ideas and provided comments especially for the ROC curves analysis; \emph{AR} performed custom Besan\c{c}on Galactic simulations and wrote the second pagraph of Section~\ref{sec:fieldmodel}; and \emph{RD} shared comments and supervized \emph{ORL}.

\software{BANYAN~$\Sigma$ (this paper), LACEwING \citep{2017AJ....153...95R}, BANYAN~II \citep{2014ApJ...783..121G}, BANYAN~I \citep{2013ApJ...762...88M}, the convergent point tool \citep{2011ApJ...727...62R}, Notability by Ginger Labs, Sublime Text.}

\bibliographystyle{apj}

\appendix
\twocolumngrid

\section{COORDINATE TRANSFORMATION OF THE BAYESIAN LIKELIHOOD}\label{app:jacobian_demo}

In this section, a change of coordinates is applied to the Bayesian likelihood $\mathcal{P}_o(\{O_i\}|H_k)$ from the direct observables frame of reference $\{O_i\}$ (sky position, proper motion, etc.) to the Galactic position ($XYZ$) and space velocity ($UVW$) reference frame $\{Q_i\}$:
\begin{align*}
	\mathcal{P}_o(\{O_i\}|H_k)\,\prod_j \mathrm{d}O_j = \mathcal{P}_q(\{Q_i\}|H_k)\,\prod_j \mathrm{d}Q_j.
\end{align*}

This change of coordinates can be expressed in the following form:
\begin{align}
	\mathcal{P}_o(\{O_i\}|H_k) &= |\bar{\bar J}|\cdot \mathcal{P}_q(\{Q_i\}|H_k),\\
	\mbox{with } J_{lm} &= \frac{\partial Q_l}{\partial O_m}\label{eqn:jacobiandef1},
\end{align}
\noindent where $\bar{\bar J}$ is a Jacobian matrix, which can be expressed as a $2\times 2$ block matrix of four $3\times 3$ sub-matrices:
\begin{align*}
	\bar{\bar J} &= \left[\begin{array}{cc}
		\boldsymbol{\mathcal A} & \boldsymbol{\mathcal B}\\
		\boldsymbol{\mathcal C} & \boldsymbol{\mathcal D}\\
	\end{array}\right],\\
	\boldsymbol{\mathcal A} &= \left[\begin{array}{ccc}
		\frac{\partial X}{\partial\alpha} & \frac{\partial X}{\partial\delta} & \frac{\partial X}{\partial\varpi}\\
		\frac{\partial Y}{\partial\alpha} & \frac{\partial Y}{\partial\delta} & \frac{\partial Y}{\partial\varpi}\\
		\frac{\partial Z}{\partial\alpha} & \frac{\partial Z}{\partial\delta} & \frac{\partial Z}{\partial\varpi}\\
	\end{array}\right],\\
	\boldsymbol{\mathcal B} &= \left[\begin{array}{ccc}
		\frac{\partial X}{\partial\mu_\alpha} & \frac{\partial X}{\partial\mu_\delta} & \frac{\partial X}{\partial\nu}\\
		\frac{\partial Y}{\partial\mu_\alpha} & \frac{\partial Y}{\partial\mu_\delta} & \frac{\partial Y}{\partial\nu}\\
		\frac{\partial Z}{\partial\mu_\alpha} & \frac{\partial Z}{\partial\mu_\delta} & \frac{\partial Z}{\partial\nu}\\
	\end{array}\right],\\
	\boldsymbol{\mathcal C} &= \left[\begin{array}{ccc}
		\frac{\partial U}{\partial\alpha} & \frac{\partial U}{\partial\delta} & \frac{\partial U}{\partial\varpi}\\
		\frac{\partial V}{\partial\alpha} & \frac{\partial V}{\partial\delta} & \frac{\partial V}{\partial\varpi}\\
		\frac{\partial W}{\partial\alpha} & \frac{\partial W}{\partial\delta} & \frac{\partial W}{\partial\varpi}\\
	\end{array}\right],\\
	\boldsymbol{\mathcal D} &= \left[\begin{array}{ccc}
		\frac{\partial U}{\partial\mu_\alpha} & \frac{\partial U}{\partial\mu_\delta} & \frac{\partial U}{\partial\nu}\\
		 \frac{\partial V}{\partial\mu_\alpha} & \frac{\partial V}{\partial\mu_\delta} & \frac{\partial V}{\partial\nu}\\
		 \frac{\partial W}{\partial\mu_\alpha} & \frac{\partial W}{\partial\mu_\delta} & \frac{\partial W}{\partial\nu}\\
	\end{array}\right].
\end{align*}

From the definition of Galactic coordinates \citep{1987AJ.....93..864J}:
\begin{align}
	\left(X,Y,Z\right) = \varpi\boldsymbol\lambda\left(\alpha,\delta\right)\label{eqn:defgalcoord}
\end{align}
\noindent it is apparent that $\boldsymbol{\mathcal{B}} = \boldsymbol{0}$. The determinant of $\bar{\bar J}$ can be obtained from the following property of block matrices, and further simplified:
\begin{align*}
	|\bar{\bar J}| &= |\boldsymbol{\mathcal{A}}\boldsymbol{\mathcal{D}}-\boldsymbol{0}\boldsymbol{\mathcal{C}}|,\\
	&= |\boldsymbol{\mathcal{A}}\boldsymbol{\mathcal{D}}|,\\
	&= |\boldsymbol{\mathcal{A}}|\cdot|\boldsymbol{\mathcal{D}}|.
\end{align*}

Using the definition of Galactic coordinates in Equation~\eqref{eqn:defgalcoord} , the sub-matrix $\boldsymbol{\mathcal{A}}$ can be simplified to:
\begin{align*}
	\boldsymbol{\mathcal{A}} = &\left[\begin{array}{ccc}
		\varpi\frac{\partial\lambda_0}{\partial\alpha} & \varpi\frac{\partial\lambda_0}{\partial\delta} & \lambda_0\\
		\varpi\frac{\partial\lambda_1}{\partial\alpha} & \varpi\frac{\partial\lambda_1}{\partial\delta} & \lambda_1\\
		\varpi\frac{\partial\lambda_2}{\partial\alpha} & \varpi\frac{\partial\lambda_2}{\partial\delta} & \lambda_2\\
	\end{array}\right],\\
	|\boldsymbol{\mathcal{A}}| =\ &\varpi^2\lambda_0\left(\frac{\partial\lambda_1}{\partial\alpha}\frac{\partial\lambda_2}{\partial\delta}-\frac{\partial\lambda_1}{\partial\delta}\frac{\partial\lambda_2}{\partial\alpha}\right)\\
	&-\varpi^2\lambda_1\left(\frac{\partial\lambda_0}{\partial\alpha}\frac{\partial\lambda_2}{\partial\delta}-\frac{\partial\lambda_0}{\partial\delta}\frac{\partial\lambda_2}{\partial\alpha}\right)\\
	&+\varpi^2\lambda_2\left(\frac{\partial\lambda_0}{\partial\alpha}\frac{\partial\lambda_1}{\partial\delta}-\frac{\partial\lambda_0}{\partial\delta}\frac{\partial\lambda_1}{\partial\alpha}\right),\\
	|\boldsymbol{\mathcal{A}}| =\ &\varpi^2 f\left(\alpha,\delta\right),
\end{align*}
\noindent where $f\left(\alpha,\delta\right)$ is a function of sky coordinates only.

From the definition of space velocity (see Section~\ref{sec:coordchange}):
\begin{align*}
	\left(U,V,W\right) = &\varpi\boldsymbol{N}\left(\alpha,\delta,\mu_\alpha,\mu_\delta\right) + \nu\boldsymbol{M}\left(\alpha,\delta\right),
\end{align*}
\noindent where the term $\cos\delta$ has been omitted in $\mu_\alpha\cos\delta$, the sub-matrix $\bar{\bar D}$ can be simplified to:
\begin{align*}
	\boldsymbol{\mathcal{D}} = &\left[\begin{array}{ccc}
		\varpi\frac{\partial N_0}{\partial\mu_\alpha} & \varpi\frac{\partial N_0}{\partial\mu_\delta} & M_0\\
		\varpi\frac{\partial N_1}{\partial\mu_\alpha} & \varpi\frac{\partial N_1}{\partial\mu_\delta} & M_1\\
		\varpi\frac{\partial N_2}{\partial\mu_\alpha} & \varpi\frac{\partial N_2}{\partial\mu_\delta} & M_2\\
	\end{array}\right],\\
	|\boldsymbol{\mathcal{D}}| =\ &\varpi^2M_0\left(\frac{\partial N_1}{\partial\mu_\alpha}\frac{\partial N_2}{\partial\mu_\delta}-\frac{\partial N_1}{\partial\mu_\delta}\frac{\partial N_2}{\partial\mu_\alpha}\right)\\
	&-\varpi^2M_1\left(\frac{\partial N_0}{\partial\mu_\alpha}\frac{\partial N_2}{\partial\mu_\delta}-\frac{\partial N_0}{\partial\mu_\delta}\frac{\partial N_2}{\partial\mu_\alpha}\right)\\
	&+\varpi^2M_2\left(\frac{\partial N_0}{\partial\mu_\alpha}\frac{\partial N_1}{\partial\mu_\delta}-\frac{\partial N_0}{\partial\mu_\delta}\frac{\partial N_1}{\partial\mu_\alpha}\right),\\
	|\boldsymbol{\mathcal{D}}| =\ &\varpi^2 g\left(\alpha,\delta\right).
\end{align*}
\noindent where $g\left(\alpha,\delta\right)$ is a function of sky coordinates only. The fact that $g$ does not depend on the proper motion components arises from the fact that all components of $\boldsymbol N$, defined in Section~\ref{sec:coordchange}, depend linearly on the proper motion components. It follows that:
\begin{align*}
	|\bar{\bar J}| &\propto\varpi^4.
\end{align*}

\section{SOLVING THE MARGINALIZATION INTEGRALS}\label{app:solving}

In this Section, the analytical solution to the marginalization integrals of Equation~\eqref{eqn:princint}, over distance and radial velocity, is developed. The index $k$ that referring to hypothesis $H_k$ is ignored in this section for simplicity. The Bayesian likelihood in the Galactic frame of reference $\{Q_i\}$ described in Equation~\eqref{eqn:bayesianlikelihood_q} can be inserted in Equation~\eqref{eqn:princint} to obtain:
\begin{align*}
	\mathcal{P}(\{O_i\}|H) &= \int_{0}^{\infty} \int_{-\infty}^{\infty}\varpi^4\frac{e^{-\frac{1}{2}\mathcal{M}^2}}{\sqrt{\left(2\pi\right)^6\left|\bar{\bar{\Sigma}}\right|}}\,\mathrm{d}\nu\,\mathrm{d}\varpi,
\end{align*}
\noindent and the Mahalanobis distance $\mathcal{M}$ defined in Equation~\eqref{eqn:mahal_dev2} can be used to develop it further:
\begin{align}
	\mathcal{P}(\{O_i\}|H) &= C_0 \int_{0}^{\infty}I_0\left(\varpi\right)\,\varpi^4e^{-\frac{1}{2}\left<\bar\Gamma,\bar\Gamma\right>\varpi^2+\left<\bar\Gamma,\bar\tau\right>\varpi}\,\mathrm{d}\varpi,\label{eqn:intw}\\
	I_0\left(\varpi\right) &= \int_{-\infty}^\infty e^{-\frac{1}{2}\left<\bar\Omega,\bar\Omega\right>\nu^2+\left(\left<\bar\Omega,\bar\tau\right>-\left<\bar\Omega,\bar\Gamma\right>\varpi\right)\nu}\,\mathrm{d}\nu,\label{eqn:I0}\\
	C_0 &= \frac{e^{-\frac{1}{2}\left<\bar\tau,\bar\tau\right>}}{\sqrt{\left(2\pi\right)^6\left|\bar{\bar\Sigma}\right|}}.\notag
\end{align}
Equation~\eqref{eqn:I0} can be solved with the identity:
\begin{align*}
	\int_{-\infty}^\infty e^{-ax^2-bx}\,\mathrm{d}x = \sqrt{\frac{\pi}{a}}e^{b^2/4a},
\end{align*}
\noindent with $a = \frac{1}{2}\left<\bar\Omega,\bar\Omega\right>$ and $b = -\left<\bar\Omega,\bar\tau\right> + \left<\bar\Omega,\bar\Gamma\right>\varpi$.

The term in $b^2$ can be developed into a second-degree polynomial in $\varpi$:
\begin{align*}
	\frac{b^2}{4a} = \frac{1}{2}\frac{\left<\bar{\Omega},\bar{\Gamma}\right>^2}{\left<\bar\Omega,\bar\Omega\right>}\varpi^2-\frac{\left<\bar{\Omega},\bar{\Gamma}\right>\left<\bar{\Omega},\bar\tau\right>}{\left<\bar\Omega,\bar\Omega\right>}\varpi+\frac{1}{2}\frac{\left<\bar{\Omega},\bar\tau\right>^2}{\left<\bar\Omega,\bar\Omega\right>},
\end{align*}
\noindent which can be inserted back into Equation~\eqref{eqn:intw}:
\begin{align}
	P(\{O_i\}|H) &= \frac{\left|\bar\Omega\right|^{-1}e^{-\zeta}}{\sqrt{\left(2\pi\right)^5\left|\bar{\bar\Sigma}\right|}}\int_{0}^{\infty}\varpi^4e^{-\beta\varpi^2-\gamma\varpi}\mathrm{d}\varpi,\label{eqn:secondtolast}\\
	\mathrm{where\ } \beta &= \frac{\left<\bar\Gamma,\bar\Gamma\right>}{2} - \frac{1}{2}\frac{\left<\bar{\Omega},\bar{\Gamma}\right>^2}{\left<\bar\Omega,\bar\Omega\right>},\notag\\
	\gamma &= \frac{\left<\bar{\Omega},\bar{\Gamma}\right>\left<\bar{\Omega},\bar\tau\right>}{\left<\bar\Omega,\bar\Omega\right>} - \left<\bar\Gamma,\bar\tau\right>.\notag\\
	\zeta &= \frac{\left<\bar\tau,\bar\tau\right>}{2}-\frac{1}{2}\frac{\left<\bar{\Omega},\bar\tau\right>^2}{\left<\bar\Omega,\bar\Omega\right>}.\notag
\end{align}
The solution to this integral is given by \cite{Erdelyi:1955wk,Gradshteyn:2014uy}:
\begin{align*}
	\int_{0}^{\infty} x^n &e^{-\beta x^2-\gamma x}\ \mathrm{d}x = \frac{n!\ e^{\gamma^2/8\beta}}{(2\beta)^{(n+1)/2}}\ \mathcal{D}_{-(n+1)}\left(\gamma/\sqrt{2\beta}\right),
\end{align*}
\noindent where $\mathcal{D}_m(x)$ is a parabolic cylinder function \citep{Magnus:zYVRWLJr}.

The $m=-5$ ($n=4$) corresponds to the required integral, and the corresponding parabolic cylinder function can be developed as:
\begin{align*}
	\mathcal{D}_{-5}(x) =& \frac{e^{x^2/4}}{24}\bigg(\sqrt\frac{\pi}{2}\left(x^4+6x^2+3\right)\erfc\left(\tfrac{x}{\sqrt{2}}\right)\\
	&-\left(x^3+5x\right)e^{-x^2/2}\bigg).
\end{align*}

Equation~\eqref{eqn:secondtolast} becomes:
\begin{align}
	\mathcal{P}(\{O_i\}|H) &= \frac{3}{4}\frac{\mathcal{D}_{-5}\left(\gamma/\sqrt{2\beta}\right)e^{\gamma^2/8\beta-\zeta}}{\left|\bar\Omega\right|\sqrt{\pi^5\beta^5\left|\bar{\bar\Sigma}\right|}}\label{eqn:firstsol}.
\end{align}

The term in $e^{x^2/4}$ in the definition of $\mathcal{D}_{-5}(x)$ can become very large for typical values of $x$, making the numerical computation of $\mathcal{D}_{-5}(x)$ unstable. To avoid this problem, a modified parabolic cylinder function $\mathcal{D}^\prime_{-5}(x)$ can be defined so that the large term is combined with the exponential term in Equation~\eqref{eqn:firstsol}:
\begin{align}
	\mathcal{P}(\{O_i\}|H) &= \frac{1}{32}\frac{\mathcal{D}^\prime_{-5}\left(\gamma/\sqrt{2\beta}\right)e^{\gamma^2/4\beta-\zeta}}{\left|\bar\Omega\right|\sqrt{\pi^5\beta^5\left|\bar{\bar\Sigma}\right|}},\notag\\
	\mathcal{D}^\prime_{-5}(x) =& 24\,e^{-x^2/4}\,\mathcal{D}_{-5}(x).\notag
\end{align}
\noindent which completes the analytical solution of the Bayesian likelihood. The $1/32$ factor will be ignored here because it will disappear in the marginalization of the Bayesian likelihood (Equation~\ref{eqn:bayesrule}), and other multiplicative factors independent of the Bayesian hypothesis have already been ignored in determining the Jacobian of the coordinate transformation (see Appendix~\ref{app:jacobian_demo}).

\section{DETERMINING THE OPTIMAL RADIAL VELOCITY AND DISTANCE}\label{app:optimald}

The optimal radial velocity $\nu_\mathrm{o}$ and distance $\varpi_\mathrm{o}$ that maximize the value of the non-marginalized Bayesian likelihood $\mathcal{P}_o(\{O_i\}|H)$ can be obtained by solving the system of equations:
\begin{align*}
	\left.\frac{\partial \ln\mathcal{P}_o(\{O_i\}|H)}{\partial \nu}\right|_{\nu=\nu_\mathrm{o},\varpi=\varpi_\mathrm{o}} &= 0,\label{eqn:derivnu}\\
	\left.\frac{\partial \ln\mathcal{P}_o(\{O_i\}|H)}{\partial \varpi}\right|_{\nu=\nu_\mathrm{o},\varpi=\varpi_\mathrm{o}} &= 0,
\end{align*}
\noindent which can be developed with Equation~\eqref{eqn:mahal_dev2}:
\begin{align*}
	0 &= \left<\bar\Omega,\bar\Omega\right>\nu_\mathrm{o} + \left<\bar\Omega,\bar\Gamma\right>\varpi_\mathrm{o} - \left<\bar\Omega,\bar\tau\right>,\\
	0 &= \left<\bar\Gamma,\bar\Gamma\right>\varpi_\mathrm{o}^2 + \left<\bar\Omega,\bar\Gamma\right>\varpi_\mathrm{o}\nu_\mathrm{o} - \left<\bar\Gamma,\bar\tau\right>\varpi_\mathrm{o} - 4.
\end{align*}

This system of equations has two solutions :
\begin{align*}
	\varpi_\mathrm{o} &= \frac{-\gamma \pm \sqrt{\gamma^2 + 32\beta}}{4\beta},\\
	\nu_\mathrm{o} &= \frac{4+\left<\bar\Gamma,\bar\tau\right>\varpi_\mathrm{o}-\left<\bar\Gamma,\bar\Gamma\right>\varpi_\mathrm{o}^2}{\left<\bar\Omega,\bar\Gamma\right>\varpi_\mathrm{o}},\\
\end{align*}
\noindent where $\gamma$ and $\beta$ are defined in Equations~\eqref{eqn:gamma} and \eqref{eqn:beta}.

Any combination of $\gamma$ and $\beta$ that respects the following inequality:
\begin{align*}
	\sqrt{1+32\beta/\gamma^2} &> 1,\\
	\,\mathrm{i.e.,}\ \beta &> 0,
\end{align*}
\noindent will yield an unphysical negative distance for the negative root of $\varpi_\mathrm{o}$. Since multivariate Gaussians have $\beta > 0$ by definition, the inequality is always respected. As a consequence, only the positive root of $\varpi_\mathrm{o}$ has a physical meaning.

Error bars on the optimal radial velocity $\sigma_\nu$ and distance $\sigma_\varpi$ can be defined by measuring the characteristic width of $\mathcal{P}_o(\{O_i\}|H)$ along $\nu$ and $\varpi$ in the vicinity of $(\nu_\mathrm{o},\varpi_\mathrm{o})$. The effect of the Jacobian term $\varpi^4$ will be ignored to obtain an analytical approximation of $(\sigma_\nu,\sigma_\varpi)$. 

The relation between the expectancy $E(x)$ of a variable and the characteristic width of a Gaussian function $G(x)$ can be used to determine $\sigma_\nu$ and $\sigma_\varpi$:
\begin{align*}
	\sigma_x = \sqrt{E(x^2)-E(x)^2},\\
	E(x) = \int_{-\infty}^{\infty} x G(x)\,\mathrm{d}x.
\end{align*}

In the case of $\sigma_\nu$, this yields:
\begin{align*}
E(\nu) &= \frac{\int_{-\infty}^{\infty} \nu e^{-\beta_\nu\nu^2-\gamma_\nu\nu}\,\mathrm{d}\nu}{\int_{-\infty}^{\infty} e^{-\beta_\nu\nu^2-\gamma_\nu\nu}\,\mathrm{d}\nu},\\
&= \frac{\gamma_\nu}{2\beta_\nu},\\
E(\nu^2) &= \frac{\int_{-\infty}^{\infty} \nu^2 e^{-\beta_\nu\nu^2-\gamma_\nu\nu}\,\mathrm{d}\nu}{\int_{-\infty}^{\infty} e^{-\beta_\nu\nu^2-\gamma_\nu\nu}\,\mathrm{d}\nu},\\
&= \frac{2\beta_\nu + \gamma_\nu^2}{4\beta_\nu^2},\\
\beta_\nu &= \frac{\left<\bar\Omega,\bar\Omega\right>}{2},\\
\gamma_\nu &= \varpi_0\left<\bar\Omega,\bar\Gamma\right>-\left<\bar\Omega,\bar\tau\right>,
\end{align*}
\noindent leading to:
\begin{align*}
	\sigma_\nu &= \frac{1}{\sqrt{2\beta_\nu}},\\
	\sigma_\nu &= |\bar\Omega|^{-1}.
\end{align*}

The case of $\sigma_\varpi$ requires the introduction of a new variable $\varpi^\prime$ that is defined in the range $]-\infty,\infty[$ and matches the distance $\varpi^\prime = \varpi$ for $\varpi^\prime \geq 0$. Assuming that $\varpi_\mathrm{o} >> \sigma_\varpi$ will ensure that the Bayesian likelihood $\mathcal{P}_0(\nu,\varpi^\prime) \approx 0$ for all negative values of $\varpi^\prime$. It follows that:
\begin{align*}
E(\varpi) &\approx \frac{\int_{-\infty}^{\infty} \varpi^\prime\,e^{-\beta_\varpi\varpi^2-\gamma_\varpi\varpi}\,\mathrm{d}\varpi^\prime}{\int_{-\infty}^{\infty} e^{-\beta_\varpi\varpi^2-\gamma_\varpi\varpi}\,\mathrm{d}\varpi^\prime}\\
&= \frac{\gamma_\varpi}{2\beta_\varpi},\\
E(\varpi^2) &\approx \frac{2\beta_\varpi + \gamma_\varpi^2}{4\beta_\varpi^2},\\
\beta_\varpi &= \frac{\left<\bar\Gamma,\bar\Gamma\right>}{2},\\
\gamma_\varpi &= \nu_\mathrm{o}\left<\bar\Omega,\bar\Gamma\right>-\left<\bar\Gamma,\bar\tau\right>,
\end{align*}
\noindent leading to:
\begin{align*}
	\sigma_\varpi &\approx |\bar\Gamma|^{-1}.
\end{align*}

The optimal distance and radial velocity do not correspond exactly to the statistical distance and radial velocities defined in the BANYAN~II formalism \citep{2014ApJ...783..121G}. The latter are obtained by maximizing the Bayesian likelihood in one dimension after the other dimension was marginalized. The optimal distance and radial velocity maximize the Bayesian probability of a given hypothesis as a couple, whereas the BANYAN~II statistical distance maximizes the Bayesian probability when radial velocity is treated as an unknown parameter, and vice versa.

\ifdisplaylongtables
	\input{table_bonafide.tex}
	\input{table_excluded.tex}
	\input{table_designations.tex}
\else
	\input{table_excluded_short.tex}
	\input{table_designations_short.tex}
	\global\pdfpageattr\expandafter{\the\pdfpageattr/Rotate 0}
\fi

\clearpage
\maxdeadcycles=200
\ifdisplaylongtables
	\newcount\tmpnum
	\def\storedata#1#2{\tmpnum=0 \edef\tmp{\string#1}\storedataA#2\end}
	\def\storedataA#1{\advance\tmpnum by1
	   \ifx\end#1\else
	      \expandafter\def\csname data:\tmp:\the\tmpnum\endcsname{#1}%
	      \expandafter\storedataA\fi
	}
	\def\getdata[#1]#2{\csname data:\string#2:#1\endcsname}
	
	\storedata\mydata{{118TAU}{ABDMG}{BPMG}{CAR}{CARN}{CBER}{COL}{CRA}{EPSC}{ETAC}{HYA}{IC2391}{IC2602}{LCC}{OCT}{PL8}{PLE}{ROPH}{TAU}{THA}{THOR}{UCL}{UMA}{USCO}{UCRA}{XFOR}}
	\newcount\ii
	\loop
	\ifnum\ii<26
	\advance\ii by 1
	\begin{figure*}[p]
	 	\centering
	 	\includegraphics[width=0.98\textwidth]{\getdata[\the\ii]\mydata_params.pdf}
		\caption{Multivariate Gaussian model of \getdata[\the\ii]\mydata.}
	\end{figure*}
	\repeat
\fi
\listofchanges
\end{document}

%% file: table_young_pars3.tex
\startlongtable
\tabletypesize{\normalsize}
\begin{deluxetable*}{lcccccccccccc}
\tablecolumns{13}
\tablecaption{General characteristics and Bayesian priors of young associations.\label{tab:young_pars3}}
\tablehead{\colhead{Asso.} & \colhead{$N_k$\tablenotemark{a}} & \multicolumn{4}{c}{$\ln\alpha_k$\tablenotemark{b}} & \colhead{$\left<\varpi\right>$\tablenotemark{c}} & \colhead{$\left<\nu\right>$\tablenotemark{d}} & \colhead{$S_{\rm spa}$\tablenotemark{e}} & \colhead{$S_{\rm kin}$\tablenotemark{f}} & \colhead{Age} & \colhead{Age}\\
\cline{3-6}
\colhead{} & \colhead{} & \colhead{$\mu$} & \colhead{$\mu,\nu$} & \colhead{$\mu,\varpi$} & \colhead{$\mu,\nu,\varpi$} & \colhead{(pc)} & \colhead{(\kms)} & \colhead{(pc)} & \colhead{(\kms)} & \colhead{(Myr)} & \colhead{Ref.} }
\startdata
118TAU & 10 & -17.22 & -18.60 & -21.37 & -22.66 & $100 \pm 10$ & $14 \pm 2$ & 3.4 & 2.1 & $\sim$\,10 & 1\\
ABDMG & 48 & -14.11 & -15.39 & -16.56 & -17.60 & $30_{-10}^{+20}$ & $10_{-20}^{+10}$ & 19.0 & 1.4 & $149_{-19}^{+51}$ & 2\\
$\beta$PMG & 42 & -13.57 & -14.77 & -17.39 & -18.24 & $30_{-10}^{+20}$ & $10 \pm 10$ & 14.8 & 1.4 & $24 \pm 3$ & 2\\
CAR & 7 & -13.41 & -14.82 & -18.45 & -19.15 & $60 \pm 20$ & $20 \pm 2$ & 11.8 & 0.8 & $45_{-7}^{+11}$ & 2\\
CARN & 13 & -15.51 & -16.85 & -17.64 & -18.55 & $30 \pm 20$ & $15_{-10}^{+7}$ & 14.0 & 2.1 & $\sim$\,200 & 3\\
CBER & 40 & -13.70 & -15.09 & -22.32 & -23.43 & $85_{-5}^{+4}$ & $-0.1 \pm 0.8$ & 3.6 & 0.5 & $562_{-84}^{+98}$ & 4\\
COL & 23 & -13.08 & -14.10 & -17.74 & -18.34 & $50 \pm 20$ & $21_{-8}^{+3}$ & 15.8 & 0.9 & $42_{-4}^{+6}$ & 2\\
CRA & 12 & -17.55 & -19.07 & -21.89 & -22.89 & $139 \pm 4$ & $-1 \pm 1$ & 1.5 & 1.7 & 4--5 & 5\\
EPSC & 25 & -17.47 & -18.59 & -22.38 & -22.79 & $102 \pm 4$ & $14 \pm 3$ & 2.8 & 1.8 & $3.7_{-1.4}^{+4.6}$ & 6\\
ETAC & 16 & -20.19 & -21.36 & -25.75 & -26.22 & $95 \pm 1$ & $20 \pm 3$ & 0.6 & 2.0 & $11 \pm 3$ & 2\\
HYA & 177 & -20.02 & -21.54 & -22.14 & -23.57 & $42 \pm 7$ & $39_{-4}^{+3}$ & 4.5 & 1.2 & $750 \pm 100$ & 7\\
IC2391 & 16 & -18.05 & -18.90 & -21.55 & -21.55 & $149 \pm 6$ & $15 \pm 3$ & 2.2 & 1.4 & $50 \pm 5$ & 8\\
IC2602 & 17 & -15.33 & -16.60 & -22.50 & -22.60 & $146 \pm 5$ & $17 \pm 3$ & 1.8 & 1.1 & $46_{-5}^{+6}$ & 9\\
LCC & 82 & -13.13 & -14.27 & -17.76 & -17.77 & $110 \pm 10$ & $14 \pm 5$ & 11.6 & 2.2 & $15 \pm 3$ & 10\\
OCT & 14 & -11.74 & -11.56 & -13.85 & -11.29 & $130_{-20}^{+30}$ & $8_{-9}^{+8}$ & 22.4 & 1.3 & $35 \pm 5$ & 11\\
PL8 & 11 & -13.28 & -14.54 & -19.37 & -19.75 & $130 \pm 10$ & $22 \pm 2$ & 5.0 & 1.1 & $\sim$\,60 & 12\\
PLE & 190 & -18.72 & -20.06 & -20.61 & -21.46 & $134 \pm 9$ & $6 \pm 2$ & 4.1 & 1.4 & $112 \pm 5$ & 13\\
ROPH & 186 & -17.49 & -19.04 & -24.10 & -25.55 & $131 \pm 1$ & $-6.3 \pm 0.2$ & 0.7 & 1.6 & $<$\,2 & 14\\
TAU & 122 & -10.39 & -11.37 & -17.04 & -17.99 & $120 \pm 10$ & $16 \pm 3$ & 10.7 & 3.6 & 1--2 & 15\\
THA & 39 & -16.48 & -17.78 & -19.58 & -20.25 & $46_{-6}^{+8}$ & $9_{-6}^{+5}$ & 9.1 & 0.8 & $45 \pm 4$ & 2\\
THOR & 35 & -14.05 & -15.57 & -20.22 & -21.12 & $96 \pm 2$ & $19 \pm 3$ & 3.9 & 2.1 & $22_{-3}^{+4}$ & 2\\
TWA & 23 & -16.99 & -18.22 & -20.42 & -20.93 & $60 \pm 10$ & $10 \pm 3$ & 6.6 & 1.5 & $10 \pm 3$ & 2\\
UCL & 103 & -11.70 & -13.10 & -15.91 & -16.19 & $130 \pm 20$ & $5 \pm 5$ & 17.4 & 2.5 & $16 \pm 2$ & 10\\
UCRA & 10 & -15.85 & -16.87 & -20.24 & -20.48 & $147 \pm 7$ & $-1 \pm 3$ & 4.5 & 1.8 & $\sim$\,10 & 16\\
UMA & 9 & -23.14 & -24.01 & -26.44 & -27.13 & $25.4_{-0.7}^{+0.8}$ & $-12 \pm 3$ & 1.2 & 1.3 & $414 \pm 23$ & 17\\
USCO & 84 & -12.77 & -13.71 & -17.62 & -17.96 & $130 \pm 20$ & $-5 \pm 4$ & 9.9 & 2.8 & $10 \pm 3$ & 10\\
XFOR & 11 & -19.37 & -20.80 & -23.43 & -23.72 & $100 \pm 6$ & $19 \pm 2$ & 2.6 & 1.3 & $\sim$\,500 & 18\\
\enddata
\tablenotetext{a}{Number of bona fide members included in the kinematic model.}
\tablenotetext{b}{Bayesian prior ensuring a recovery rate of $\approx$50--90\% for a treshold $P = 90$\%, depending on input observables.}
\tablenotetext{c}{Peak of distance distribution and $\pm$1$\sigma$ range.}
\tablenotetext{d}{Peak of radial velocity distribution and $\pm$1$\sigma$ range.}
\tablenotetext{e}{Characteristic spatial scale in $XYZ$ space.}
\tablenotetext{f}{Characteristic kinematic scale in $UVW$ space.}
\tablecomments{See Sections~\ref{sec:members} and \ref{sec:ymgmodels} for more detail. The full names of young associations are: 118~Tau (118TAU), AB~Doradus (ABDMG), $\beta$~Pictoris ($\beta$PMG), Carina (CAR), Carina-Near (CARN), Coma Berenices (CBER), Columba (COL), Corona~Australis (CRA), $\epsilon$~Chamaeleontis (EPSC), $\eta$~Chamaeleontis (ETAC), the Hyades cluster (HYA), Lower Centaurus Crux (LCC), Octans (OCT), Platais~8 (PL8), the Pleiades cluster (PLE), $\rho$~Ophiuchi (ROPH), the Tucana-Horologium association (THA), 32~Orionis (THOR), TW~Hya (TWA), Upper Centaurus Lupus (UCL), Upper~CrA (UCRA), the core of the Ursa~Major cluster (UMA), Upper~Scorpius (USCO), Taurus (TAU) and $\chi^1$~For (XFOR).}
\tablerefs{(1)~\citealt{mamajek118tau}; (2)~\citealt{2015MNRAS.454..593B}; (3)~\citealt{2006ApJ...649L.115Z}; (4)~\citealt{2014AA...566A.132S}; (5)~\citealt{2012MNRAS.420..986G}; (6)~\citealt{2013MNRAS.435.1325M}; (7)~\citealt{2015ApJ...807...24B}; (8)~\citealt{2004ApJ...614..386B}; (9)~\citealt{2010MNRAS.409.1002D}; (10)~\citealt{2016MNRAS.461..794P}; (11)~\citealt{2015MNRAS.447.1267M}; (12)~\citealt{1998AJ....116.2423P}; (13)~\citealt{2015ApJ...813..108D}; (14)~\citealt{2008hsf2.book..351W}; (15)~\citealt{1995ApJS..101..117K}; (16)~this paper; (17)~\citealt{2015AAS...22511203J}; (18)~\citealt{2010AA...514A..81P}.}
\end{deluxetable*}

%% file: table_bonafide_short.tex
\startlongtable
\tabletypesize{\normalsize}
\begin{longrotatetable}
\global\pdfpageattr\expandafter{\the\pdfpageattr/Rotate 90}
\begin{deluxetable*}{llccr@{\;$\pm$\;}lr@{\;$\pm$\;}lr@{\;$\pm$\;}lr@{\;$\pm$\;}ll}
\tablecolumns{13}
\tablecaption{Literature compilation of bona fide members. \label{tab:bonafide}}
\tablehead{\colhead{Main} & \colhead{Spectral} & \colhead{R.A.} & \colhead{Decl.} & \multicolumn{2}{c}{$\mu_\alpha\cos\delta$}  & \multicolumn{2}{c}{$\mu_\delta$} & \multicolumn{2}{c}{Rad. Vel.} & \multicolumn{2}{c}{Distance} & \colhead{} \\
\colhead{Designation} & \colhead{Type} & \colhead{(hh:mm:ss)} & \colhead{(dd:mm:ss)} & \multicolumn{2}{c}{(\masyr)} & \multicolumn{2}{c}{(\masyr)} & \multicolumn{2}{c}{(\kms)} & \multicolumn{2}{c}{(pc)} & \colhead{References}}
\startdata
\sidehead{\Large\textbf{AB~Doradus}}
2MASS~J00192626+4614078 & M8\,$\beta$ & 00:19:26.26 & 46:14:07.8 & $119.4$ & $0.9$ & $-75.4$ & $0.9$ & $-20$ & $3$ & $39$ & $2$ & 1,2,3,2\\
BD+54~144~A & F8\,V & 00:45:51.06 & 54:58:39.1 & $96.40$ & $0.03$ & $-73.97$ & $0.04$ & $-15$ & $2$ & $50.3$ & $0.9$ & 4,5,6,5\\
| BD+54~144~B & K3 & 00:45:51.23 & 54:58:40.8 & \multicolumn{2}{c}{$\cdots$} & \multicolumn{2}{c}{$\cdots$} & \multicolumn{2}{c}{$\cdots$} & \multicolumn{2}{c}{$\cdots$} & 7,--,--,--\\
2MASS~J00470038+6803543 & L6--L8\,$\gamma$ & 00:47:00.39 & 68:03:54.4 & $385$ & $1$ & $-201$ & $1$ & $-20$ & $1$ & $12.2$ & $0.3$ & --,2,8,2\\
G~132--51~B & M2.6 & 01:03:42.23 & 40:51:13.6 & $132$ & $5$ & $-164$ & $5$ & $-10.6$ & $0.3$ & $30$ & $2$ & 9,9,9,9\\
| G~132--50 & M0 & 01:03:40.30 & 40:51:26.7 & $126.9$ & $0.1$ & $-166.18$ & $0.09$ & \multicolumn{2}{c}{$\cdots$} & $33.0$ & $0.6$ & 10,5,--,5\\
| G~132--51~C & M3.8 & 01:03:42.44 & 40:51:13.3 & \multicolumn{2}{c}{$\cdots$} & \multicolumn{2}{c}{$\cdots$} & $-10.9$ & $0.4$ & \multicolumn{2}{c}{$\cdots$} & 9,--,9,--\\
HIP~6276 & G0\,V & 01:20:32.38 & -11:28:05.8 & $111.44$ & $0.06$ & $-136.95$ & $0.05$ & $8.3$ & $0.4$ & $35.1$ & $0.4$ & 11,5,12,5\\
G~269--153~A & M4.3 & 01:24:27.85 & -33:55:11.4 & $180$ & $20$ & $-110$ & $20$ & $19$ & $3$ & $25$ & $1$ & 9,13,9,9\\
| G~269--153~B & M4.6 & 01:24:27.96 & -33:55:09.9 & \multicolumn{2}{c}{$\cdots$} & \multicolumn{2}{c}{$\cdots$} & $18$ & $1$ & \multicolumn{2}{c}{$\cdots$} & 9,--,9,--\\
\enddata
\tablerefs{The references to this table are listed in Table~\ref{tab:refs}.}
\end{deluxetable*}
\end{longrotatetable}
\global\pdfpageattr\expandafter{\the\pdfpageattr/Rotate 0}

%% file: table_bonafide_refs.tex
\startlongtable
\tabletypesize{\scriptsize}
\begin{deluxetable*}{l}
\tablecolumns{1}
\tablecaption{References for literature compilation of bona fide members. \label{tab:refs}}
\tablehead{\colhead{References}}
\startdata
\multicolumn{1}{m{0.8\textwidth}}{(1)~\citealt{2015ApJS..219...33G}; (2)~\citealt{2016ApJ...833...96L}; (3)~\citealt{2009ApJ...705.1416R}; (4)~\citealt{1964PLPla..28....1J}; (5)~\citealt{2016AA...595A...2G}; (6)~\citealt{2006ARep...50..733B}; (7)~\citealt{2004ARAA..42..685Z}; (8)~\citealt{2016ApJS..225...10F}; (9)~\citealt{2012ApJ...758...56S}; (10)~\citealt{2013AJ....145..102L}; (11)~\citealt{2013ApJ...762...88M}; (12)~\citealt{2007AJ....133.2524W}; (13)~\citealt{2003AJ....125..984M}; (14)~\citealt{2014ApJ...788...81M}; (15)~\citealt{2010AJ....140..119S}; (16)~\citealt{2007AA...474..653V}; (17)~\citealt{2008KFNT...24..480I}; (18)~\citealt{2002AA...382..118T}; (19)~\citealt{1994AJ....107.1556G}; (20)~\citealt{2006AstL...32..759G}; (21)~\citealt{1994BICDS..44....3B}; (22)~\citealt{1992AA...258..217E}; (23)~\citealt{2012AstL...38..331A}; (24)~\citealt{1975mcts.book.....H}; (25)~\citealt{2011ApJ...732...61Z}; (26)~\citealt{2010ApJ...723..684B}; (27)~\citealt{2011AAS...21743412C}; (28)~\citealt{2013AJ....146..134K}; (29)~\citealt{2000AA...355L..27H}; (30)~\citealt{2006AA...460..695T}; (31)~\citealt{2004AAS...205.4815Z}; (32)~\citealt{2006AJ....132..161G}; (33)~\citealt{1995AA...294..744M}; (34)~\citealt{2007AN....328..889K}; (35)~\citealt{2010AA...520A..15M}; (36)~\citealt{2001MNRAS.328...45M}; (37)~\citealt{2004AJ....128..463R}; (38)~\citealt{2017AJ....153...75K}; (39)~\citealt{2015ApJ...802L..10T}; (40)~\citealt{2004AJ....127.3553K}; (41)~\citealt{2012ApJS..201...19D}; (42)~\citealt{2015ApJ...808L..20G}; (43)~\citealt{2014AJ....147...94D}; (44)~\citealt{2008AA...488..401R}; (45)~\citealt{1995ApJS...99..135A}; (46)~\citealt{2001ApJ...562L..87Z}; (47)~\citealt{2010AJ....139.2184Z}; (48)~\citealt{2014AJ....147...85R}; (49)~\citealt{2005ApJS..159..141V}; (50)~\citealt{2007AA...475..519H}; (51)~\citealt{2003ApJ...599..342S}; (52)~\citealt{2015Sci...350...64M}; (53)~\citealt{2004AJ....127.3043Z}; (54)~\citealt{2013ApJ...772...79A}; (55)~\citealt{2009AJ....137.3632L}; (56)~\citealt{2007AstL...33..571B}; (57)~\citealt{2002AJ....123.3356G}; (58)~\citealt{2013AA...555A.107B}; (59)~\citealt{2009ApJ...698..242T}; (60)~\citealt{2011MNRAS.411..117K}; (61)~\citealt{1984ApJS...55..657C}; (62)~\citealt{2013ApJ...777L..20L}; (63)~\citealt{2016ApJ...819..133A}; (64)~\citealt{2017AJ....154...69S}; (65)~\citealt{2003AJ....126.2048G}; (66)~\citealt{2013MNRAS.428.3104E}; (67)~\citealt{2003AJ....125.1980K}; (68)~\citealt{1978PASP...90..429L}; (69)~\citealt{1967IAUS...30...57E}; (70)~\citealt{2002AA...384..180F}; (71)~\citealt{1987ApJS...65..581G}; (72)~\citealt{2010AJ....139..919M}; (73)~\citealt{2006ApJ...649L.115Z}; (74)~\citealt{2003ApJ...582.1011S}; (75)~\citealt{2004AA...418..989N}; (76)~\citealt{2015AA...573A.126D}; (77)~\citealt{2010MNRAS.403.1949K}; (78)~\citealt{1988AJ.....96..172G}; (79)~\citealt{1949ApJ...109..231J}; (80)~\citealt{1953GCRV..C......0W}; (81)~\citealt{1998AA...331...81P}; (82)~\citealt{1986AJ.....92..139S}; (83)~\citealt{2006MNRAS.367.1699H}; (84)~\citealt{1995AAS..110..367N}; (85)~\citealt{2001AA...373..159C}; (86)~\citealt{1986AJ.....91..144S}; (87)~\citealt{2009MNRAS.396.1895C}; (88)~\citealt{2010AA...523A..71G}; (89)~\citealt{1991ApJS...76..813S}; (90)~\citealt{2001AJ....121.2148G}; (91)~\citealt{2004MNRAS.349.1069K}; (92)~\citealt{2007MNRAS.374..664C}; (93)~\citealt{2004AA...424..727P}; (94)~\citealt{1965ApJ...141..177M}; (95)~\citealt{1989ApJS...71..245K}; (96)~\citealt{2010AJ....140.1433B}; (97)~\citealt{1973ARAA..11...29M}; (98)~\citealt{2001AA...373..625P}; (99)~\citealt{2014AJ....148..108B}; (100)~\citealt{2009AA...498..949M}; (101)~\citealt{1974PASP...86...70C}; (102)~\citealt{1991AJ....101.1495M}; (103)~\citealt{2012AA...546A..61D}; (104)~\citealt{1969PASP...81..643C}; (105)~\citealt{2009ApJ...694.1085V}; (106)~\citealt{1995AJ....109..780T}; (107)~\citealt{1935ApJ....81..187A}; (108)~\citealt{1989ApJS...69..301G}; (109)~\citealt{1965ApJ...142..681K}; (110)~\citealt{2013MNRAS.430.2154P}; (111)~\citealt{1985ApJS...59..229A}; (112)~\citealt{1962ApJ...136..793W}; (113)~\citealt{2013AJ....146..143M}; (114)~\citealt{2007AA...463..671R}; (115)~\citealt{1955ApJ...121...32N}; (116)~\citealt{1969AJ.....74..375C}; (117)~\citealt{2000ApJ...535..959Z}; (118)~\citealt{2007AJ....133..439C}; (119)~\citealt{1975AAS...19...91L}; (120)~\citealt{2015ApJ...798...73G}; (121)~\citealt{2005ESASP.560..571G}; (122)~\citealt{2014AJ....147..146K}; (123)~\citealt{2006ApJ...644..525M}; (124)~\citealt{1982mcts.book.....H}; (125)~\citealt{2014AA...563A.121D}; (126)~\citealt{2014AA...568A..26E}; (127)~\citealt{2017ApJS..228...18G}; (128)~\citealt{2008AA...489..825T}; (129)~\citealt{2013ApJS..208....9P}; (130)~\citealt{2013ApJ...762..118W}; (131)~\citealt{1999ApJ...512L..63W}; (132)~\citealt{2003AJ....125..825T}; (133)~\citealt{2011ApJ...727....6S}; (134)~\citealt{2010ApJ...714...45L}; (135)~\citealt{2016ApJ...833...95D}; (136)~\citealt{2010AJ....140.1486L}; (137)~\citealt{2012ApJ...757..163S}; (138)~\citealt{2003ApJ...593L.109M}; (139)~\citealt{2013AJ....145...44Z}; (140)~\citealt{2007ApJ...669L..45G}; (141)~\citealt{2011ApJ...727...62R}; (142)~\citealt{2005ApJ...634.1385M}; (143)~\citealt{2007AJ....134.2340K}; (144)~\citealt{1997JApA...18..161Y}; (145)~\citealt{2008AJ....135..209M}; (146)~\citealt{2004MNRAS.355..363L}; (147)~\citealt{2011AJ....142...15G}; (148)~\citealt{2004ApJ...609..917L}; (149)~\citealt{2013AA...551A..46L}; (150)~\citealt{2017MNRAS.468.1198B}; (151)~\citealt{2015AJ....150..101Z}; (152)~\citealt{2016arXiv161206924S}; (153)~\citealt{2010AJ....139.2440R}; (154)~\citealt{2000AA...353..186A}; (155)~\citealt{2009AJ....137.3487M}; (156)~\citealt{1976AJ.....81..245E}; (157)~\citealt{1978mcts.book.....H}; (158)~\citealt{1955AnCap..18....0J}; (159)~\citealt{2015IAUS..314...21M}; (160)~\citealt{2013MNRAS.435.1325M}; (161)~\citealt{2008hsf2.book..757T}; (162)~\citealt{1999AA...341L..79T}; (163)~\citealt{2012ApJ...747L..23K}; (164)~\citealt{2007AA...467.1147G}; (165)~\citealt{2000ApJ...544..356M}; (166)~\citealt{1999AAS..137..451G}; (167)~\citealt{2014AJ....148...47H}; (168)~\citealt{2004ApJ...608..809G}; (169)~\citealt{2004ApJ...602..816L}; (170)~\citealt{2006AJ....132..866R}; (171)~E.~Bubar et al., in prep.; (172)~\citealt{1997AA...328..187C}; (173)~\citealt{1998AAS..132..173L}; (174)~\citealt{2017ApJ...838..150K}; (175)~\citealt{2008ApJS..176..216A}; (176)~\citealt{2004ApJS..155..175A}; (177)~\citealt{2006AJ....132.2665S}; (178)~\citealt{1969ApJ...157..313H}; (179)~\citealt{2016MNRAS.461..794P}; (180)~\citealt{2012AJ....144....8S}; (181)~\citealt{2011ApJ...738..122C}; (182)~\citealt{2006MNRAS.371..252L}; (183)~\citealt{1977AJ.....82..431L}; (184)~\citealt{2011ApJS..193....4M}; (185)~\citealt{1988AJ.....96..297W}; (186)~\citealt{2012AA...538L...3R}; (187)~\citealt{2000AA...359..181W}; (188)~\citealt{1977ApJ...214..747H}; (189)~\citealt{2003ApJ...582.1109W}; (190)~\citealt{1987AJ.....93..907H}; (191)~\citealt{1998AA...334..592S}; (192)~\citealt{2012AA...539A.151A}; (193)~\citealt{2012ApJS..202....7X}; (194)~\citealt{1986ApJ...309..275H}; (195)~\citealt{1993AJ....105.1519P}; (196)~\citealt{1997MNRAS.291..658D}; (197)~\citealt{1986AJ.....91..575H}; (198)~\citealt{1997AA...321..497H}; (199)~\citealt{1988PASP..100..815A}; (200)~\citealt{1983ApJ...269..229M}; (201)~\citealt{2008AA...487..965S}; (202)~\citealt{2003ApJ...583..334H}; (203)~\citealt{2009AARv..17..251S}; (204)~\citealt{1994ApJ...427..961H}; (205)~\citealt{2010AA...515A..91M}; (206)~\citealt{1990ApJ...360..639H}; (207)~\citealt{1997AJ....113.1841M}; (208)~\citealt{1949ApJ...110..424J}; (209)~\citealt{2012yCat.1322....0Z}; (210)~\citealt{2013ApJ...771..110M}; (211)~\citealt{2014ApJ...784..126E}; (212)~\citealt{1979ApJS...41..743C}; (213)~\citealt{2001AA...378..116M}; (214)~\citealt{1999AA...351..954D}; (215)~\citealt{1999AA...352L..95G}; (216)~\citealt{2012AJ....144..130B}; (217)~\citealt{1970MmRAS..72..233H}; (218)~\citealt{1969MNRAS.144...31B}; (219)~\citealt{1997AAS..123..329K}; (220)~\citealt{2002AJ....124.1670M}; (221)~\citealt{2000AA...356..541K}; (222)~\citealt{2011AJ....142..140E}; (223)~\citealt{2007ApJ...667..308C}; (224)~\citealt{2007ApJ...657..338P}; (225)~\citealt{1949ApJ...109...92S}; (226)~\citealt{1992AAS...92..481B}; (227)~\citealt{2005AJ....130.1733W}; (228)~\citealt{2007MNRAS.379.1599L}; (229)~\citealt{2016ApJ...816...69A}; (230)~\citealt{2015AA...579A..66M}; (231)~\citealt{2006MNRAS.372.1879K}; (232)~\citealt{2010AA...521A..66R}; (233)~\citealt{2017AA...597A..90D}; (234)~\citealt{2006AA...458..173S}; (235)~\citealt{1998MNRAS.300..733M}; (236)~\citealt{2006AJ....131.3016S}; (237)~\citealt{2013MNRAS.431.3222L}; (238)~\citealt{1980AJ.....85..438R}; (239)~\citealt{2012AA...539A..41O}; (240)~\citealt{1988mcts.book.....H}; (241)~\citealt{2012ApJ...745...56D}; (242)~\citealt{1998AA...333..619P}; (243)~\citealt{2017arXiv171000909D}; (244)~\citealt{2011AJ....141..187S}; (245)~\citealt{2009AA...504..981B}; (246)~\citealt{1999AJ....117.2381P}; (247)~\citealt{1994AJ....107..692W}; (248)~\citealt{1981ApJS...45..437A}; (249)~\citealt{2001AJ....121.1040P}; (250)~\citealt{1976AAS...26..241C}; (251)~\citealt{2006ApJ...651L..49C}; (252)~\citealt{1980ApJS...44..489B}; (253)~\citealt{2006MNRAS.370..629D}; (254)~\citealt{2009ApJS..180..117A}; (255)~\citealt{2017AA...598A..48G}; (256)~\citealt{2016AN....337..863M}; (257)~\citealt{2015ApJ...807...27C}; (258)~\citealt{1964RA......6..535M}; (259)~\citealt{1984AAS...57..217B}; (260)~\citealt{1978PASP...90..201A}; (261)~\citealt{1956ApJ...123...54M}; (262)~\citealt{1989ApJS...70..623G}; (263)~\citealt{1946AnLei..19b...1B}; (264)~\citealt{1997AJ....114.1544W}; (265)~\citealt{2017yCat.1340....0Z}; (266)~\citealt{2003AA...410..269M}; (267)~\citealt{2007AA...464..687C}; (268)~\citealt{2007AA...475..959F}; (269)~\citealt{2003AJ....126.2971V}; (270)~\citealt{1996AA...310..228C}; (271)~\citealt{1988ApSS.148..163G}; (272)~\citealt{2011AA...532A..10M}; (273)~\citealt{2013MNRAS.431.1005D}; (274)~\citealt{2010PASP..122.1437P}; (275)~\citealt{2013MNRAS.435.1376M}; (276)~\citealt{2014ASPC..485..223B}; (277)~\citealt{2012AstL...38..331A}; (278)~\citealt{2006ARep...50..733B}; (279)~\citealt{2007AN....328..889K}; (280)~\citealt{2007AJ....133.2524W}; (281)~\citealt{2006AA...460..695T}; (282)~\citealt{2011AJ....141..187S}; (283)~\citealt{2012arXiv1207.6212C}; (284)~\citealt{2009AA...498..949M}.}
\enddata
\end{deluxetable*}

%% file: table_avg_observables.tex
\tabletypesize{\small}
\begin{deluxetable}{lcccccc}
\tablecolumns{7}
\tablecaption{Adopted average distances and radial velocities for young associations with partial kinematics. \label{tab:partialkin}}
\tablehead{\colhead{Association} & \colhead{$\nu_{\rm avg}$} & \colhead{$\sigma_\nu$} & \colhead{$N_{\nu}$} & \colhead{$\varpi_{\rm avg}$} & \colhead{$\sigma_\varpi$} & \colhead{$N_{\varpi}$}\\
\colhead{} & \colhead{(\kms)} & \colhead{(\kms)} & \colhead{} & \colhead{(pc)} & \colhead{(pc)} & \colhead{}}
\startdata
EPSC & $\cdots$ & $\cdots$ & $\cdots$ & 102.3 & 5.7 & 8\\
ETAC & 20.0 & 3.1 & 14 & 94.4 & 2.0 & 14\\
THOR& $\cdots$ & $\cdots$ & $\cdots$ & 96.2 & 3.5 & 4\\
XFOR & 18.8 & 1.4 & 5 & $\cdots$ & $\cdots$ & $\cdots$\\
PL8 & 21.9 & 2.0 & 4 & $\cdots$ & $\cdots$ & $\cdots$\\
ROPH & -6.3 & 0.3 & \tablenotemark{a} & 131.0 & 3.0 & \tablenotemark{a}\\
CRA & -0.4 & 1.2 & 11 & 139.4 & 6.1 & 3\\
UCRA & -2.5 & 2.4 & 9 & 148.0 & 3.0 & 4\\
TAU & 17.0 & 2.9 & 119 & 126 & 16 & 30\\
118TAU & 14.7 & 1.1 & 8 & 112.4 & 5.6 & 6\\
\enddata
\tablenotetext{a}{Average observables taken from \cite{2008AN....329...10M}.}
\end{deluxetable}

%% file: table_xyzuvw_cuts.tex
\startlongtable
\tabletypesize{\normalsize}
\begin{deluxetable}{ll}
\tablecolumns{2}
\tablecaption{Visual selection cuts applied to moving groups with noticeable outliers.\label{tab:xyzuvwcuts}}
\tablehead{\colhead{Association} & \colhead{Rejection criterion}}
\startdata
ABDMG & $V > -24$\,\kms\\
$\beta$PMG & $V > -13$\,\kms\\
 & $W > -4$\,\kms\\
CBER & $W < -3$\,\kms\\
ETAC & $Z < -37$\,pc\\
HYA & $V < -22$\,\kms\\
LCC & $U < -15$\,\kms\\  
PLE & $U > 0$\,\kms\\
THA & $Z > -20$\,pc\\
 & $U < 12.5$\,\kms\\
 & $ -4\,\kms < W < 2$\,\kms\\
THOR & $U < -25$\,\kms\\
USCO & $U > 10$\,\kms\\
\enddata
\end{deluxetable}

%% file: table_new_bonafide_objects.tex
\tabletypesize{\small}
\begin{deluxetable}{ll}
\tablecolumns{2}
\tablecaption{New bona fide members with full kinematics compiled in this work. \label{tab:newbonafide}}
\tablehead{\colhead{Designation} & \colhead{Ref.\tablenotemark{a}}}
\startdata
\sidehead{\textbf{AB~Doradus}}
HS~Psc & \cite{2014ApJ...788...81M}\\
HD~201919 & \cite{2014ApJ...788...81M}\\
\sidehead{\textbf{$\boldsymbol\beta$~Pictoris}}
AF~Psc & \cite{2017AJ....154...69S}\\
2MASS~J16572029-5343316 & \cite{2014ApJ...788...81M}\\
CD--31~16041 & \cite{2014ApJ...788...81M}\\
2MASS~J19560438-3207376 & \cite{2014ApJ...788...81M}\\
2MASS~J22424896-7142211 & \cite{2014ApJ...788...81M}\\
BD--13~6424 & \cite{2014ApJ...788...81M}\\
\sidehead{\textbf{Columba}}
GJ~1284 & \cite{2014ApJ...788...81M}\\
\sidehead{\textbf{Tucana-Horologium}}
2MASS~J02303239--4342232 & \cite{2014AJ....147..146K}\\
2MASS~J04000382--2902165 & \cite{2014AJ....147..146K}\\
2MASS~J04000395--2902280 & \cite{2014AJ....147..146K}\\
2MASS~J04021648--1521297 & \cite{2014AJ....147..146K}\\
CD--34~521 & \cite{2014ApJ...788...81M}\\
CD--53~544 & \cite{2014ApJ...788...81M}\\
CD--58~553 & \cite{2014ApJ...788...81M}\\
CD--35~1167 & \cite{2014ApJ...788...81M}\\
CD--44~1173 & \cite{2014ApJ...788...81M}\\
2MASS~J04480066-5041255 & \cite{2014ApJ...788...81M}\\
2MASS~J05332558-5117131 & \cite{2014ApJ...788...81M}\\
2MASS~J23261069-7323498 & \cite{2014ApJ...788...81M}\\
\sidehead{\textbf{$\boldsymbol\chi^{\boldsymbol 1}$~For}}
CD--37~1263\tablenotemark{b} & \cite{2014AA...564A..79D}\\
HD~17864 & B.~S.~Alessi, priv. comm.\\
\enddata
\tablenotetext{a}{Reference that designated this object as a candidate member of the young association.}
\tablenotetext{b}{The age of CD--37~1263 has not been investigated here; verifying its youth is still necessary to confirm that it is a true bona fide member.}
\tablecomments{This table lists objects that were designated as candidate members and that we confirm as bona fide members with full kinematics from compiling their missing measurements. See Section~\ref{sec:members} for more detail.}
\end{deluxetable}

%% file: table_young_pars.tex
\startlongtable
\tabletypesize{\small}
\begin{deluxetable*}{lccc@{\extracolsep{5pt}}ccc@{\extracolsep{5pt}}ccc@{\extracolsep{5pt}}ccc}
\tablecolumns{13}
\tablecaption{Parameters for the central location and variances of the multivariate Gaussian models of young associations.\label{tab:young_pars1}}
\tablehead{\colhead{Asso.} & \colhead{$\left<X\right>$} & \colhead{$\left<Y\right>$} & \colhead{$\left<Z\right>$} & \colhead{$\left<U\right>$} & \colhead{$\left<V\right>$} & \colhead{$\left<W\right>$} & \colhead{$\Sigma_{00}^{1/2}$} & \colhead{$\Sigma_{11}^{1/2}$} & \colhead{$\Sigma_{22}^{1/2}$} & \colhead{$\Sigma_{33}^{1/2}$} & \colhead{$\Sigma_{44}^{1/2}$} & \colhead{$\Sigma_{55}^{1/2}$}\\
\cline{2-4}
\cline{5-7}
\cline{8-10}
\cline{11-13}
\colhead{} & \multicolumn{3}{c}{(pc)} & \multicolumn{3}{c}{(\kms)} & \multicolumn{3}{c}{(pc)} & \multicolumn{3}{c}{(\kms)}}
\startdata
118TAU & -102.3 & -4.8 & -9.9 & -12.8 & -19.1 & -9.2 & 12.7 & 2.4 & 1.8 & 2.1 & 2.8 & 1.6\\
ABDMG & -6.0 & -7.2 & -8.8 & -7.2 & -27.6 & -14.2 & 21.4 & 20.3 & 16.3 & 1.4 & 1.0 & 1.8\\
$\beta$PMG & 4.1 & -6.7 & -15.7 & -10.9 & -16.0 & -9.0 & 29.3 & 14.0 & 9.0 & 2.2 & 1.2 & 1.0\\
CAR & 6.7 & -50.5 & -15.5 & -10.66 & -21.92 & -5.48 & 10.0 & 18.1 & 12.6 & 0.67 & 1.02 & 1.01\\
CARN & 0.7 & -28.1 & -4.3 & -25.3 & -18.1 & -2.3 & 7.8 & 20.8 & 17.3 & 3.2 & 1.9 & 2.0\\
CBER & -6.0 & -5.1 & 84.9 & -2.30 & -5.51 & -0.61 & 3.3 & 3.3 & 4.5 & 0.53 & 0.44 & 0.71\\
COL & -25.9 & -25.9 & -21.4 & -11.90 & -21.28 & -5.66 & 12.1 & 23.0 & 17.8 & 1.04 & 1.29 & 0.75\\
CRA & 132.45 & -0.21 & -42.43 & -3.7 & -15.7 & -8.8 & 3.71 & 0.75 & 2.04 & 1.3 & 2.2 & 2.2\\
EPSC & 49.9 & -84.8 & -25.6 & -9.9 & -19.3 & -9.7 & 2.5 & 3.6 & 4.0 & 1.6 & 2.2 & 2.0\\
ETAC & 33.65 & -81.36 & -34.81 & -10.0 & -22.3 & -11.7 & 0.65 & 0.98 & 0.71 & 1.6 & 2.8 & 1.8\\
HYA & -38.5 & 0.8 & -15.8 & -42.27 & -18.79 & -1.47 & 7.4 & 4.4 & 2.9 & 2.01 & 0.94 & 1.10\\
IC2391 & 1.9 & -148.1 & -18.0 & -23.04 & -14.89 & -5.48 & 1.3 & 6.4 & 1.4 & 1.10 & 3.40 & 0.78\\
IC2602 & 47.4 & -137.6 & -12.6 & -8.22 & -20.60 & -0.58 & 1.5 & 5.4 & 1.1 & 1.18 & 2.61 & 0.65\\
LCC & 54.3 & -94.2 & 5.8 & -7.8 & -21.5 & -6.2 & 11.9 & 12.4 & 13.7 & 2.7 & 3.8 & 1.8\\
OCT & 4.0 & -96.9 & -59.7 & -13.7 & -3.3 & -10.1 & 78.3 & 25.8 & 8.8 & 2.4 & 1.3 & 1.4\\
PL8 & 10.6 & -124.5 & -13.9 & -11.01 & -22.89 & -3.59 & 7.0 & 11.6 & 4.5 & 1.15 & 1.96 & 0.74\\
PLE & -118.9 & 28.5 & -54.4 & -6.7 & -28.0 & -14.0 & 7.7 & 3.5 & 4.2 & 1.7 & 1.8 & 1.2\\
ROPH & 124.79 & -15.23 & 37.60 & -5.9 & -13.5 & -7.9 & 1.33 & 0.51 & 0.66 & 1.3 & 4.7 & 4.3\\
TAU & -116.3 & 6.7 & -35.9 & -14.3 & -9.3 & -8.8 & 11.4 & 10.8 & 10.1 & 3.1 & 4.5 & 3.4\\
THA & 5.4 & -20.1 & -36.1 & -9.79 & -20.94 & -0.99 & 19.4 & 12.4 & 3.8 & 0.87 & 0.79 & 0.72\\
THOR & -88.4 & -25.7 & -23.9 & -12.8 & -18.8 & -9.0 & 4.1 & 6.9 & 5.1 & 2.2 & 2.2 & 2.0\\
TWA & 14.4 & -47.7 & 22.7 & -11.6 & -17.9 & -5.6 & 12.2 & 9.7 & 3.9 & 1.8 & 1.8 & 1.6\\
UCL & 107.5 & -60.9 & 26.5 & -4.7 & -19.7 & -5.2 & 21.0 & 19.6 & 13.5 & 3.8 & 3.0 & 1.7\\
UCRA & 142.1 & -1.2 & -39.2 & -3.7 & -17.1 & -8.0 & 7.3 & 2.4 & 5.9 & 3.0 & 1.8 & 1.2\\
UMA & -7.5 & 9.9 & 21.9 & 14.8 & 1.8 & -10.2 & 3.1 & 1.5 & 1.1 & 1.0 & 1.2 & 2.6\\
USCO & 121.2 & -17.0 & 48.9 & -4.9 & -14.2 & -6.5 & 17.0 & 8.2 & 8.9 & 3.7 & 3.2 & 2.3\\
XFOR & -27.1 & -46.3 & -84.2 & -12.54 & -22.24 & -6.26 & 4.7 & 3.8 & 4.4 & 0.96 & 1.41 & 2.21\\
\enddata
\tablecomments{The average Galactic positions $\left<X\right>$, $\left<Y\right>$ and $\left<Z\right>$ and space velocities $\left<U\right>$, $\left<V\right>$ and $\left<W\right>$ correspond to the components of the $\bar\tau$ vector defining the center of the multivariate Gaussian model. Parameters $\Sigma_{00}$ through $\Sigma_{55}$ represent the diagonal elements of the covariance matrix, and correspond to the dispersion of Galactic positions and space velocities along the principal axes of the multivariate Gaussian, which are not necessarily aligned with the Galactic coordinates axes. See Section~\ref{sec:ymgmodels} for more detail. The covariances of the multivariate Gaussian models are given in a FITS file format in the online-only additional material.}
\end{deluxetable*}

%% file: table_riz2011.tex
\startlongtable
\tabletypesize{\small}
\begin{deluxetable}{lcc}
\tablecolumns{3}
\tablecaption{BANYAN~$\Sigma$ probabilities for new candidate members of Sco-Cen identified by \cite{2011MNRAS.416.3108R}.\label{tab:riz2011}}
\tablehead{\colhead{2MASS} & \colhead{Asso.} & \colhead{$P$}\\
\colhead{Designation} & \colhead{} & \colhead{(\%)}}
\startdata
\sidehead{\textbf{Unambiguous candidate members of USCO}}
15480330-2512562 & USCO & 96.8\\
15574880-2331383 & USCO & 99.0\\
16052655-1948066 & USCO & 99.2\\
16120593-2314445 & USCO & 98.6\\
16203056-2006518 & USCO & 99.2\\
16301246-2506548 & USCO & 97.1\\
\sidehead{\textbf{Unambiguous candidate members of UCL}}
13514960-3259387 & UCL & 96.3\\
14170338-3432122 & UCL & 96.6\\
14250102-3726493 & UCL & 97.8\\
14281043-2929299 & UCL & 97.3\\
14353043-4209281 & UCL & 99.6\\
14353149-4131026 & UCL & 99.9\\
15025928-3238357 & UCL & 96.9\\
15130106-3714480 & UCL & 99.6\\
15183199-4752307 & UCL & 97.6\\
15195653-3006249 & UCL & 95.4\\
15342085-3920572 & UCL & 99.6\\
15383263-3909384 & UCL & 99.8\\
15500707-5312351 & UCL & 96.0\\
16003131-3605164 & UCL & 98.6\\
16040712-3510367 & UCL & 96.8\\
16312294-3442153 & UCL & 95.6\\
16383094-3909083 & UCL & 95.5\\
\sidehead{\textbf{Unambiguous candidate members of LCC}}
11454479-5241258 & LCC & 95.1\\
12044525-5915117 & LCC & 99.8\\
12113912-5222065 & LCC & 98.4\\
12263615-6305571 & LCC & 99.8\\
12474326-5941194 & LCC & 99.7\\
\sidehead{\textbf{Ambiguous candidate members}}
13172895-4255587 & TWA,UCL,LCC & 79.4,10.5,5.7\\
13324248-5549391 & LCC,UCL & 89.4,7.8\\
13414477-5433339 & LCC,UCL & 76.6,21.6\\
14074081-4842144 & UCL,LCC & 80.4,18.8\\
15585013-3203082 & UCL,USCO & 74.3,23.4\\
16115069-2733098 & USCO,UCL & 49.3,47.4\\
16201552-2843008 & USCO,UCL & 77.7,20.6\\
\enddata
\tablecomments{See Section~\ref{sec:ambig} for more detail.}
\end{deluxetable}

%% file: table_young_roc.tex
\startlongtable
\tabletypesize{\footnotesize}
\setlength\tabcolsep{3pt}
\begin{longrotatetable}
\global\pdfpageattr\expandafter{\the\pdfpageattr/Rotate 90}
\begin{deluxetable*}{lcccc@{\extracolsep{5pt}}cccc@{\extracolsep{5pt}}cccc}
\tablecolumns{13}
\tablecaption{BANYAN~$\Sigma$ classification performance for different young associations as a function of input observables.\label{tab:young_roc}}
\tablehead{\colhead{Asso.} & \multicolumn{4}{c}{TPR$_{\rm crit}$} & \multicolumn{4}{c}{$\log_{10}$\,FPR$_{\rm crit}$} & \multicolumn{4}{c}{$\log_{10}$\,MCC$_{\rm crit}$}\\
\cline{2-5}
\cline{6-9}
\cline{10-13}
\colhead{} & \colhead{$\mu$} & \colhead{$\mu$,$\nu$} & \colhead{$\mu,\varpi$} & \colhead{$\mu,\nu,\varpi$} & \colhead{$\mu$} & \colhead{$\mu$,$\nu$} & \colhead{$\mu,\varpi$} & \colhead{$\mu,\nu,\varpi$} & \colhead{$\mu$} & \colhead{$\mu$,$\nu$} & \colhead{$\mu,\varpi$} & \colhead{$\mu,\nu,\varpi$}}
\startdata
118TAU & $0.5_{-0.1}^{+0.3}$ & $0.7_{-0.1}^{+0.2}$ & $0.82_{-0.05}^{+0.10}$ & $0.90_{-0.03}^{+0.07}$ & $-4.9_{-0.2}^{+0.3}$ & $-5.6 \pm 0.2$ & $-5.9_{-0.2}^{+0.3}$ & $-6.4^{+0.5}$ & $-1.39_{-0.01}^{+0.20}$ & $-0.88_{-0.02}^{+0.08}$ & $-0.70_{-0.06}^{+0.07}$ & $-0.43_{-0.01}^{+0.10}$\\
ABDMG & $0.5_{-0.1}^{+0.2}$ & $0.68_{-0.07}^{+0.10}$ & $0.82_{-0.06}^{+0.20}$ & $0.90_{-0.03}^{+0.09}$ & $-2.8_{-0.2}^{+0.4}$ & $-3.5_{-0.2}^{+0.3}$ & $-4.0_{-0.2}^{+0.3}$ & $-4.5_{-0.1}^{+0.3}$ & $-2.24_{-0.01}$ & $-1.78_{-0.05}^{+0.07}$ & $-1.43_{-0.05}^{+0.07}$ & $-1.14_{-0.06}^{+0.08}$\\
BPMG & $0.5_{-0.1}^{+0.2}$ & $0.68_{-0.07}^{+0.10}$ & $0.82_{-0.06}^{+0.10}$ & $0.90_{-0.03}^{+0.08}$ & $-2.6_{-0.2}^{+0.4}$ & $-3.2_{-0.2}^{+0.3}$ & $-4.4_{-0.1}^{+0.3}$ & $-4.8_{-0.2}^{+0.3}$ & $-2.39_{-0.02}$ & $-1.95_{-0.04}^{+0.06}$ & $-1.28_{-0.05}^{+0.06}$ & $-1.03 \pm 0.09$\\
CAR & $0.5_{-0.1}^{+0.2}$ & $0.68_{-0.09}^{+0.20}$ & $0.82_{-0.05}^{+0.10}$ & $0.90_{-0.03}^{+0.07}$ & $-2.9_{-0.2}^{+0.5}$ & $-3.6_{-0.2}^{+0.4}$ & $-5.0_{-0.2}^{+0.4}$ & $-5.3_{-0.1}^{+0.5}$ & $-2.36^{+0.06}$ & $-1.93_{-0.04}^{+0.03}$ & $-1.12_{-0.08}^{+0.10}$ & $-0.96_{-0.04}^{+0.20}$\\
CARN & $0.5_{-0.1}^{+0.2}$ & $0.68_{-0.08}^{+0.20}$ & $0.82_{-0.06}^{+0.10}$ & $0.90_{-0.03}^{+0.09}$ & $-3.3_{-0.2}^{+0.5}$ & $-4.0_{-0.2}^{+0.3}$ & $-4.5_{-0.2}^{+0.3}$ & $-4.9_{-0.1}^{+0.3}$ & $-2.18^{+0.03}$ & $-1.71_{-0.05}^{+0.03}$ & $-1.39_{-0.06}^{+0.05}$ & $-1.14_{-0.06}^{+0.10}$\\
CBER & $0.5_{-0.2}^{+0.5}$ & $0.7_{-0.1}^{+0.4}$ & $0.82_{-0.06}^{+0.20}$ & $0.90_{-0.04}^{+0.10}$ & $-4.5_{-0.2}^{+0.6}$ & $-5.2_{-0.2}^{+0.6}$ & $-6.9_{-0.3}$ & $< -7$ & $-1.40_{-0.02}^{+0.80}$ & $-0.96_{-0.03}^{+0.05}$ & $-0.19_{-0.06}^{+0.08}$ & $-0.17_{-0.01}^{+0.02}$\\
COL & $0.50_{-0.08}^{+0.10}$ & $0.68_{-0.05}^{+0.09}$ & $0.82_{-0.05}^{+0.10}$ & $0.90_{-0.03}^{+0.06}$ & $-2.7_{-0.1}^{+0.3}$ & $-3.2_{-0.1}^{+0.2}$ & $-4.5_{-0.1}^{+0.3}$ & $-4.8_{-0.1}^{+0.2}$ & $-2.46_{-0.01}$ & $-2.08_{-0.03}^{+0.05}$ & $-1.33_{-0.05}^{+0.07}$ & $-1.13_{-0.05}^{+0.08}$\\
CRA & $0.5_{-0.1}^{+0.2}$ & $0.68_{-0.09}^{+0.20}$ & $0.82_{-0.06}^{+0.20}$ & $0.90_{-0.03}^{+0.09}$ & $-5.06_{-0.09}^{+0.30}$ & $-5.8 \pm 0.1$ & $-6.55_{-0.20}$ & $< -7$ & $-1.31_{-0.06}^{+0.10}$ & $-0.81^{+0.08}$ & $-0.39_{-0.05}^{+0.04}$ & $> -0.22$\\
EPSC & $0.5_{-0.2}^{+0.3}$ & $0.7_{-0.1}^{+0.3}$ & $0.82_{-0.06}^{+0.20}$ & $0.90_{-0.03}^{+0.10}$ & $-4.9_{-0.2}^{+0.6}$ & $-5.3_{-0.2}^{+0.5}$ & $< -7$ & $-6.85_{-0.50}$ & $-1.35_{-0.01}^{+0.20}$ & $-1.00 \pm 0.05$ & $-0.43_{-0.03}^{+0.10}$ & $-0.22_{-0.20}^{+0.03}$\\
ETAC & $0.5_{-0.2}^{+0.3}$ & $0.7_{-0.1}^{+0.3}$ & $0.82_{-0.07}^{+0.20}$ & $0.90_{-0.04}^{+0.10}$ & $-6.0_{-0.2}^{+0.4}$ & $-6.37_{-0.40}^{+0.01}$ & $< -7$ & $< -7$ & $-0.85_{-0.03}^{+0.30}$ & $-0.6_{-0.1}^{+0.2}$ & $> -0.22$ & $> -0.22$\\
HYA & $0.5_{-0.1}^{+0.3}$ & $0.68_{-0.09}^{+0.20}$ & $0.82_{-0.06}^{+0.10}$ & $0.90_{-0.03}^{+0.09}$ & $-5.4_{-0.2}^{+0.6}$ & $-6.2_{-0.3}^{+0.7}$ & $-6.6_{-0.2}^{+0.3}$ & $-6.82_{-0.40}$ & $-0.67_{-0.02}^{+0.05}$ & $-0.27_{-0.06}^{+0.04}$ & $-0.12_{-0.01}^{+0.02}$ & $-0.07_{-0.05}^{+0.02}$\\
IC2391 & $0.5_{-0.1}^{+0.3}$ & $0.68_{-0.07}^{+0.20}$ & $0.82_{-0.04}^{+0.09}$ & $0.90_{-0.02}^{+0.05}$ & $-5.3_{-0.1}^{+0.3}$ & $-5.7_{-0.1}^{+0.3}$ & $-6.0^{+0.4}$ & $-6.6^{+0.3}$ & $-1.18_{-0.03}^{+0.20}$ & $-0.88^{+0.03}$ & $-0.66_{-0.02}^{+0.10}$ & $-0.36^{+0.09}$\\
IC2602 & $0.5_{-0.1}^{+0.3}$ & $0.68_{-0.08}^{+0.20}$ & $0.82_{-0.05}^{+0.10}$ & $0.90_{-0.02}^{+0.06}$ & $-5.0_{-0.1}^{+0.3}$ & $-5.42_{-0.07}^{+0.20}$ & $< -7$ & $< -7$ & $-1.35_{-0.05}^{+0.30}$ & $-1.00_{-0.01}^{+0.08}$ & $-0.28 \pm 0.02$ & $> -0.24$\\
LCC & $0.5_{-0.1}^{+0.3}$ & $0.68_{-0.09}^{+0.20}$ & $0.82_{-0.06}^{+0.10}$ & $0.90_{-0.03}^{+0.08}$ & $-3.2_{-0.2}^{+0.4}$ & $-3.5_{-0.2}^{+0.3}$ & $-4.5_{-0.2}^{+0.3}$ & $-4.6 \pm 0.2$ & $-1.91_{-0.02}^{+0.10}$ & $-1.62_{-0.03}^{+0.02}$ & $-1.06_{-0.06}^{+0.07}$ & $-0.97 \pm 0.07$\\
OCT & $0.50_{-0.06}^{+0.10}$ & $0.68_{-0.03}^{+0.05}$ & $0.82_{-0.02}^{+0.04}$ & $0.90_{-0.01}^{+0.02}$ & $-3.0_{-0.1}^{+0.2}$ & $-3.22_{-0.09}^{+0.20}$ & $-3.43_{-0.09}^{+0.20}$ & $-3.38_{-0.05}^{+0.10}$ & $-2.31^{+0.01}$ & $-2.09_{-0.03}^{+0.04}$ & $-1.91_{-0.03}^{+0.06}$ & $-1.89_{-0.02}^{+0.04}$\\
PL8 & $0.5_{-0.1}^{+0.3}$ & $0.68_{-0.08}^{+0.20}$ & $0.82_{-0.04}^{+0.09}$ & $0.90_{-0.02}^{+0.05}$ & $-4.3_{-0.1}^{+0.2}$ & $-4.6_{-0.1}^{+0.2}$ & $-5.4_{-0.1}^{+0.2}$ & $-5.7_{-0.1}^{+0.2}$ & $-1.70_{-0.03}^{+0.30}$ & $-1.38^{+0.03}$ & $-0.94 \pm 0.05$ & $-0.72_{-0.05}^{+0.08}$\\
PLE & $0.5_{-0.2}^{+0.4}$ & $0.7_{-0.1}^{+0.3}$ & $0.82_{-0.05}^{+0.10}$ & $0.90_{-0.03}^{+0.09}$ & $-5.3_{-0.3}^{+0.4}$ & $-5.7_{-0.2}^{+0.7}$ & $-6.0_{-0.4}^{+0.2}$ & $-6.4_{-0.3}^{+0.5}$ & $-0.72_{-0.01}^{+0.40}$ & $-0.43_{-0.01}^{+0.03}$ & $-0.2_{-0.1}$ & $-0.12_{-0.06}^{+0.03}$\\
ROPH & $0.5_{-0.1}^{+0.2}$ & $0.68_{-0.07}^{+0.20}$ & $0.82_{-0.07}^{+0.20}$ & $0.90_{-0.04}^{+0.10}$ & $-5.4 \pm 0.2$ & $-6.3 \pm 0.1$ & $< -7$ & $< -7$ & $-0.68^{+0.08}$ & $-0.24^{+0.05}$ & $> -0.05$ & $> -0.04$\\
TAU & $0.50_{-0.08}^{+0.10}$ & $0.68_{-0.07}^{+0.10}$ & $0.82_{-0.06}^{+0.20}$ & $0.90_{-0.03}^{+0.09}$ & $-2.5_{-0.1}^{+0.2}$ & $-3.0_{-0.1}^{+0.2}$ & $-4.2_{-0.2}^{+0.3}$ & $-4.5_{-0.2}^{+0.3}$ & $-2.18_{-0.01}^{+0.05}$ & $> -1.81$ & $-1.14_{-0.06}^{+0.07}$ & $-0.93_{-0.06}^{+0.10}$\\
THA & $0.5_{-0.1}^{+0.2}$ & $0.68_{-0.08}^{+0.20}$ & $0.82_{-0.05}^{+0.10}$ & $0.90_{-0.03}^{+0.07}$ & $-3.9_{-0.2}^{+0.5}$ & $-4.5_{-0.2}^{+0.4}$ & $-5.3_{-0.1}^{+0.3}$ & $-5.47_{-0.08}^{+0.20}$ & $-1.76_{-0.03}$ & $-1.33_{-0.06}^{+0.08}$ & $-0.84_{-0.04}^{+0.07}$ & $-0.71_{-0.03}^{+0.05}$\\
THOR & $0.5_{-0.1}^{+0.3}$ & $0.7_{-0.1}^{+0.2}$ & $0.82_{-0.06}^{+0.20}$ & $0.90_{-0.03}^{+0.10}$ & $-3.6_{-0.2}^{+0.5}$ & $-4.2_{-0.2}^{+0.4}$ & $-5.7_{-0.3}^{+0.5}$ & $-6.1_{-0.2}^{+0.5}$ & $-1.90_{-0.01}^{+0.10}$ & $-1.49_{-0.03}^{+0.01}$ & $-0.66_{-0.09}^{+0.10}$ & $-0.46_{-0.08}^{+0.10}$\\
TWA & $0.5_{-0.1}^{+0.2}$ & $0.68_{-0.09}^{+0.20}$ & $0.82_{-0.05}^{+0.10}$ & $0.90_{-0.03}^{+0.08}$ & $-4.1_{-0.2}^{+0.5}$ & $-4.6_{-0.2}^{+0.4}$ & $-5.5_{-0.1}^{+0.3}$ & $-5.8_{-0.2}^{+0.3}$ & $-1.76^{+0.03}$ & $-1.37 \pm 0.04$ & $-0.86_{-0.04}^{+0.07}$ & $-0.67_{-0.08}^{+0.10}$\\
UCL & $0.5_{-0.1}^{+0.3}$ & $0.68_{-0.09}^{+0.20}$ & $0.82_{-0.06}^{+0.10}$ & $0.90_{-0.03}^{+0.08}$ & $-2.9_{-0.2}^{+0.4}$ & $-3.2_{-0.2}^{+0.3}$ & $-3.7_{-0.2}^{+0.3}$ & $-3.9_{-0.1}^{+0.2}$ & $-2.05_{-0.01}^{+0.10}$ & $-1.75_{-0.02}$ & $-1.40_{-0.05}^{+0.09}$ & $-1.27_{-0.06}^{+0.08}$\\
UCRA & $0.5_{-0.1}^{+0.2}$ & $0.68_{-0.06}^{+0.10}$ & $0.82_{-0.03}^{+0.07}$ & $0.90_{-0.02}^{+0.03}$ & $-4.3_{-0.1}^{+0.3}$ & $-4.8_{-0.1}^{+0.2}$ & $-5.2_{-0.1}^{+0.2}$ & $-5.32_{-0.07}^{+0.20}$ & $-1.67_{-0.02}^{+0.10}$ & $-1.29_{-0.02}$ & $-1.02_{-0.05}^{+0.07}$ & $-0.92_{-0.03}^{+0.09}$\\
UMA & $0.5_{-0.1}^{+0.2}$ & $0.68_{-0.09}^{+0.20}$ & $0.82_{-0.06}^{+0.10}$ & $0.90_{-0.03}^{+0.09}$ & $< -7$ & $< -7$ & $< -7$ & $< -7$ & $-0.58_{-0.07}^{+0.09}$ & $-0.33 \pm 0.02$ & $> -0.21$ & $\sim 0$\\
USCO & $0.5_{-0.1}^{+0.2}$ & $0.68_{-0.09}^{+0.20}$ & $0.82_{-0.05}^{+0.10}$ & $0.90_{-0.03}^{+0.08}$ & $-3.7_{-0.2}^{+0.3}$ & $-4.0_{-0.1}^{+0.3}$ & $-4.5_{-0.1}^{+0.3}$ & $-4.7_{-0.1}^{+0.3}$ & $-1.70_{-0.02}^{+0.10}$ & $-1.39_{-0.01}^{+0.02}$ & $-1.04_{-0.04}^{+0.08}$ & $-0.93_{-0.04}^{+0.09}$\\
XFOR & $0.5_{-0.1}^{+0.3}$ & $0.68_{-0.08}^{+0.20}$ & $0.82_{-0.04}^{+0.10}$ & $0.90_{-0.02}^{+0.05}$ & $-5.4_{-0.1}^{+0.4}$ & $-5.9_{-0.3}^{+0.2}$ & $-6.82_{-0.30}$ & $< -7$ & $-1.15_{-0.03}^{+0.20}$ & $-0.78_{-0.08}^{+0.06}$ & $-0.29_{-0.08}^{+0.02}$ & $-0.25^{+0.02}$\\
\enddata
\tablecomments{True-positive rates (TPRs), false-positive rates (FPRs) and Matthews correlation coefficients (MCCs) are reported at the `critical' $P_t=$90\% threshold, and their reported ranges are for thresholds in the range 85--95\%. See Section~\ref{sec:performance} for more detail.}
\end{deluxetable*}
\setlength\tabcolsep{6pt}
\end{longrotatetable}
\global\pdfpageattr\expandafter{\the\pdfpageattr/Rotate 0}

%% file: table_bonafide.tex
\startlongtable
\tabletypesize{\normalsize}
\begin{longrotatetable}
\global\pdfpageattr\expandafter{\the\pdfpageattr/Rotate 90}
\begin{deluxetable*}{llccr@{\;$\pm$\;}lr@{\;$\pm$\;}lr@{\;$\pm$\;}lr@{\;$\pm$\;}ll}
\tablecolumns{13}
\tablecaption{Literature compilation of bona fide members. \label{tab:bonafide}}
\tablehead{\colhead{Main} & \colhead{Spectral} & \colhead{R.A.} & \colhead{Decl.} & \multicolumn{2}{c}{$\mu_\alpha\cos\delta$}  & \multicolumn{2}{c}{$\mu_\delta$} & \multicolumn{2}{c}{Rad. Vel.} & \multicolumn{2}{c}{Distance} & \colhead{} \\
\colhead{Designation} & \colhead{Type} & \colhead{(hh:mm:ss)} & \colhead{(dd:mm:ss)} & \multicolumn{2}{c}{(\masyr)} & \multicolumn{2}{c}{(\masyr)} & \multicolumn{2}{c}{(\kms)} & \multicolumn{2}{c}{(pc)} & \colhead{References}}
\startdata
\sidehead{\Large\textbf{AB~Doradus}}
2MASS~J00192626+4614078 & M8\,$\beta$ & 00:19:26.26 & 46:14:07.8 & $119.4$ & $0.9$ & $-75.4$ & $0.9$ & $-20$ & $3$ & $39$ & $2$ & 1,2,3,2\\
BD+54~144~A & F8\,V & 00:45:51.06 & 54:58:39.1 & $96.40$ & $0.03$ & $-73.97$ & $0.04$ & $-15$ & $2$ & $50.3$ & $0.9$ & 4,5,6,5\\
| BD+54~144~B & K3 & 00:45:51.23 & 54:58:40.8 & \multicolumn{2}{c}{$\cdots$} & \multicolumn{2}{c}{$\cdots$} & \multicolumn{2}{c}{$\cdots$} & \multicolumn{2}{c}{$\cdots$} & 7,--,--,--\\
2MASS~J00470038+6803543 & L6--L8\,$\gamma$ & 00:47:00.39 & 68:03:54.4 & $385$ & $1$ & $-201$ & $1$ & $-20$ & $1$ & $12.2$ & $0.3$ & --,2,8,2\\
G~132--51~B & M2.6 & 01:03:42.23 & 40:51:13.6 & $132$ & $5$ & $-164$ & $5$ & $-10.6$ & $0.3$ & $30$ & $2$ & 9,9,9,9\\
| G~132--50 & M0 & 01:03:40.30 & 40:51:26.7 & $126.9$ & $0.1$ & $-166.18$ & $0.09$ & \multicolumn{2}{c}{$\cdots$} & $33.0$ & $0.6$ & 10,5,--,5\\
| G~132--51~C & M3.8 & 01:03:42.44 & 40:51:13.3 & \multicolumn{2}{c}{$\cdots$} & \multicolumn{2}{c}{$\cdots$} & $-10.9$ & $0.4$ & \multicolumn{2}{c}{$\cdots$} & 9,--,9,--\\
HIP~6276 & G0\,V & 01:20:32.38 & -11:28:05.8 & $111.44$ & $0.06$ & $-136.95$ & $0.05$ & $8.3$ & $0.4$ & $35.1$ & $0.4$ & 11,5,12,5\\
\enddata
\tablerefs{The references to this table are listed in Table~\ref{tab:refs}.}
\end{deluxetable*}
\end{longrotatetable}

%% file: table_excluded.tex
\startlongtable
\tabletypesize{\normalsize}
\begin{longrotatetable}
\begin{deluxetable*}{llccr@{\;$\pm$\;}lr@{\;$\pm$\;}lr@{\;$\pm$\;}lr@{\;$\pm$\;}lll}
\tablecolumns{14}
\tablecaption{Excluded objects.\label{tab:excluded}}
\tablehead{\colhead{Main} & \colhead{Spectral} & \colhead{R.A.} & \colhead{Decl.} & \multicolumn{2}{c}{$\mu_\alpha\cos\delta$}  & \multicolumn{2}{c}{$\mu_\delta$} & \multicolumn{2}{c}{Rad. Vel.} & \multicolumn{2}{c}{Distance} & \colhead{Reason for} & \colhead{} \\
\colhead{Designation} & \colhead{Type} & \colhead{(hh:mm:ss)} & \colhead{(dd:mm:ss)} & \multicolumn{2}{c}{(\masyr)} & \multicolumn{2}{c}{(\masyr)} & \multicolumn{2}{c}{(\kms)} & \multicolumn{2}{c}{(pc)} & \colhead{Exclusion\tablenotemark{a}} & \colhead{References}}
\startdata
\sidehead{\Large\textbf{AB~Doradus}}
PX~Vir & G5\,V & 13:03:49.46 & -05:09:45.9 & $-191.1$ & $0.9$ & $-218.7$ & $0.7$ & $-9.2$ & $0.2$ & $21.7$ & $0.4$ & VIS & 1,2,3,2\\
2MASS~J15534211--2049282 & M3.4 & 15:53:42.09 & -20:49:28.6 & $-10$ & $10$ & $-20$ & $10$ & $-7$ & $2$ & $330$ & $80$ & $\sigma_{\rm kin}$ & 4,5,4,4\\
HR~7214 & A4\,V & 19:03:32.25 & 01:49:07.6 & $22.6$ & $0.2$ & $-68.5$ & $0.3$ & $-23$ & $2$ & $54.9$ & $0.9$ & MST$_{0}$ & 6,2,6,2\\
LQ~Peg & K8\,V & 21:31:01.86 & 23:20:05.2 & $134.6$ & $0.1$ & $-144.9$ & $0.1$ & $-17$ & $1$ & $24.2$ & $0.1$ & VIS & 1,7,8,7\\
HIP~107948 & M3\,Ve & 21:52:10.51 & 05:37:33.7 & $106$ & $2$ & $-147$ & $1$ & $-15$ & $2$ & $31$ & $5$ & $\sigma_{\rm kin}$ & 9,9,9,2\\
\sidehead{\Large\textbf{$\boldsymbol\beta$~Pictoris}}
LP~353--51 & M3\,V\,e & 02:23:26.75 & 22:44:05.1 & $98.5$ & $0.2$ & $-112.5$ & $0.1$ & $13.0$ & $0.4$ & $27.1$ & $0.3$ & VIS & 1,7,10,7\\
HD~15115 & F4\,IV & 02:26:16.33 & 06:17:32.4 & $87.97$ & $0.04$ & $-50.35$ & $0.03$ & $0.8$ & $0.1$ & $48$ & $1$ & VIS & 11,7,12,7\\
EPIC~211046195 & M8.5 & 03:35:02.09 & 23:42:35.6 & $50$ & $10$ & $-60$ & $10$ & $16$ & $2$ & $42$ & $2$ & VIS & 4,5,4,4\\
\enddata
\end{deluxetable*}
\end{longrotatetable}

%% file: table_designations.tex
\startlongtable
\tabletypesize{\normalsize}
\begin{longrotatetable}
\begin{deluxetable*}{lllll}
\tablecolumns{5}
\tablecaption{List of bona fide members designations. \label{tab:designations}}
\tablehead{\colhead{Main} & \colhead{2MASS} & \colhead{AllWISE} & \colhead{Gaia} & \colhead{Other}}
\startdata
\sidehead{\Large\textbf{AB~Doradus}}
2MASS~J00192626+4614078 & J00192626+4614078 & J001926.39+461406.8 & $\cdots$ & $\cdots$\\
BD+54~144~A & J00455088+5458402 & J004551.02+545839.6 & 417565757132068096 & HD~4277~A,~HIP~3589~A\\
| BD+54~144~B & $\cdots$ & $\cdots$ & $\cdots$ & HD~4277~B,~HIP~3589~B\\
2MASS~J00470038+6803543 & J00470038+6803543 & J004701.09+680352.2 & 529737830321549952 & $\cdots$\\
G~132--51~B & J01034210+4051158 & J010342.24+405114.2 & 374400889126932096 & $\cdots$\\
| G~132--50 & J01034013+4051288 & J010340.24+405127.4 & 374400957846408192 & $\cdots$\\
| G~132--51~C & $\cdots$ & $\cdots$ & 374400893422315648 & $\cdots$\\
HIP~6276 & J01203226--1128035 & J012032.34--112805.2 & 2470272808484339200 & CD--12~243\\
\enddata
\end{deluxetable*}
\end{longrotatetable}
\global\pdfpageattr\expandafter{\the\pdfpageattr/Rotate 0}

%% file: table_excluded_short.tex
\startlongtable
\tabletypesize{\normalsize}
\begin{longrotatetable}
\global\pdfpageattr\expandafter{\the\pdfpageattr/Rotate 90}
\begin{deluxetable*}{llccr@{\;$\pm$\;}lr@{\;$\pm$\;}lr@{\;$\pm$\;}lr@{\;$\pm$\;}lll}
\tablecolumns{14}
\tablecaption{Objects listed as bona fide members that were excluded from the construction of kinematic models.\label{tab:excluded}}
\tablehead{\colhead{Main} & \colhead{Spectral} & \colhead{R.A.} & \colhead{Decl.} & \multicolumn{2}{c}{$\mu_\alpha\cos\delta$}  & \multicolumn{2}{c}{$\mu_\delta$} & \multicolumn{2}{c}{Rad. Vel.} & \multicolumn{2}{c}{Distance} & \colhead{Reason for} & \colhead{} \\
\colhead{Designation} & \colhead{Type} & \colhead{(hh:mm:ss)} & \colhead{(dd:mm:ss)} & \multicolumn{2}{c}{(\masyr)} & \multicolumn{2}{c}{(\masyr)} & \multicolumn{2}{c}{(\kms)} & \multicolumn{2}{c}{(pc)} & \colhead{Exclusion\tablenotemark{a}} & \colhead{References}}
\startdata
\sidehead{\Large\textbf{AB~Doradus}}
PX~Vir & G5\,V & 13:03:49.46 & -05:09:45.9 & $-191.1$ & $0.9$ & $-218.7$ & $0.7$ & $-9.2$ & $0.2$ & $21.7$ & $0.4$ & VIS & 1,2,3,2\\
2MASS~J15534211--2049282 & M3.4 & 15:53:42.09 & -20:49:28.6 & $-10$ & $10$ & $-20$ & $10$ & $-7$ & $2$ & $330$ & $80$ & $\sigma_{\rm kin}$ & 4,5,4,4\\
HR~7214 & A4\,V & 19:03:32.25 & 01:49:07.6 & $22.6$ & $0.2$ & $-68.5$ & $0.3$ & $-23$ & $2$ & $54.9$ & $0.9$ & MST$_{0}$ & 6,2,6,2\\
LQ~Peg & K8\,V & 21:31:01.86 & 23:20:05.2 & $134.6$ & $0.1$ & $-144.9$ & $0.1$ & $-17$ & $1$ & $24.2$ & $0.1$ & VIS & 1,7,8,7\\
HIP~107948 & M3\,Ve & 21:52:10.51 & 05:37:33.7 & $106$ & $2$ & $-147$ & $1$ & $-15$ & $2$ & $31$ & $5$ & $\sigma_{\rm kin}$ & 9,9,9,2\\
\sidehead{\Large\textbf{$\boldsymbol\beta$~Pictoris}}
LP~353--51 & M3\,V\,e & 02:23:26.75 & 22:44:05.1 & $98.5$ & $0.2$ & $-112.5$ & $0.1$ & $13.0$ & $0.4$ & $27.1$ & $0.3$ & VIS & 1,7,10,7\\
HD~15115 & F4\,IV & 02:26:16.33 & 06:17:32.4 & $87.97$ & $0.04$ & $-50.35$ & $0.03$ & $0.8$ & $0.1$ & $48$ & $1$ & VIS & 11,7,12,7\\
EPIC~211046195 & M8.5 & 03:35:02.09 & 23:42:35.6 & $50$ & $10$ & $-60$ & $10$ & $16$ & $2$ & $42$ & $2$ & VIS & 4,5,4,4\\
\enddata
\end{deluxetable*}
\end{longrotatetable}

%% file: table_designations_short.tex
\startlongtable
\tabletypesize{\normalsize}
\begin{longrotatetable}
\begin{deluxetable*}{lllll}
\tablecolumns{5}
\tablecaption{List of bona fide members designations. \label{tab:designations}}
\tablehead{\colhead{Main} & \colhead{2MASS} & \colhead{AllWISE} & \colhead{Gaia} & \colhead{Other}}
\startdata
\sidehead{\Large\textbf{AB~Doradus}}
2MASS~J00192626+4614078 & J00192626+4614078 & J001926.39+461406.8 & $\cdots$ & $\cdots$\\
BD+54~144~A & J00455088+5458402 & J004551.02+545839.6 & 417565757132068096 & HD~4277~A,~HIP~3589~A\\
| BD+54~144~B & $\cdots$ & $\cdots$ & $\cdots$ & HD~4277~B,~HIP~3589~B\\
2MASS~J00470038+6803543 & J00470038+6803543 & J004701.09+680352.2 & 529737830321549952 & $\cdots$\\
G~132--51~B & J01034210+4051158 & J010342.24+405114.2 & 374400889126932096 & $\cdots$\\
| G~132--50 & J01034013+4051288 & J010340.24+405127.4 & 374400957846408192 & $\cdots$\\
| G~132--51~C & $\cdots$ & $\cdots$ & 374400893422315648 & $\cdots$\\
HIP~6276 & J01203226--1128035 & J012032.34--112805.2 & 2470272808484339200 & CD--12~243\\
G~269--153~A & J01242767--3355086 & J012427.84--335510.0 & 5015892473055253120 & $\cdots$\\
| G~269--153~B & $\cdots$ & $\cdots$ & 5015892473055253248 & $\cdots$\\
\enddata
\end{deluxetable*}
\end{longrotatetable}
\global\pdfpageattr\expandafter{\the\pdfpageattr/Rotate 0}